\documentclass{article} 
\usepackage[top=1in, bottom=1in, left=1in, right=1in]{geometry}
\usepackage{amsthm}
\usepackage{amsmath}
\usepackage{amssymb}
\usepackage{graphicx}
\usepackage{float}
\usepackage{subcaption}
\usepackage{color}
\usepackage{natbib}
\title{Function-on-Function Regression for the Identification of Epigenetic Regions Exhibiting Windows of Susceptibility to Environmental Exposures} 

\author{Michele Zemplenyi\thanks{Department of Biostatistics, Harvard T.H. Chan School of Public Health, Boston, Massachusetts, USA} \and
	Mark J. Meyer\thanks{Department of Mathematics and Statistics, Georgetown University, Washington D.C., USA} \and
	Andres Cardenas\thanks{Division of Environmental Health Sciences, School of Public Health, University of California, Berkeley, California, USA} \and
	Marie-France Hivert\thanks{Division of Chronic Disease Research Across the Lifecourse, Department of Population Medicine, Harvard Medical School and Harvard Pilgrim Health Care Institute, Boston, Massachusetts, USA} \thanks{Diabetes Unit, Massachusetts General Hospital, Boston, Massachusetts, USA} \and
	Sheryl L. Rifas-Shiman\footnotemark[4] \and
	Heike Gibson\thanks{Department of Environmental Health, Harvard T.H. Chan School of Public Health, Boston, Massachusetts} \and
	Itai Kloog\thanks{Department of Geography and Environmental Development, Faculty of Humanities and Social Sciences, Ben-Gurion University, Beer-Sheva, Israel} \and
	Joel Schwartz\footnotemark[6] \thanks{Channing Division of Network Medicine, Department of Medicine, Brigham and Women’s Hospital, Harvard Medical School, Boston, Massachusetts, USA} \and
	Emily Oken\footnotemark[4] \and
	Dawn L. DeMeo\thanks{Center for Chest Diseases, Brigham and Women's Hospital, Boston, Massachusetts, USA } \and
	Diane R. Gold\footnotemark[6] \footnotemark[8] \and
	Brent A. Coull\footnotemark[1]}

\begin{document}

\maketitle
\let\thefootnote\relax\footnotetext{\emph{Please address correspondence to Michele Zemplenyi; email: mzemplenyi@g.harvard.edu.}}
\section*{Abstract}
The ability to identify time periods when individuals are most susceptible to exposures, as well as the biological mechanisms through which these exposures act, is of great public health interest. Growing evidence supports an association between prenatal exposure to air pollution and epigenetic marks, such as DNA methylation, but the timing and gene-specific effects of these epigenetic changes are not well understood. Here, we present the first study that aims to identify prenatal windows of susceptibility to air pollution exposures in cord blood DNA methylation. In particular, we propose a function-on-function regression model that leverages data from nearby DNA methylation probes to identify epigenetic regions that exhibit windows of susceptibility to ambient particulate matter less than 2.5 microns (PM$_{2.5}$). By incorporating the covariance structure among both the multivariate DNA methylation outcome and the time-varying exposure under study, this framework yields greater power to detect windows of susceptibility and greater control of false discoveries than methods that model probes independently. We compare our method to a distributed lag model approach that models DNA methylation in a probe-by-probe manner, both in simulation and by application to motivating data from the Project Viva birth cohort. In two epigenetic regions selected based on prior studies of air pollution effects on epigenome-wide methylation, we identify windows of susceptibility to PM$_{2.5}$ exposure near the beginning and middle of the third trimester of pregnancy. \\
\\
\textbf{Keywords}: functional data analysis, wavelet regression, windows of susceptibility, epigenetics

\section{Introduction}

Recent epidemiological evidence supports the hypothesis that exposures during fetal development and in early life can lead to a variety of adverse birth and child health outcomes. The fetal \emph{in utero} environment can be altered by external factors such as the mother's diet or toxins to which she is exposed, thereby influencing the early development of a child at a time of heightened susceptibility. The National Institutes of Health, through its 2016 Environmental influences on Child Health Outcomes (ECHO) initiative, highlighted the importance of not only identifying child health outcomes associated with environmental exposures, but also of identifying sensitive developmental windows during which an exposure has increased association with a child's health outcomes.  Understanding when these windows of susceptibility occur and how they coincide with environmental exposures may shed light on the underlying biological pathways through which exposures act.  Ultimately, this can lead to the development of interventions that mitigate risks due to an exposure. 

One proposed biological pathway by which prenatal exposures could contribute to subsequent adverse health outcomes involves DNA methylation at cytosine-phosphate-guanine (CpG) sites.  DNA methylation is an epigenetic modification that can regulate gene expression. Previous studies have found associations between DNA methylation levels and environmental exposures such as particulate air pollution (\citealp{Gruzieva2019}; \citealp{Soberanes2012}; \citealp{Baccarelli2009}) and lead (\citealp{Bollati2010}; \citealp{Schneider2013}). \citet{Lee2018PFPE} found evidence linking \emph{in utero} PM$_{2.5}$ exposure to both hypermethylation of the \emph{GSTP1} gene in nasal epithelia and impaired early childhood lung function. In a cohort of elderly men in the Normative Aging Study, \citet{Lepeule2012} found an association between lower DNA methylation levels in several gene promoter regions and reduced lung function. While a large number of epigenetic studies based on adult populations have been published, due to the inherent difficulties and health risks of interrogating the epigenome of a developing child, relatively few prenatal epigenetic studies have been conducted. Thus, our understanding of which prenatal exposures affect the epigenome, which epigenetic regions are affected, and which time periods are most sensitive remains limited. 

Previous work to identify windows of susceptibility during pregnancy often involved regressing the outcome of interest on trimester-average exposures (TAEs). Either separate models for each of the three TAEs were fit, or a single model that jointly estimated the associations for each TAE was constructed (\citealp{Shah2011}; \citealp{Dadvand2013}). However, \citet{Wilson2017} demonstrated that trimester-specific effect estimates obtained using separate regression models can be biased and identify incorrect windows of susceptibility.  \citet{Wilson2017} showed that, while a joint model suffers less from these issues, a distributed lag model (DLM) significantly outperforms either of the TAE methods in this regard. A DLM models the association between an outcome and a finely sampled time-varying exposure by assuming that their relationship varies smoothly over time (\citealp{Schwartz2000}; \citealp{Zanobetti2000}). In the context of prenatal exposure, the DLM regresses a health outcome measured after the exposure period of interest against exposure measured at frequent, regular intervals throughout pregnancy. DLMs have been used to identify windows of susceptibility during which air pollution is associated with disrupted neurodevelopment (\citealp{Chiu2016}), childhood asthma (\citealp{Lavigne2019}; \citealp{Bose2017}; \citealp{Hsu2015}), reduced lung function (\citealp{Bose2018}), sleep disruption (\citealp{Bose2019}), and lower birth weights (\citealp{Wu2018}; \citealp{Darrow2011}). \citet{Warren2019} recently proposed critical window variable selection as an alternative to the DLM in the context of identifying windows during pregnancy when PM$_{2.5}$ is associated with an increased risk of preterm birth.

While DLMs and methods like critical window variable selection efficiently model the multivariate nature of exposure, further efficiency gains can be made by taking into account the dependence among multivariate outcomes. In mammalian genomes, DNA methyltransferase enzymes can co-methylate adjacent CpG sites, resulting in blocks of CpG sites with similar methylation statuses and genomic functionality (\citealp{Guo2017}). \citet{LeeKyu2017} proposed a Bayesian variable selection method for multivariate methylation outcomes that, through leveraging information on the covariance structure of outcomes, increases power to detect associations while maintaining a low false discovery rate. Additionally, \citet{LeeMorris2016} used the wavelet-based functional mixed model approach introduced by \cite{MorrisCarroll2006} to model DNA methylation outcomes jointly, thereby capturing correlations among neighboring probes as well as across samples. This yielded gains in efficiency and the ability to detect differentially methylated regions (DMRs).

\citet{LeeMorris2016} focused on detecting DMRs associated with a scalar exposure, such as cancer status.  Here, we consider the functional exposure setting where our goal is not only to find DMRs, but to simultaneously identify time periods during which an exposure is associated with these DMRs. In particular,  we are interested in the association between two functions: 1) cord-blood DNA methylation levels measured at birth as a function of CpG site position in the genome, and 2) maternal air pollution exposure as a function of time during pregnancy. To characterize the association surface between these two functions, we propose the use of Bayesian function-on-function regression (FFR), a recent extension of Morris and Carroll's [2006] wavelet-based functional model developed by \citet{Meyer2015}. FFR transforms both the DNA methylation and air pollution profiles to a basis space, fits a regression model in that space, and then performs an inverse transformation to present results on the original methylation scale.  The basis transformation affords the ability to capture spatial and time-varying correlations within the two functions, while the Bayesian approach simultaneously allows for the smoothing of the association surface via the prior specification as well as strict control for multiple testing. 

We fit the model using a Monte-Carlo Markov Chain (MCMC) procedure and then perform statistical inference while accounting for multiple testing using a Bayesian False Discovery Rate (BFDR) procedure for functional regression and a simultaneous band score (SimBaS) (\citealp{Meyer2015}). We demonstrate the efficiency gains and better control of false discoveries attained by the functional approach relative to DLMs applied on a site-by-site basis in simulation. We then take two biologically-interesting and significant sites reported in \citealp{Gruzieva2019}, the largest analysis to date of particulate matter exposure and DNA methylation in infants, and perform the first analysis of prenatal windows of susceptibility driving these associations. We perform the analysis using DNA methylation data from 412 mother-child pairs enrolled in the Project Viva birth cohort and daily PM$_{2.5}$ measurements recorded over the third trimester of pregnancy, a critical period in fetal somatic growth, as well as in neural, lung, endocrine, and immune system development (\citealp{Hill2019}). We identify two windows of susceptibility for CpG probes in the \emph{FAM13A} gene region and another window of susceptibility in the third trimester for a subset of CpG probes in the \emph{NOTCH4} region. While we present the functional model in the context of a specific outcome measure (DNA methylation) and exposure (air pollution), the method is flexible enough to analyze other types of outcome and predictor functions that vary spatially and/or temporally.

\section{Methylation and Exposure Data in Project Viva}
In this section we briefly describe the pre-birth cohort, DNA methylation data, and air pollution data sets to which we apply our method.

\subsection{Description of Project Viva}
Project Viva is a longitudinal study designed to examine the effect of maternal diet and other lifestyle factors during pregnancy on the mother's and child's health. Pregnant women were enrolled at their initial obstetric visit at Harvard Vanguard Medical Associates in Massachusetts from 1999-2002. Of the 2,128 mother-child pairs enrolled in the cohort, 485 had cord blood DNA methylation measurements that passed quality control. For a more detailed description of the Project Viva cohort see \citet{Oken2015}.

\subsection{DNA methylation data} \label{DNAmethylation}
Umbilical vein cord blood DNA was extracted using the Qiagen Puregene Kit (Valencia, CA) and bisulfite converted using the EZ DNA Methylation-Gold Kit (Zymo Research, Irvine, CA). Samples were randomly allocated to chips and plates and analyzed using Infinium HumanMethylation450 BeadChip arrays (Illumina, San Diego, CA) that probe approximately 485,000 CpG sites at a single nucleotide resolution. 

We adjusted for sample plate as technical batch by directly including the batch number as a scalar covariate in the FFR and DLM regression models. We modelled the logit-transformed percentage DNA methylation value, or M-values, where the percentage methylation value for an individual CpG site is the percentage of methylated cytosines over the sum of methylated and un-methylated cytosines at the 5C position for that probe (\citealp{Du2010}).  
 
We focused our analysis on two regions encompassing CpG sites identified in closely related work by \citealp{Gruzieva2019}. From their meta-analysis of the associations between prenatal exposure to particulate matter and DNA methylation in nine birth cohorts, one of which was Project Viva, \citealp{Gruzieva2019} identified 20 CpGs that were significantly associated with either prenatal PM$_{2.5}$ or PM$_{10}$ exposure. Two of these CpGs mapped to \emph{FAM13A} and \emph{NOTCH4}, genes previously associated with COPD and asthma, respectively (\citealp{Hobbs2017}; \citealp{Hancock2009}; \citealp{Li2013}). We selected CpG probes annotated to these two genes for our analysis.

\subsection{PM$_{2.5}$ data}
 Particulate matter with diameter less than 2.5 $\mu$m (PM$_{2.5}$) is released by vehicles and other industrial processes via the combustion of solid and liquid fuels. Estimated daily ambient PM$_{2.5}$ levels at the home addresses of the mothers enrolled in Project Viva were obtained using a hybrid satellite-based model that integrated remote sensing data and spatio-temporal land-use and meteorology data (\citealp{Kloog2011}; \citealp{Kloog2014}). In Project Viva we have demonstrated associations of estimated residential third trimester PM$_{2.5}$ or its black carbon component with subsequent reduced fetal growth measures at birth (\citealp{Fleisch2015}); childhood executive function and behavior (\citealp{Harris2016}); and allergen sensitization (\citealp{Sordillo2019}). Since recruiting for Project Viva began in 1999, but the satellite technology necessary to make daily predictions only became available in 2000, we do not have complete daily 1-by-1 km-resolved PM$_{2.5}$ exposure estimates throughout pregnancy for all mother-child pairs. Because of the known importance of the third trimester in fetal development, and in the interest of retaining a large number of subjects, we limited analysis to the 412 mothers for whom we had daily PM$_{2.5}$ measurements at their residential addresses for the last 90 days prior to delivery.  While analyzing the entirety of gestation would be preferable, exploring the last trimester of pregnancy does not preclude us from finding biologically meaningful windows of susceptibility; previous studies have found associations between prenatal air pollution in the third trimester and newborn health outcomes such as systolic blood pressure (\citealp{vanRossem2015}) and fetal growth (\citealp{Lamichhane2018}).

\section{Methods}
\subsection{Function-on-Function regression model}
Here, we describe the FFR model introduced by Meyer et al. (2015) in the context of identifying regions of the genome that exhibit windows of susceptibility to an exposure of interest. Suppose for each of $i = 1, \dots, n$ individuals we observe two functions: (1) the DNA methylation profile $y_i(s)$ on a common grid of CpG sites $s = 1, \dots, S$, and (2) the air pollution exposure profile over time, $x_i(t)$, $t = 1, \dots, T$.  For our application, $x_i(t)$ represents daily ambient PM$_{2.5}$ levels at each mother's residence. 

A FFR model to regress $y_i(s)$ on the functional predictor $x_i(t)$  is given by
\begin{equation} \label{fof}
	y_i(s) = \alpha(s) + \int_{t \in T} x_i(t) \beta(t, s) dt + e_i(s),
\end{equation}
	where we assume observation-specific Gaussian process errors $e_i(s) \sim \mathcal{GP}(0, \Sigma_e)$. The target of interest in Model (\ref{fof}) is the two-dimensional surface $\beta(t, s)$ that characterizes the association between exposure at any given time and DNA methylation at any given CpG site.

  If we stack row vectors by subject, then $\boldsymbol{Y}$ and $\boldsymbol{X}$ represent $n \times S$ and $n \times T$ matrices of observed DNA methylation and exposure profiles respectively. We can then represent Model (\ref{fof}) in matrix form as
	\begin{equation}\label{discrete}
	\boldsymbol{Y} = \boldsymbol{X\beta} + \boldsymbol{E},
	\end{equation}
	where  $\boldsymbol{\beta}$ is a $T \times S$ matrix of functional effects and $\boldsymbol{E}$ is a $n \times S$ matrix of model errors. The intercept $\alpha(s)$ can be incorporated into $\boldsymbol{\beta}$, but in practice and without loss of generality, we center and scale both $y_i(s)$ and $x_i(t)$ such that $\alpha(s)$ is zero in Model \ref{fof}.
	
	One possible approach to fitting Model (\ref{discrete}) is to fit each column of $\boldsymbol{Y}$ independently using a DLM. This approach regresses DNA methylation  at a particular CpG site against lagged exposure values over time for each site $s$ separately:
	
	\begin{equation} \label{dlm}
	y_{i,s} = \alpha + \int_{t \in T} x_i(t) \beta(t) dt + e_i.
\end{equation}
	
	This site-by-site DLM approach fails to borrow information across nearby, correlated CpG sites, likely reducing the efficiency of the method relative to a joint approach. Instead, we fit Model (\ref{discrete}) jointly by using a basis function transform approach  (\citealp{LeeMorris2016}; \citealp{Meyer2015}; \citealp{MorrisCarroll2006}). This involves first transforming $y(s)$ and $x(t)$ from the data space into a basis space. We fit the model in the basis space and then transform the parameter estimates back to the data space to conduct inference.

	\subsection{Discrete Wavelet Transform in Functional Regression}
    While a number of basis functions could be used to represent the observed functions, including splines, principal components (PCs), or Fourier series, we use wavelets as the transformation for both the DNA methylation and air pollution datasets. Wavelets have previously been used in genomic settings for the purposes of denoising high-throughput DNA copy number data (\citealp{HsuWavelets2005}), detecting histone modification enrichments (Mitra and Song, 2012), and identifying nucleosome position (Nguyen et al., 2013). For equally spaced data, such as daily air pollution measurements, the discrete wavelet transform (DWT) maps the data to the wavelet space in linear time. For unequally spaced data, like CpG site positions across chromosomes, we have a choice for how to perform the wavelet basis transform. We choose to perform the transformation treating the positions as if they were equally spaced for two reasons. First, \citet{MorrisCarroll2006} showed that the wavelet-based functional mixed model can flexibly estimate a complex covariance structure such as one that might arise from unequally spaced measurements. Second, \citet{Sardy1999}  compared four different approaches for handling unequally spaced data and found that the method that treats data as if it were evenly spaced performed as well as more computationally-expensive methods that account for unequal spacing.  
    Therefore, we proceed by treating CpG sites as if they were equally spaced as others have previously done for CpG site (\citealp{LeeMorris2016}) and DNA copy number data (\citet{Hsu2015}). 

	First, we apply the discrete wavelet transform (DWT) to each row of $\boldsymbol{Y}$, giving an $n \times S^*$ matrix of wavelet basis coefficients, $\boldsymbol{Y^*}$, which represents the methylation data in the wavelet space. Each wavelet coefficient is double-indexed by $(j,k)$ with frequencies indexed by $j$ and locations indexed by $k$. For the exposure data we similarly perform the DWT on each row of $\boldsymbol{X}$, giving an $n \times T^*$ matrix $\boldsymbol{X^*}$. Applying the DWT to $\boldsymbol{Y}$ is equivalent to post-multiplication by the $S \times S^*$ wavelet transform matrix $\boldsymbol{\Omega}'$, $\boldsymbol{Y^*} = \boldsymbol{Y}\boldsymbol{\Omega}'$, where $\boldsymbol{\Omega}'$ contains the wavelet basis functions evaluated on the grid $S$. Similarly, if the $T \times T^*$ matrix $\boldsymbol{\Phi}'$ contains the wavelet basis functions evaluated on the grid $T$, then applying the DWT to $\boldsymbol{X}$ is equivalent to  $\boldsymbol{X^*} = \boldsymbol{X}\boldsymbol{\Phi}'$.  
	
	Our transformation strategy differs slightly from earlier implementations. Since dimension reduction in $\boldsymbol{X^*}$ has important computational benefits, others have performed an additional PCA step on the wavelet space exposure data before fitting the model \citep{Meyer2015}. However, we forego the computational advantage of compressing the exposure data since simulations showed that compression can distort the signal in certain areas of the association surface for our setting. 
	
	After transforming both the response and exposure data, we arrive at our model in the wavelet space: 
	\begin{equation} \label{basis}
		\boldsymbol{Y^*} = \boldsymbol{X^*\beta^*} + \boldsymbol{E^*},
	\end{equation}
	where $\boldsymbol{E^*} \sim \mathcal{MN}(\boldsymbol{0}, \boldsymbol{I}, \boldsymbol{C^*})$ and $\boldsymbol{I}$ is the appropriately sized identity matrix. The whitening property of the wavelet transform, discussed in \citet{Johnstone1997}, allows us to assume that wavelet coefficients within a given curve are independent across $j$ and $k$. Thus, we assume a diagonal structure for $\boldsymbol{C^*}$, the between-column covariance of the methylation data in the wavelet space (\citealp{MorrisCarroll2006}). Importantly, this assumed independence in the wavelet space does not imply independence in the data space; in fact, heterogeneous variances across the wavelet scales and locations $(j,k)$ induce correlations in the data space (\citealp{MorrisCarroll2006}).   
	
	The independence assumption in the wavelet space allows us to view Model (\ref{basis}) as $S^*$ separate models, one for each column of $\boldsymbol{Y^*}$. Thus, the model for each column (equivalently, Y-space wavelet coefficient) is
	\begin{equation}\label{column_model}
		\boldsymbol{y^*}_{(j,k)} =\boldsymbol{X^*}\beta^*_{(j,k)} + \boldsymbol{e^*}_{(j,k)},
	\end{equation}  
	where $\boldsymbol{y^*}_{(j,k)}$ and $\boldsymbol{e^*}_{(j,k)}$ are $n \times 1$, $\boldsymbol{X^*}$ is $n \times T^*$, and $\beta^*_{(j,k)}$ is $T^* \times 1$. We fit Model (\ref{column_model}) in a similar fashion to \citet{Meyer2015}. We place a spike-and-slab prior on each $\beta^*_{(p,j,k)}$, where $p$ indexes the wavelet coefficients of the transformed exposure data, $p = 1, \dots, T^*$:
	\begin{equation}
		\beta^*_{(p,j,k)} \sim \gamma_{(p,j,k)}\mathcal{N}(0, \tau_{p,j}) + (1-\gamma_{(p,j,k)})d_0, \; \; \gamma_{(p,j,k)} \sim \text{Bern}(\pi_{p,j})).
	\end{equation}
This prior is a mixture of a normal distribution and a point-mass at zero, $d_0$, with regularization parameters $\tau_{p,j}$ and $\pi_{p,j}$ estimated using an Empirical Bayes-type approach. This prior specification is consistent with other wavelet-based functional models (\citealp{MorrisCarroll2006}; \citealp{Malloy2010}).

	We generate posterior samples for the coefficient surface $\boldsymbol{\beta^*}$ in the wavelet space and then project back to the data space through two inverse discrete wavelet transforms, $\boldsymbol{\beta} = \boldsymbol{\Phi{'}\beta^*\Omega}$, which ultimately yields posterior samples of the discretized coefficient surface $\boldsymbol{\beta}$. We then perform inference on $\boldsymbol{\beta}$ as described in Section 2.4. We perform computations in MATLAB (version 2017a). Code for running the above model will be available soon through the BayesFMM Github repository.
	
	\subsection{Incorporation of Scalar Covariates}
		Model (\ref{fof}) can be extended to include scalar covariates, scalar-by-function interactions, and subject-specific random effect functions \citep{Meyer2015}. In the Viva analysis, we adjust for scalar covariates $\{w_a, a = 1, \dots, q\}$ using 
	\begin{equation}\label{fofscalar}
		y_i(s) = \alpha(s) + \sum_{a=1}^{q} w_{ia}\gamma_a(s) + \int_{t \in T} x_i(t) \beta(t, s) dt + e_i(s),
	\end{equation}	
	  where $\gamma_{a}(s)$ are functional coefficients for scalar predictors. These coefficients are typically of less interest than the coefficient surface $\beta(t,s)$. Because they are not functional data, we do not transform scalar covariates into the wavelet space prior to fitting the model. 
	  	  
	\subsection{Posterior Functional Inference} 
	In order to account for the multiple comparisons that occur when testing coefficients corresponding to all sites and exposure times within an analysis, we use two posterior functional inference procedures. The first is the Bayesian false discovery rate (BFDR), which was originally proposed by \citet{Muller2006} and subsequently extended to the functional regression setting (\citealp{Morris2008}; \citealp{Malloy2010}; \citealp{Meyer2015}). The second approach uses joint credible bands to detect significantly differentially methylated loci while controlling the experiment-wise error rate (\citealp{Meyer2015}).
	
	\subsubsection{Bayesian FDR}
	Interest focuses on detecting a biologically meaningful effect size of at least $\delta$. For $M$ MCMC samples, let $\beta^{(m)}(t,s)$ be one posterior sample of the coefficient surface for sample $m$, $m = 1, \dots, M$. We can compute the point-wise posterior probability of a $\delta$-sized intensity change on the $T \times S$ grid determined by each time-point $t = 1, \dots, T$ and genomic loci $s = 1. \dots, S$ via
	\[
	p(t,s) = Pr\{|\beta(t,s)| > \delta | \mathbf{Y}\} \approx \frac{1}{M}\sum_{m=1}^{M} I\{|\beta^{(m)}(t,s)| > \delta \},
	\]
	where $I$ is the indicator function. High values of $p(t,s)$ provide stronger evidence for a true discovery at a specific point on the surface; correspondingly, the quantity $1- p(t,s)$ can be interpreted as a local false discovery rate at a given location (\citealp{Morris2008}). A location on the coefficient surface is flagged as significant if $p(t,s) > \nu_\alpha$, where $\nu_\alpha$ is a threshold and $\alpha$ is a pre-specified global FDR-bound. The threshold $\nu_\alpha$ is constructed such that on average we expect at most $\alpha$ percent of flagged sites to be false positives. We do this by first sorting $\{p(t,s), t = 1, \dots, T, s= 1, \dots, S\}$ in descending order across all locations to obtain $\{p_{(r)}, r=1,\dots,R\}$, where $R = TS$. We then define $\lambda = \max[r^*: \frac{1}{r^*} \sum_{r=1}^{r^*} (1-p_{(r)}) \leq \alpha]$ and set the cutoff for flagging significant coefficients as $\nu_\alpha = p_{(\lambda)}$. 

	\subsubsection{Joint Credible Bands and Simultaneous Band Scores} 
	 In applications where there is not consensus on a biologically meaningful value for $\delta$, it is useful to consider constructing joint credible bands. Joint credible bands have the benefit of not requiring specification of $\delta$, while still providing a means of controlling the experiment-wise error rate. A $100(1-\alpha)$\% credible band of $\beta(t,s)$ must satisfy
	 \begin{equation}
	 	Pr\{L(t,s) \leq \beta(t,s) \leq U(t,s) \; \forall \; t \in T, s\in S\} \geq 1-\alpha,
	 \end{equation}
where $L(t,s)$ and $U(t,s)$ are the lower and upper bounds of the band respectively (\citealp{Ruppert2003}). If we assume approximate normality of the posterior samples, an interval satisfying this constraint is 
\begin{equation}
	I_\alpha(t,s) = \widehat{\beta}(t,s) \pm q_{(1-\alpha)}[SD\{\widehat{\beta}(t,s)\}].
\end{equation}
Here, $\widehat{\beta}(t,s)$ is the mean for a given position $(t,s)$ taken over all $M$ MCMC samples and $SD\{\widehat{\beta}(t,s)\})$ is the standard deviation over all $M$ MCMC samples divided by the factor $A(M, \rho_{MCMC})$, where $\rho_{MCMC}$ is an estimate of the lag autocorrelation in the samples, and $A$ is the bias correction factor described in \citet{Anderson1971} (\citealp{Meyer2015}; \citealp{LeeMorris2016}). The variable $q_{(1-\alpha)}$ is the $(1-\alpha)$ sample quantile taken over $M$ samples of 
\begin{equation}
	Z^{(m)} = \max_{t \in T, \; s \in S} \left|\frac{\beta^{(m)}(t,s)-\hat{\beta}(t,s)}{SD\{\hat{\beta}(t,s)\}}\right|.
\end{equation}
Suppose one constructs joint credible bands for a range of different values of $\alpha$ and finds for each location $(t,s)$ the minimum level $\alpha$ for which the $(1-\alpha)$\% joint credible band excludes zero. \citet{Meyer2015} refers to this minimum $\alpha$ level, $p_{SimBaS}(t,s) = \min\{\alpha: 0 \notin I_\alpha(t,s)\}$, as the simultaneous band score (SimBaS) for each location $(t,s)$. For any specific $\alpha$, we  flag all $(t,s)$ for which $p_{SimBaS}(t,s) < \alpha$ as significant, meaning that the $(1-\alpha)$\% joint credible interval at those locations does not include zero.

\section{Simulation}
\begin{figure}
\begin{tabular}{cc}
	\includegraphics[width=0.49\linewidth]{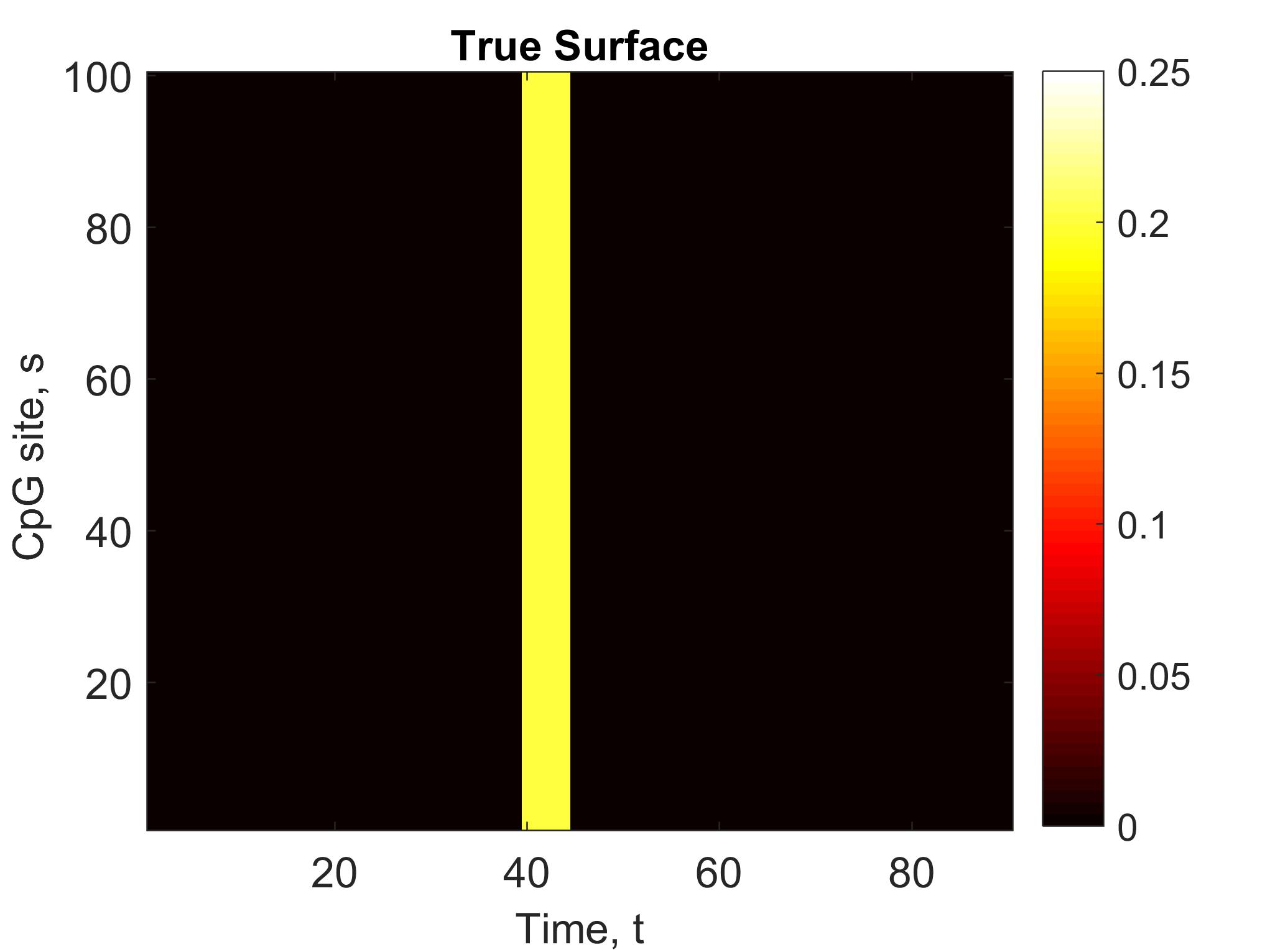} &
	\includegraphics[width=0.49\linewidth]{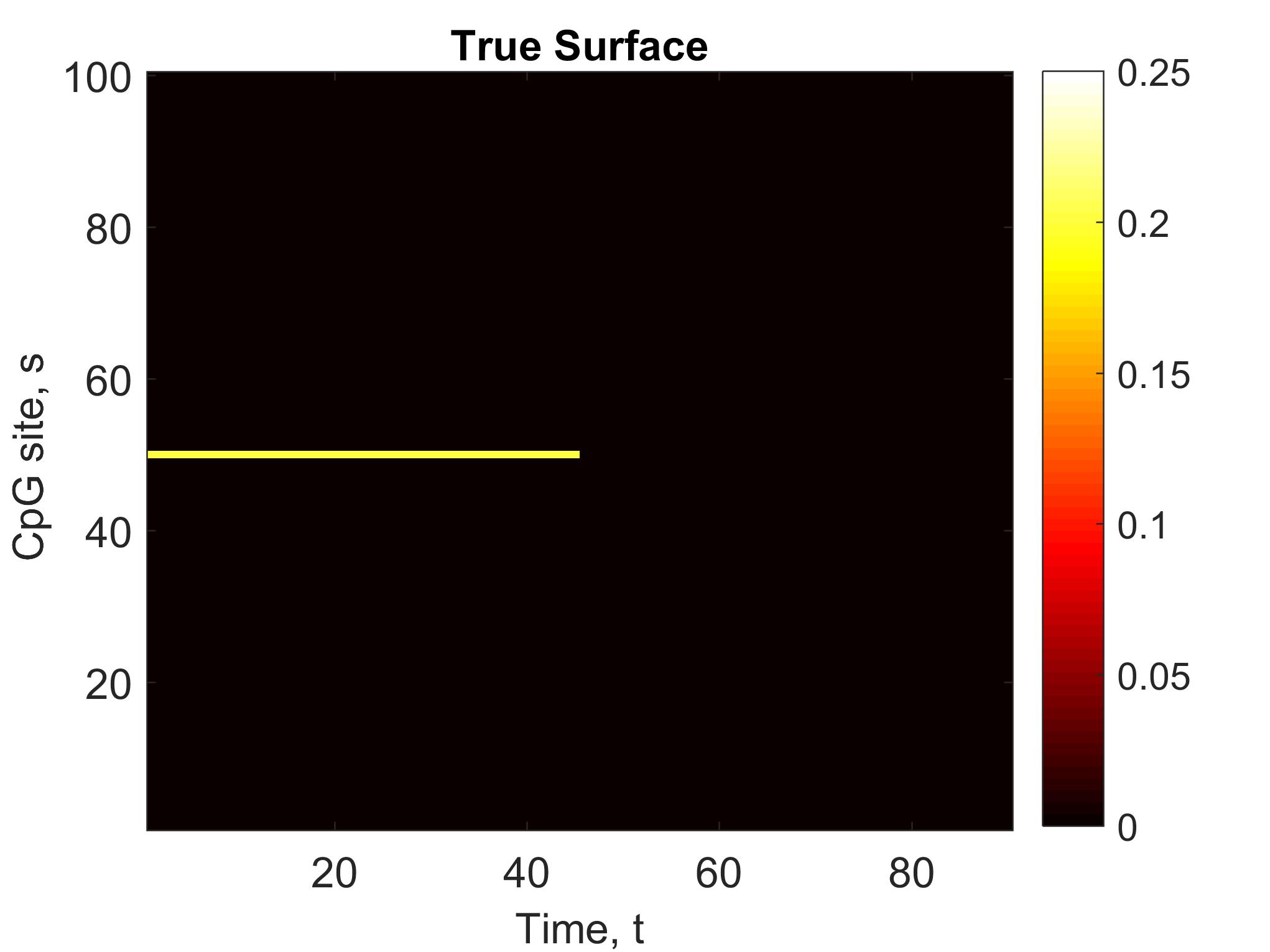} 
\end{tabular}		
\caption{True association surfaces $\boldsymbol{\beta}$. Left: vertical band setting where $\boldsymbol{\beta} = 0.2$ in the region $T \times S = \{(T,\,S): T \in \{40, \dots, 44\}, S \in \{1, \dots, 100\} \}$ and  $\boldsymbol{\beta} = 0$ everywhere else. Right: horizontal band setting where $\boldsymbol{\beta} = 0.2$ in the region $T \times S = \{(T,\,S): T \in \{1, \dots, 45\}, S = 50 \}$ and  $\boldsymbol{\beta} = 0$ everywhere else. }
\label{fig:trueSurface}
\end{figure}

	We performed two simulation studies to compare the operating characteristics of the FFR and site-by-site DLM approaches in the context of pre-birth studies of air pollution and DNA methylation. For the first study, we set the true surface of association, $\boldsymbol{\beta}$, over a $T = 90$ by $S = 100$ grid to be $\boldsymbol{\beta} = 0.2$ in the region $T \times S = \{(T,\,S): T \in \{40, \dots, 44\}, S \in \{1, \dots, 100\} \}$ and  $\boldsymbol{\beta} = 0$ everywhere else. This corresponds to a ``vertical band" of association (Figure \ref{fig:trueSurface}) representing a biologically plausible association between air pollution exposure and DNA methylation in which changes in air pollution halfway through the study period are associated with changes in DNA methylation in a genomic region spanning 100 CpG sitess.  Next, we randomly sampled PM$_{2.5}$ pollutant profiles from individuals in the Project Viva cohort and used these randomly sampled profiles as the exposure curves $\boldsymbol{x}_i$. Motivated by the 412 mother-child pairs for whom we had data in the Project Viva cohort, we used a sample size of $N=400$ subjects for the simulations.  Using measured exposure data, instead of simulating exposure curves, allowed us to capture a realistic correlation structure among daily pollutant exposures during pregnancy. (See the Appendix for a sample of the observed PM$_{2.5}$ exposure curves used in the simulation study and data application.) We then generated response curves $\boldsymbol{y}_i$ using $\boldsymbol{y}_i = \boldsymbol{x}_i\boldsymbol{\beta} + \boldsymbol{e}_i$, where $\boldsymbol{y}_i$ is a row vector of length 100, $\boldsymbol{x}_i$ is a row vector of length 90, $\boldsymbol{\beta}$ is the 90 $\times$ 100 matrix described above, and $\boldsymbol{e}_i$ is a row vector of length 100. For the model errors $\boldsymbol{e}_i$, we generated data using Gaussian Processes with auto-regressive 1 covariance structures which were then scaled by factors of $\sigma^2_e \in \{4,16,64\}$ to create three scenarios of varying noise levels. We defined the signal-to-noise ratio, STNR, as the pointwise effect in the non-zero region of the surface ($\boldsymbol{\beta} = 0.2$) expressed as a proportion of $\sigma_e$. Using this definition, the three STNRs we considered were STNR $\in \{0.10, 0.05, 0.025\}$. 
	
	\noindent For each scenario, we generated 100 simulated data sets and fit the FFR to obtain 100  $\boldsymbol{\widehat{\beta}}_\text{FFR}$ estimates of the association surface. To fit the models, we transformed the data to the wavelet space using Daubechies wavelets with six levels of decomposition, four vanishing moments, and zero-padding. We drew 2000 posterior samples and discarded the first 1000 samples.
	
\begin{figure}
\begin{tabular}{ccc}
	\includegraphics[width=0.33\linewidth]{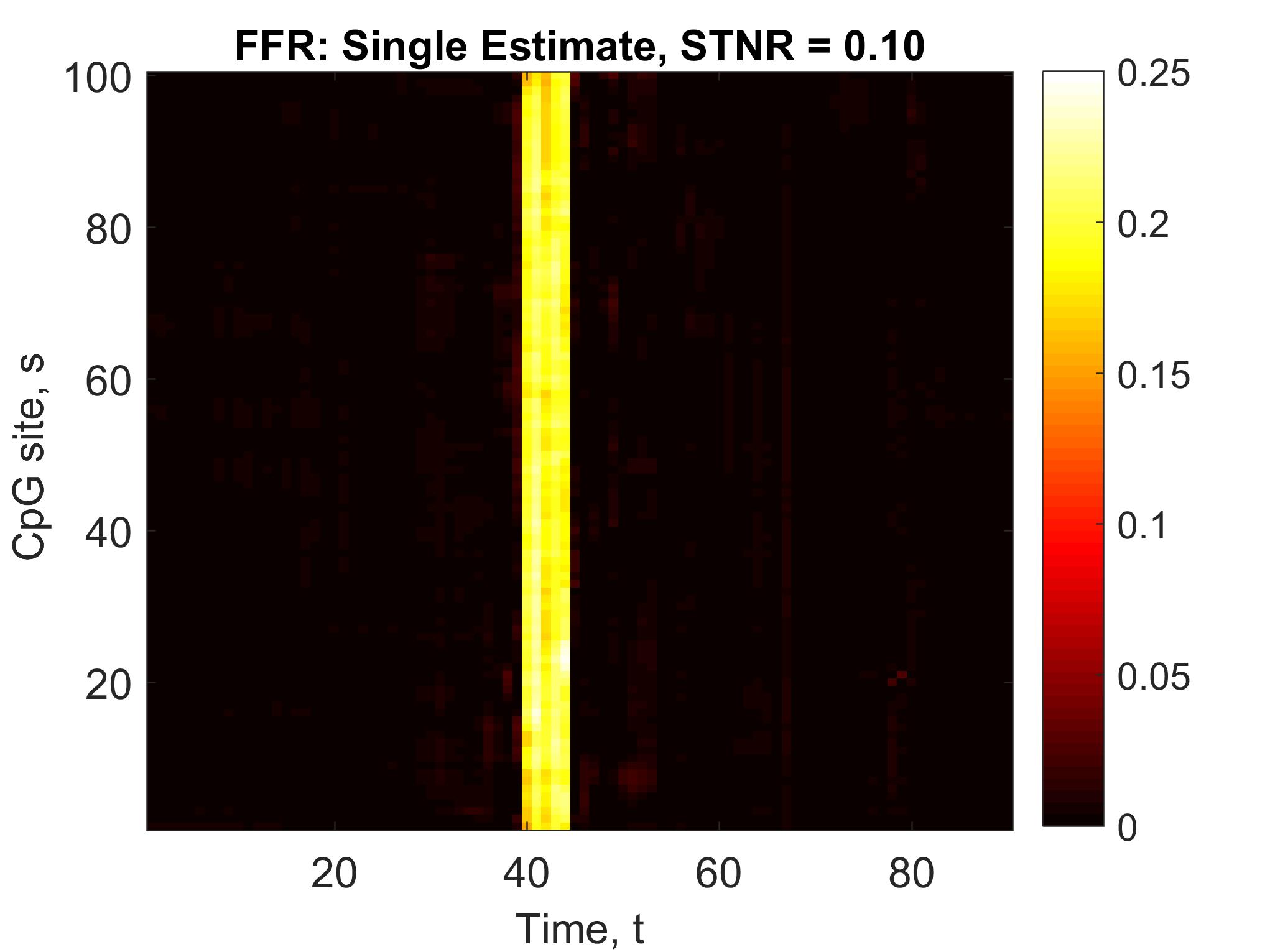} &
	\includegraphics[width=0.33\linewidth]{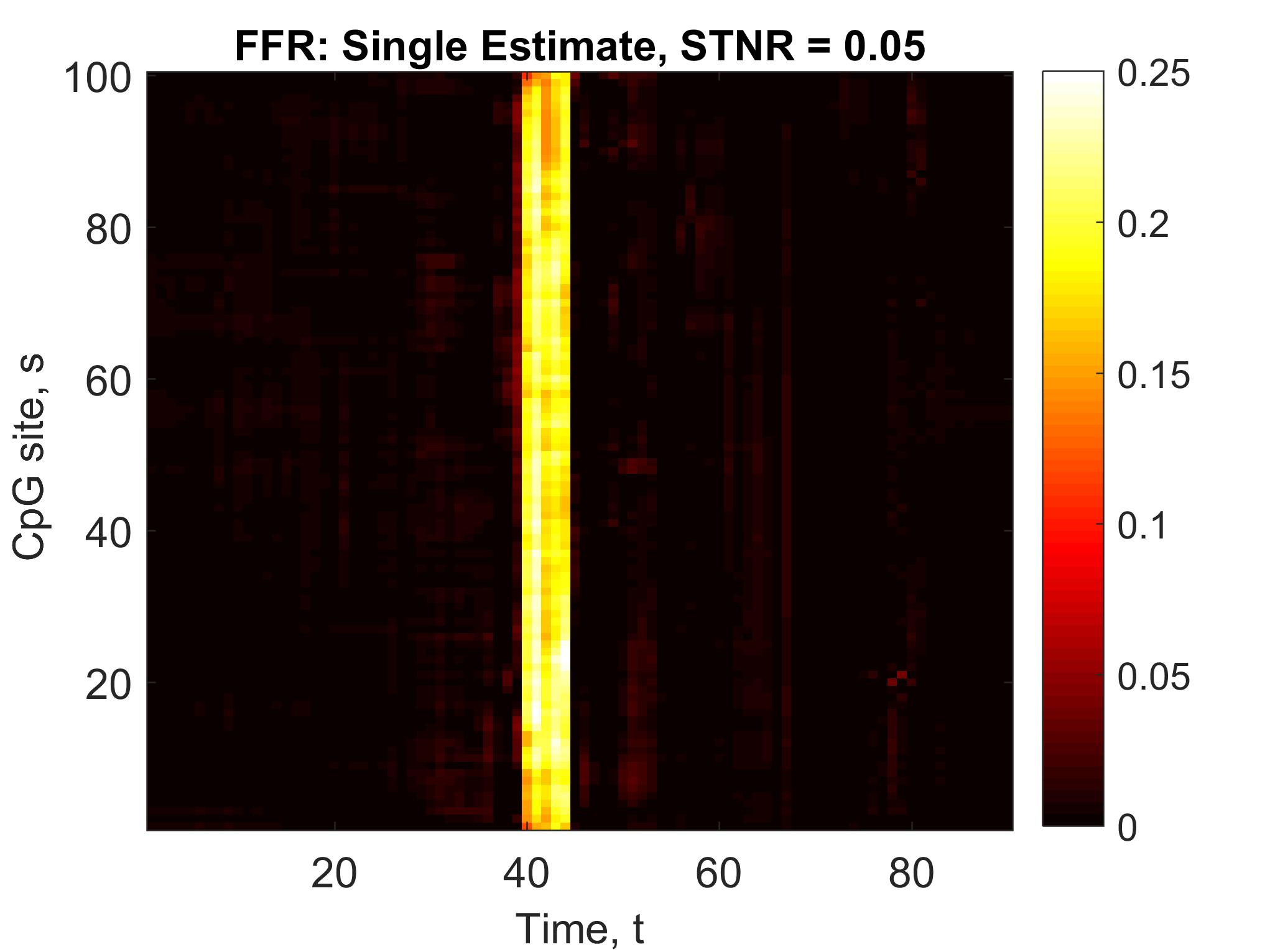} &
	\includegraphics[width=0.33\linewidth]{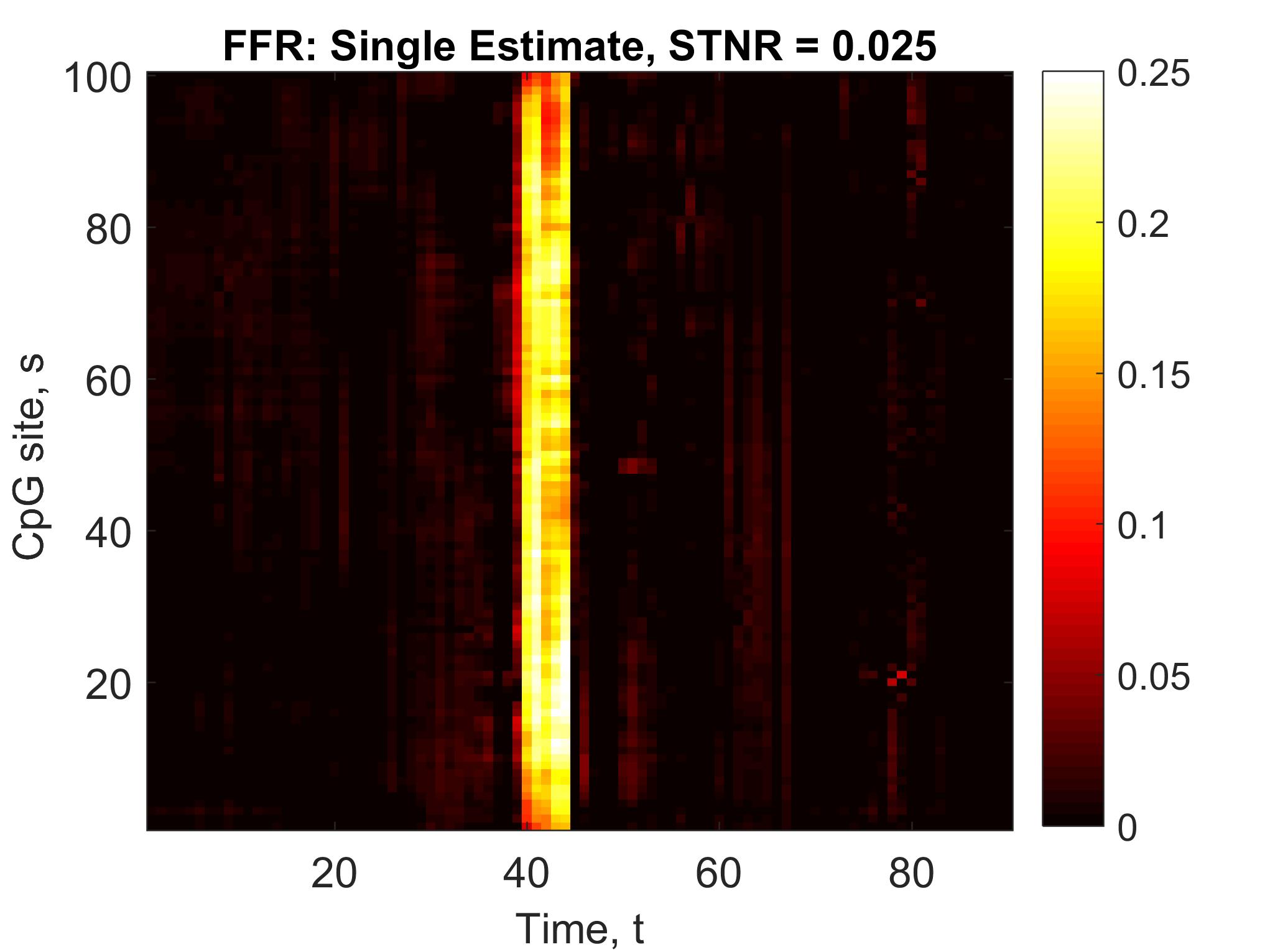} \\
	\includegraphics[width=0.33\linewidth]{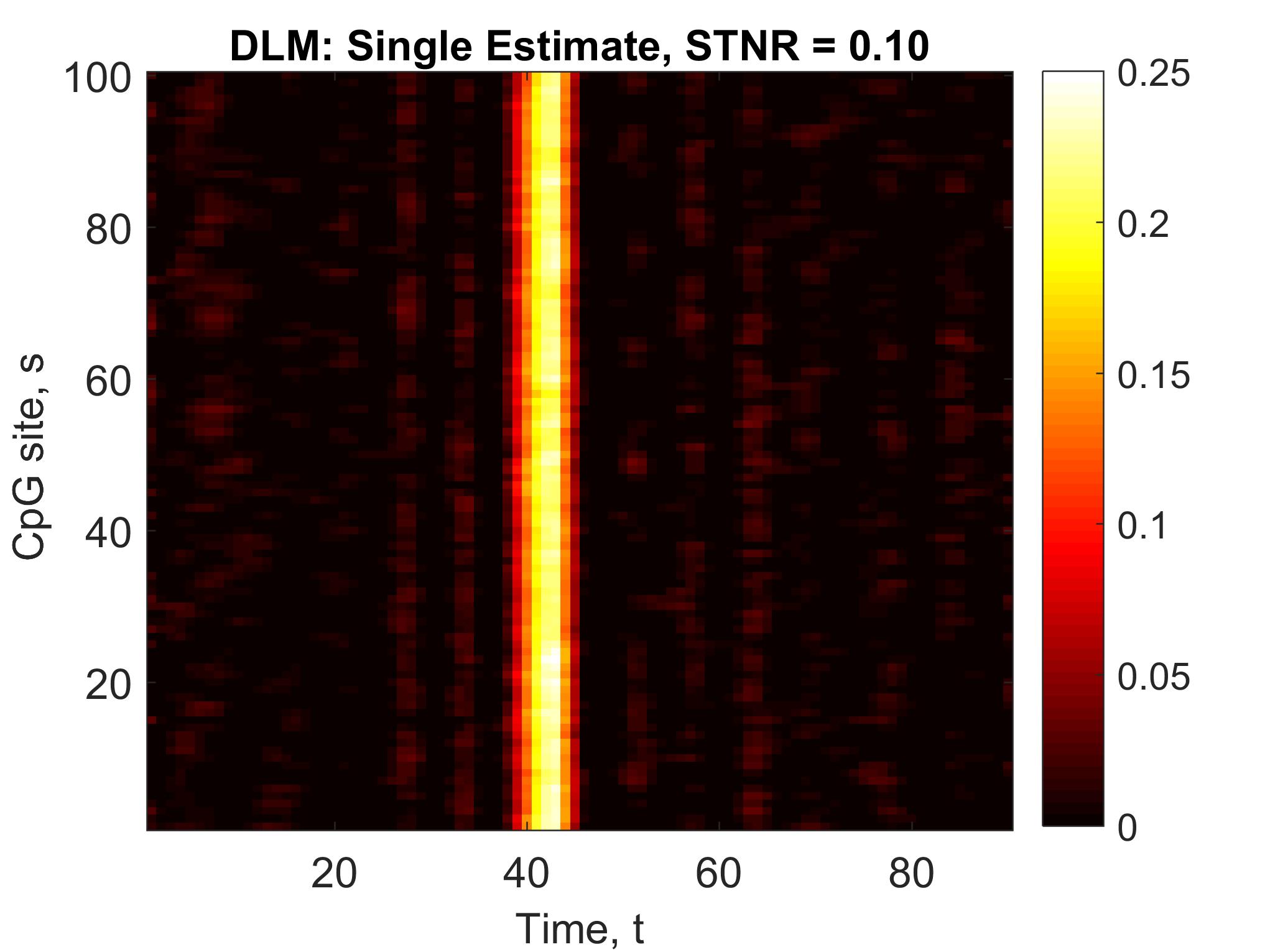} &
	\includegraphics[width=0.33\linewidth]{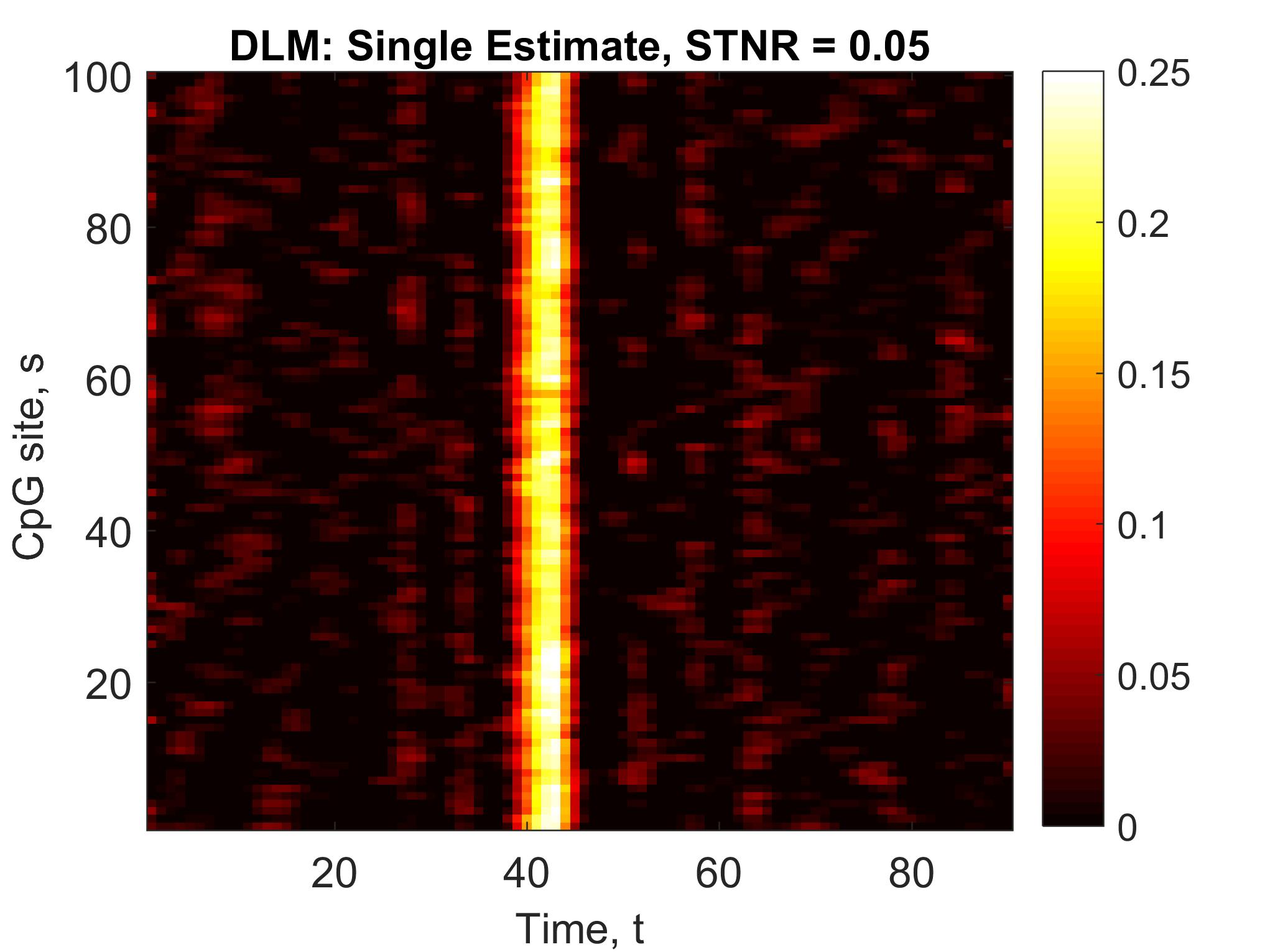} &
	\includegraphics[width=0.33\linewidth]{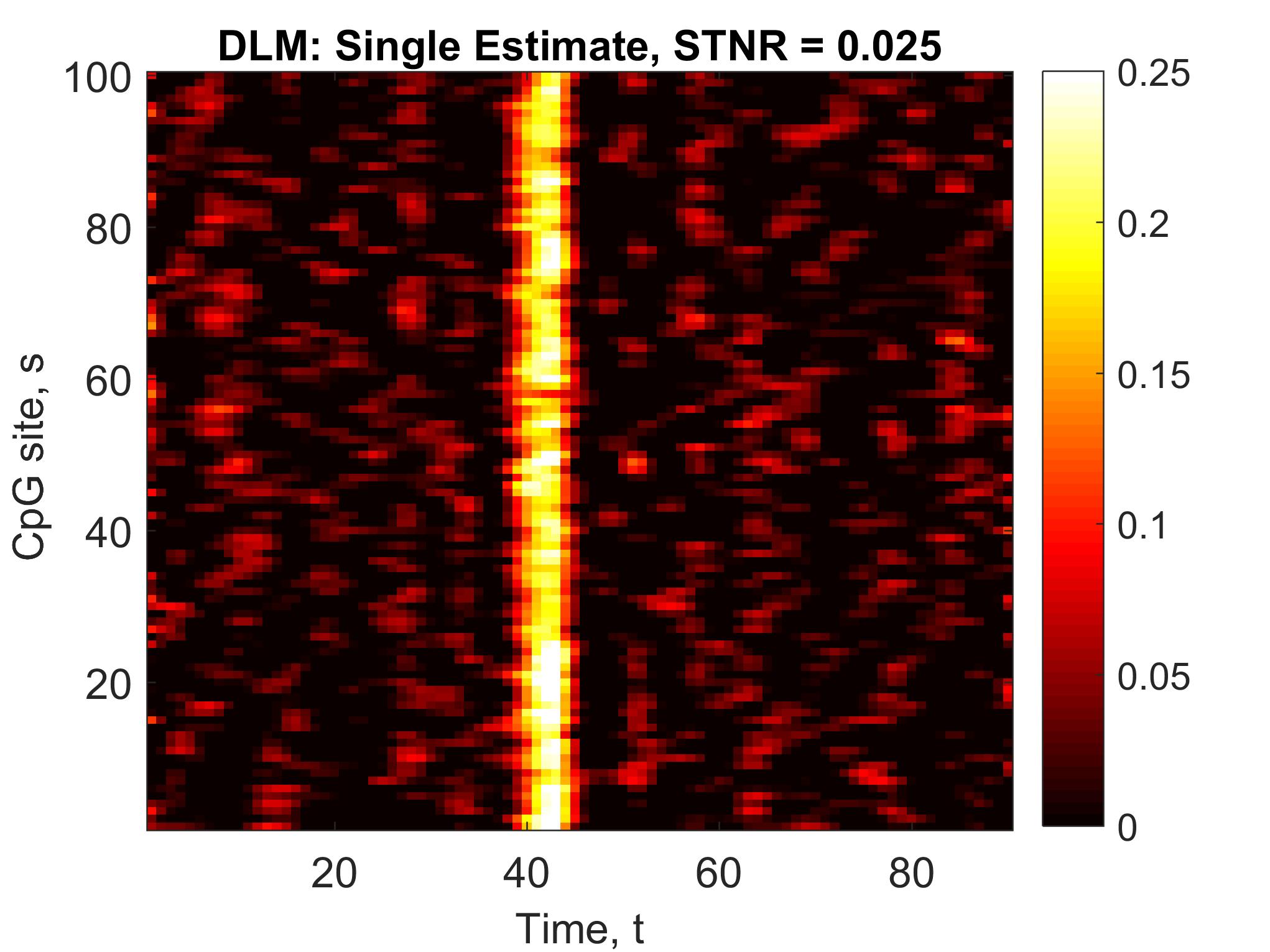}
\end{tabular}		
\caption{Heat maps of a single estimated association surface for the FFR method (top panel) and site-by-site DLM method (bottom panel).}
\label{fig:singleEstimate}
\end{figure}

	Fitting the site-by-site DLM to the simulated data sets required additional steps. Recall that the FFR takes a functional response and functional exposure as inputs and estimates a two-dimensional surface of association, $\boldsymbol{\beta}(t,s)$, whereas the DLM takes a scalar response and functional exposure as inputs and estimates a curve of association $\boldsymbol{\beta}(t)$ for each site. To compare the FFR and DLM approaches, we fit separate DLMs for each of the $S = 1, \dots, 100$ probes and concatenated the results, in essence stacking the DLM-estimated curves one behind the other to create a surface, $\boldsymbol{\widehat{\beta}}_\text{DLM}$, analogous to the FFR-estimated surface, $\boldsymbol{\widehat{\beta}}_\text{FFR}$. We used the \texttt{regimes} R package to fit the DLMs. Note that the key difference between $\boldsymbol{\widehat{\beta}}_\text{FFR}$ and $\boldsymbol{\widehat{\beta}}_\text{DLM}$ is that we fit  $\boldsymbol{\widehat{\beta}}_\text{FFR}$ using information from all sites simultaneously, whereas we constructed $\boldsymbol{\widehat{\beta}}_\text{DLM}$ using a separate model fit for each site. Figure \ref{fig:singleEstimate} displays heat maps of the estimated association surfaces for both methods across the three STNR scenarios. Results for the estimated surfaces averaged over 100 datasets can be found in the Appendix.

Figure \ref{fig:singleEstimate} shows that both methods estimated the true surface relatively well even in situations where the magnitude of the noise was 10-40 times larger than that of $\boldsymbol{\beta}$ in the signal region. However, visually we see that FFR produced sharper estimates of the association surface. In particular, the edges of the vertical band are clearly delineated and the null region is more accurately estimated in the FFR heat maps than in the DLM heat maps. 

\begin{figure}
\begin{tabular}{ccc}
	\includegraphics[width=0.33\linewidth]{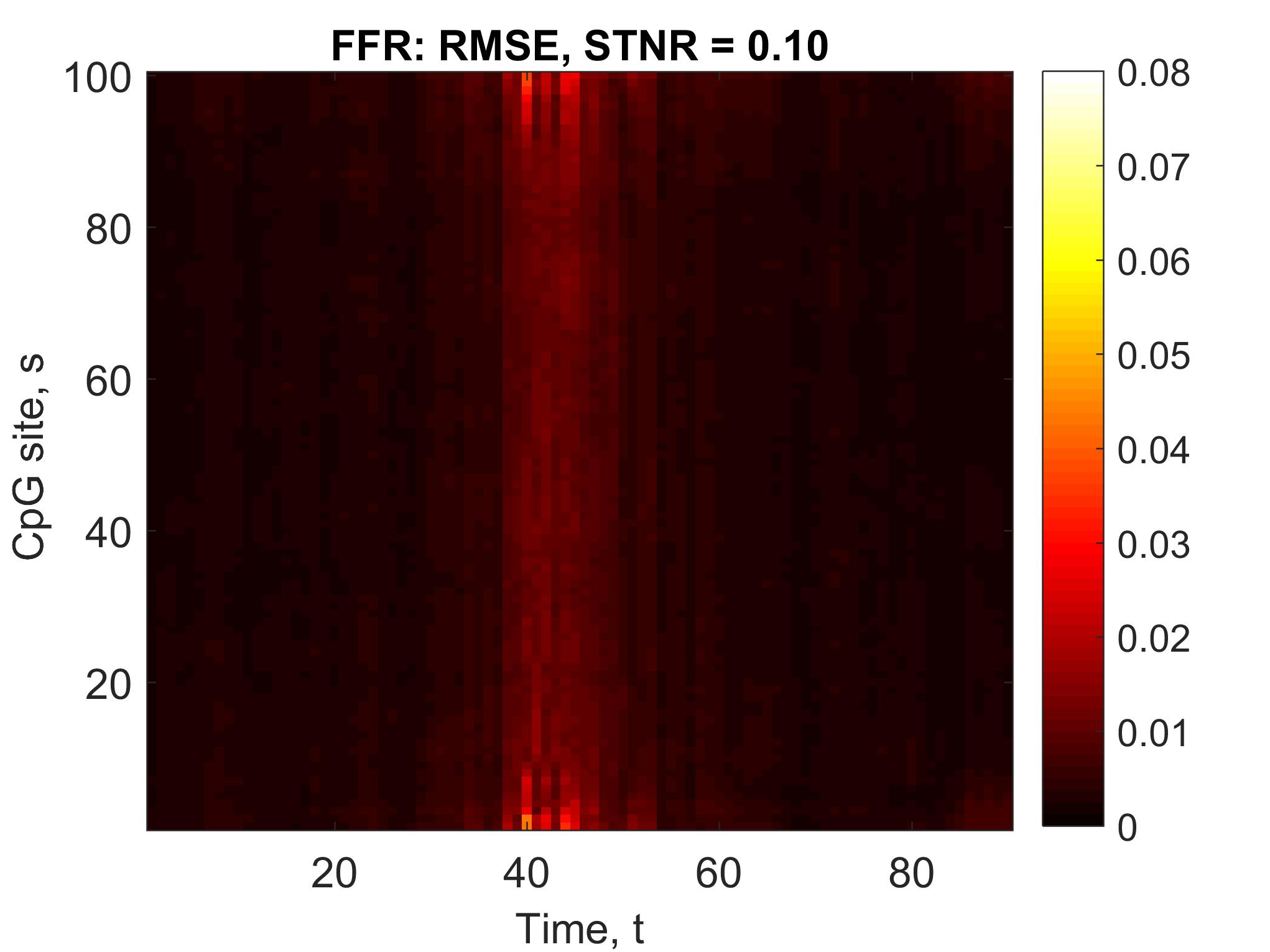} &
	\includegraphics[width=0.33\linewidth]{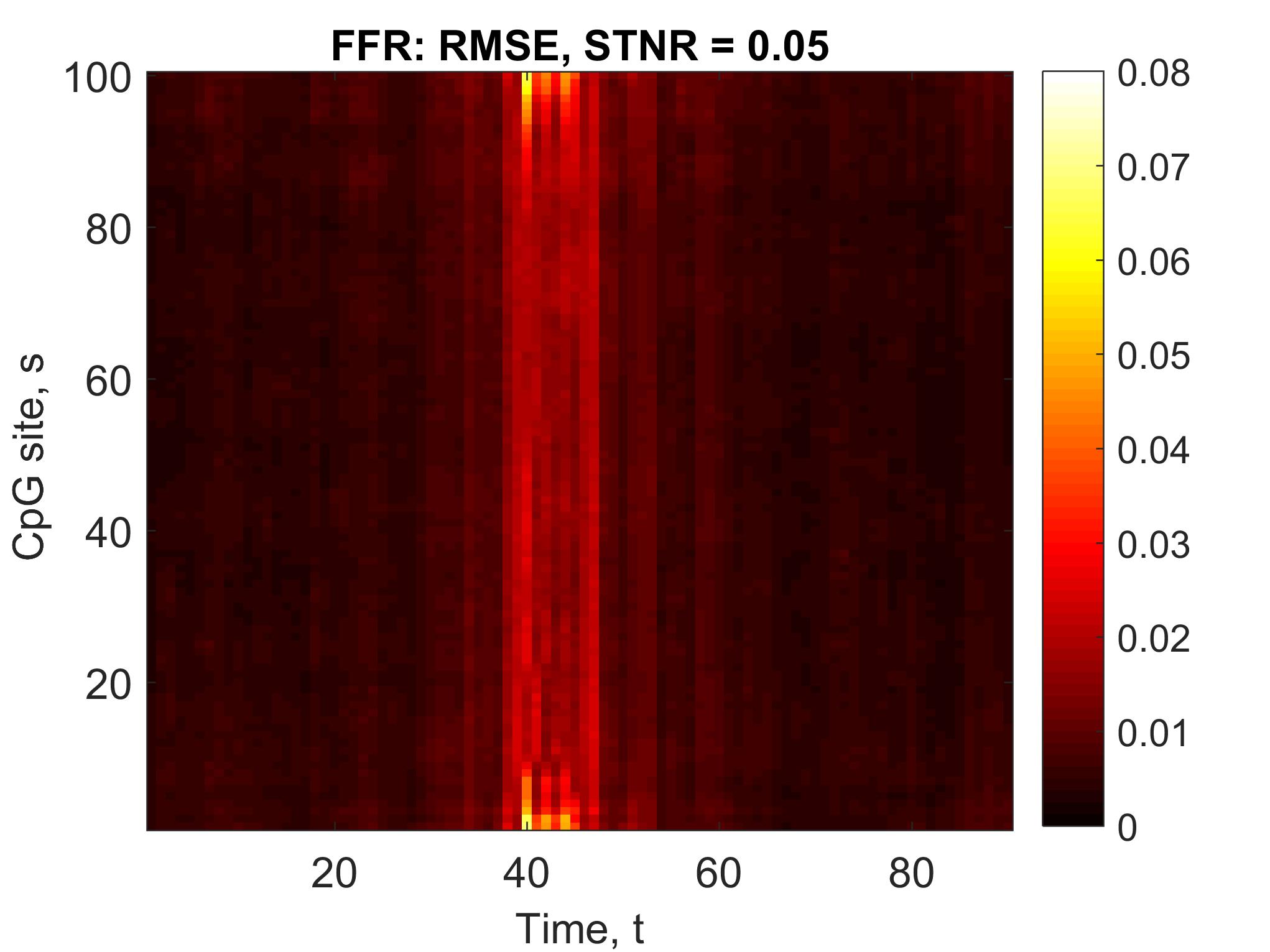} &
	\includegraphics[width=0.33\linewidth]{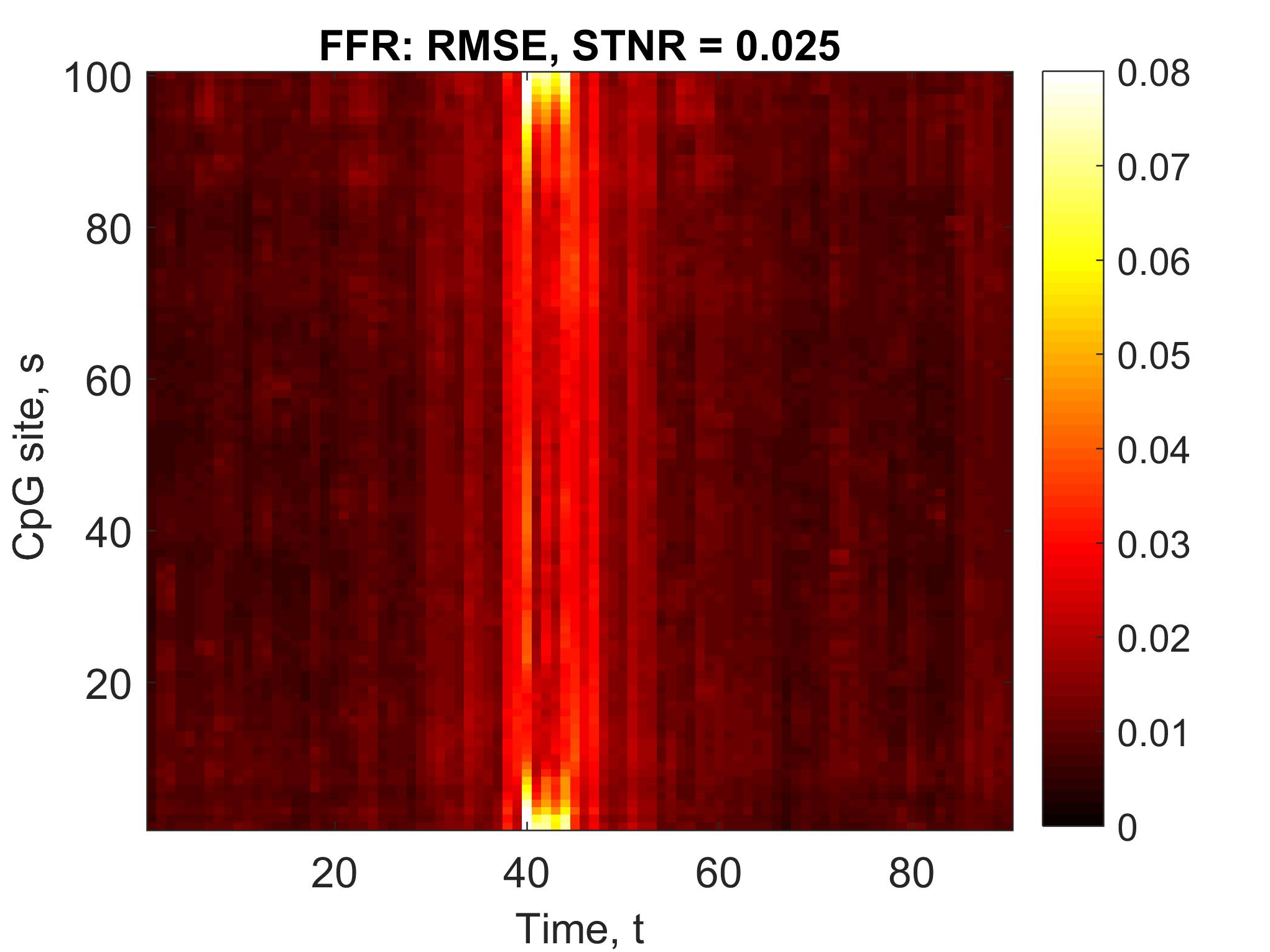} \\
	\includegraphics[width=0.33\linewidth]{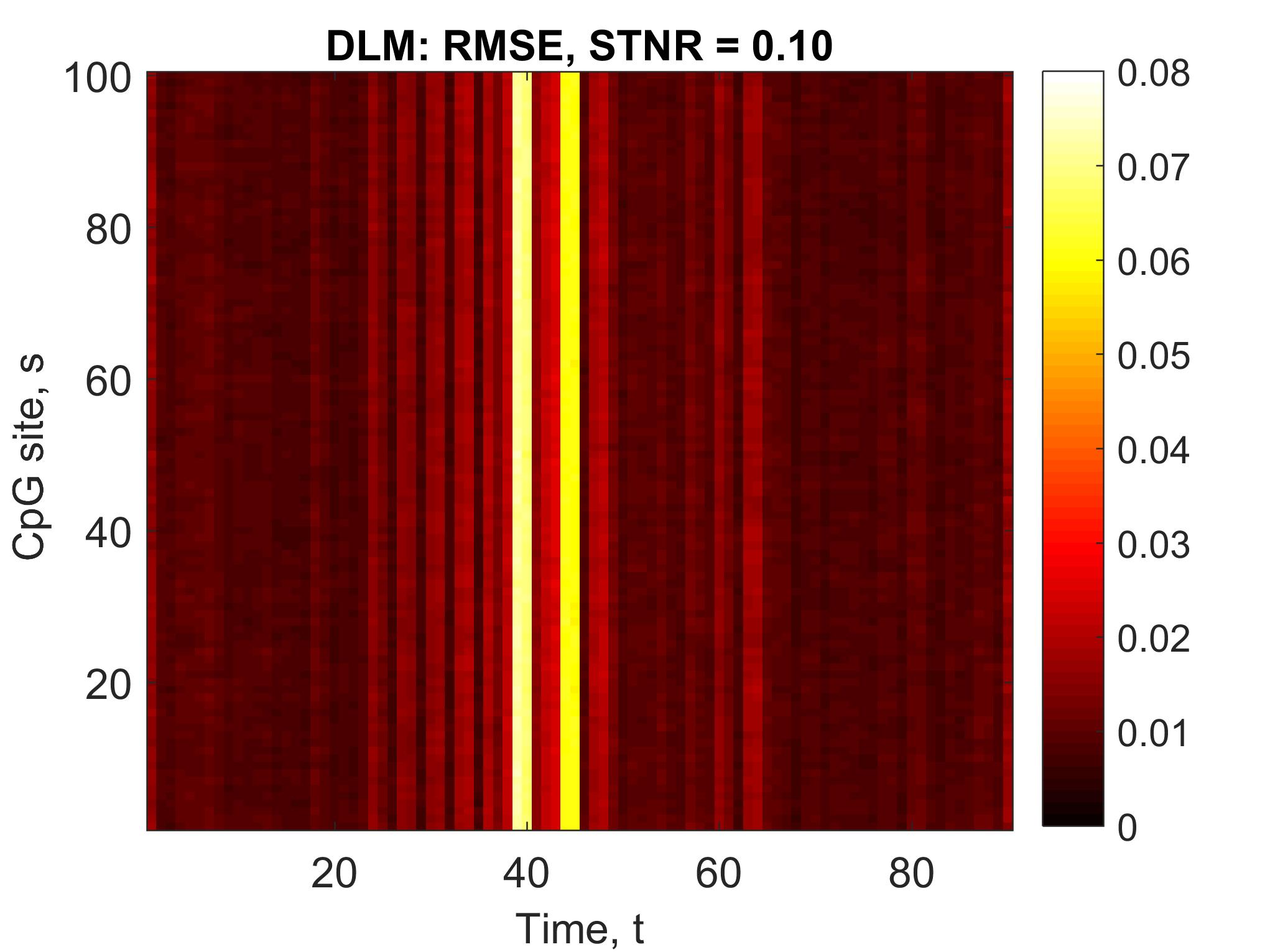} &
	\includegraphics[width=0.33\linewidth]{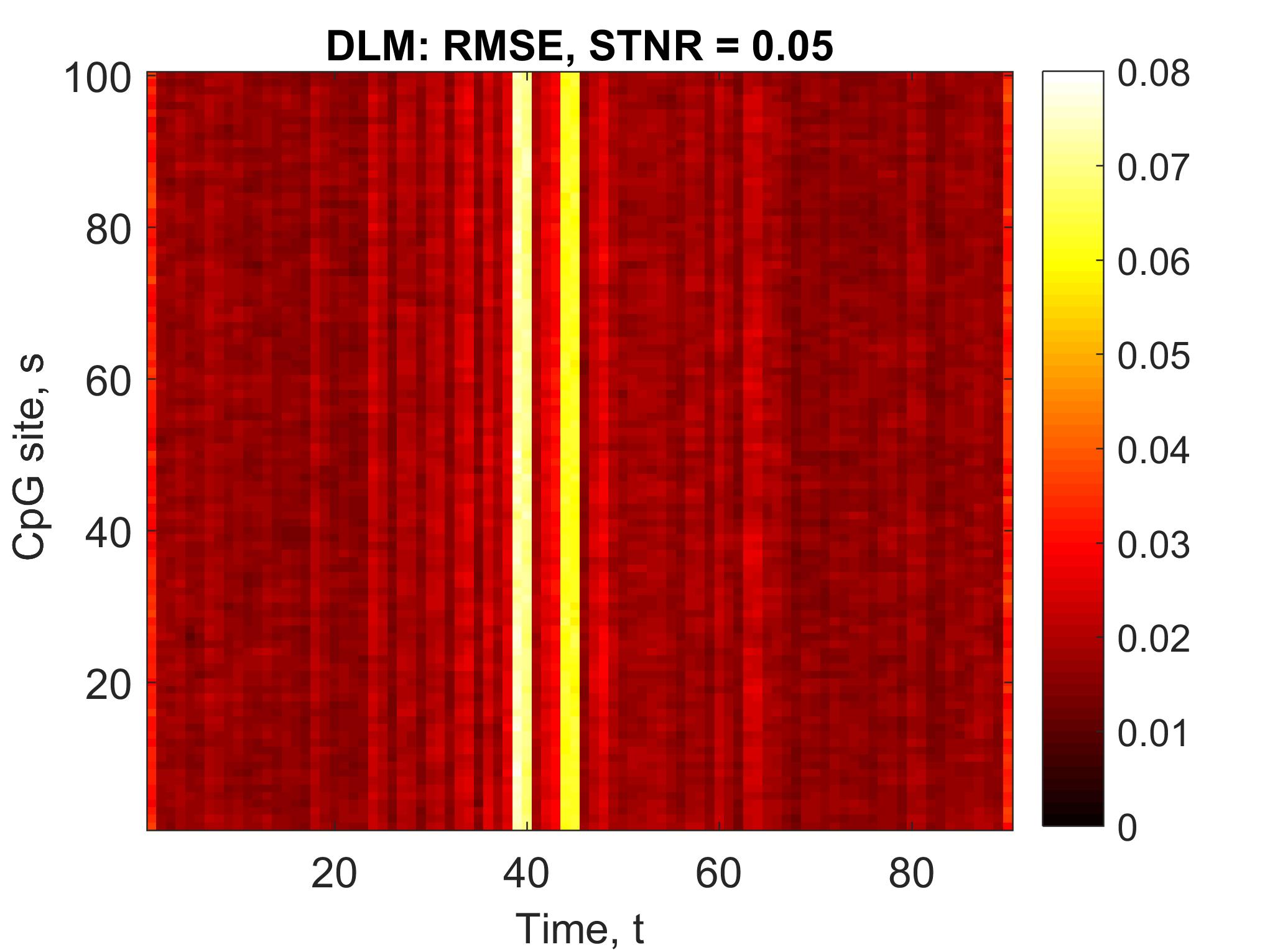} &
	\includegraphics[width=0.33\linewidth]{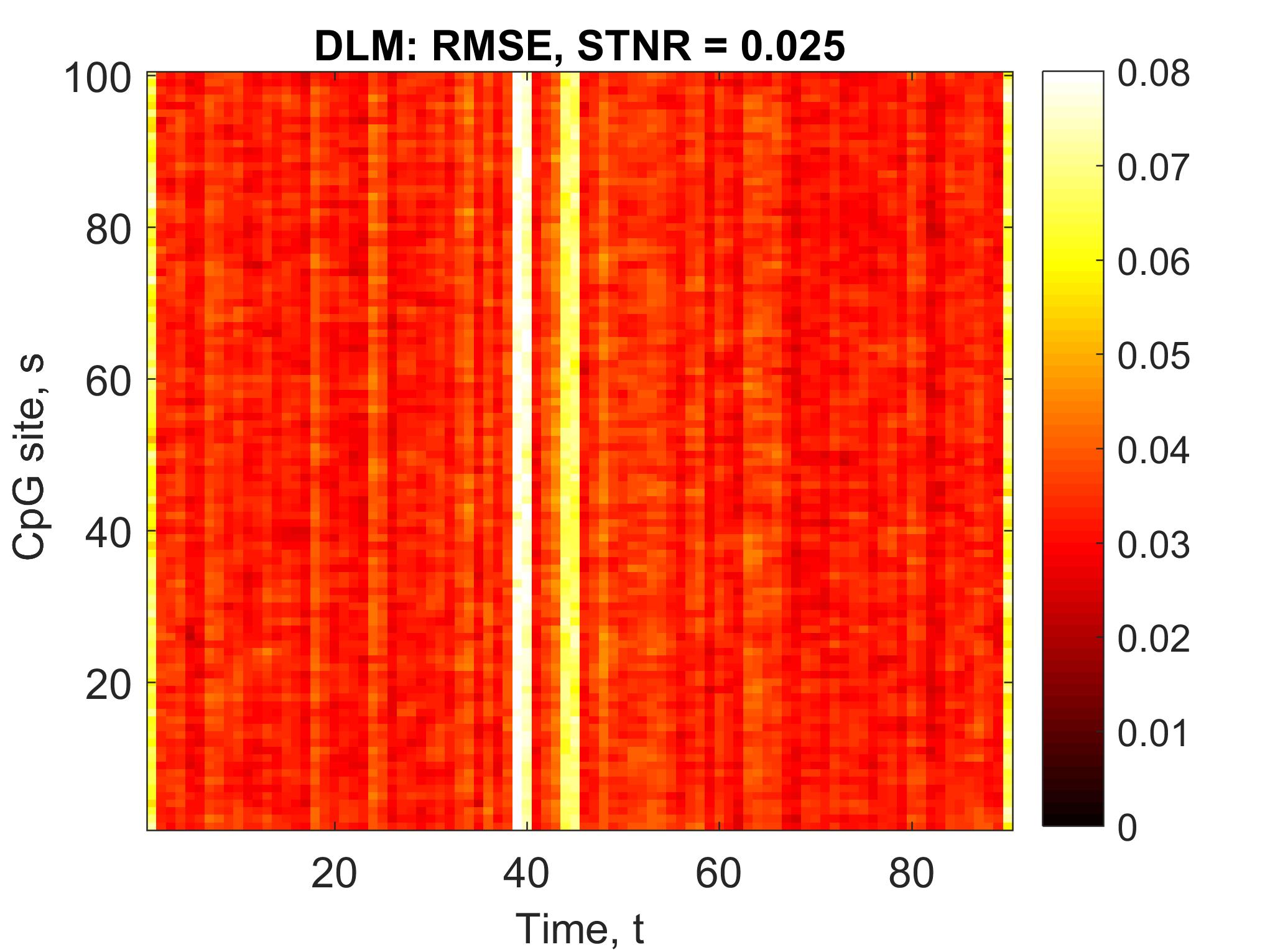}
\end{tabular}		
\caption{Heat maps displaying the RMSE averaged over 100 simulations for the FFR method (top panel) and DLM method (bottom panel).}
\label{fig:rmse}
\end{figure}

Figure \ref{fig:rmse} shows the root mean square error (RMSE) for all scenarios. The site-by-site DLM estimates had higher RMSE throughout the surface at each STNR level. Relative to the site-by-site DLM analysis, the FFR method reduced the sum of the RMSE over the entire surface by 68\%, 63\%, and 65\% for the STNR = 0.10, 0.05, and 0.025 scenarios, respectively. The null regions saw the largest gains in efficiency from the joint approach, while the top and bottom edges of the region of interest had the smallest gains. These differences in efficiency gains are due to the fact that spatial smoothing is less effective for sites at the boundary of the signal region. Sites in the interior of the region borrow information from a greater number of sites and therefore gain more from a joint-modeling approach. 

\begin{figure}
\begin{tabular}{ccc}
	\includegraphics[width=0.33\linewidth]{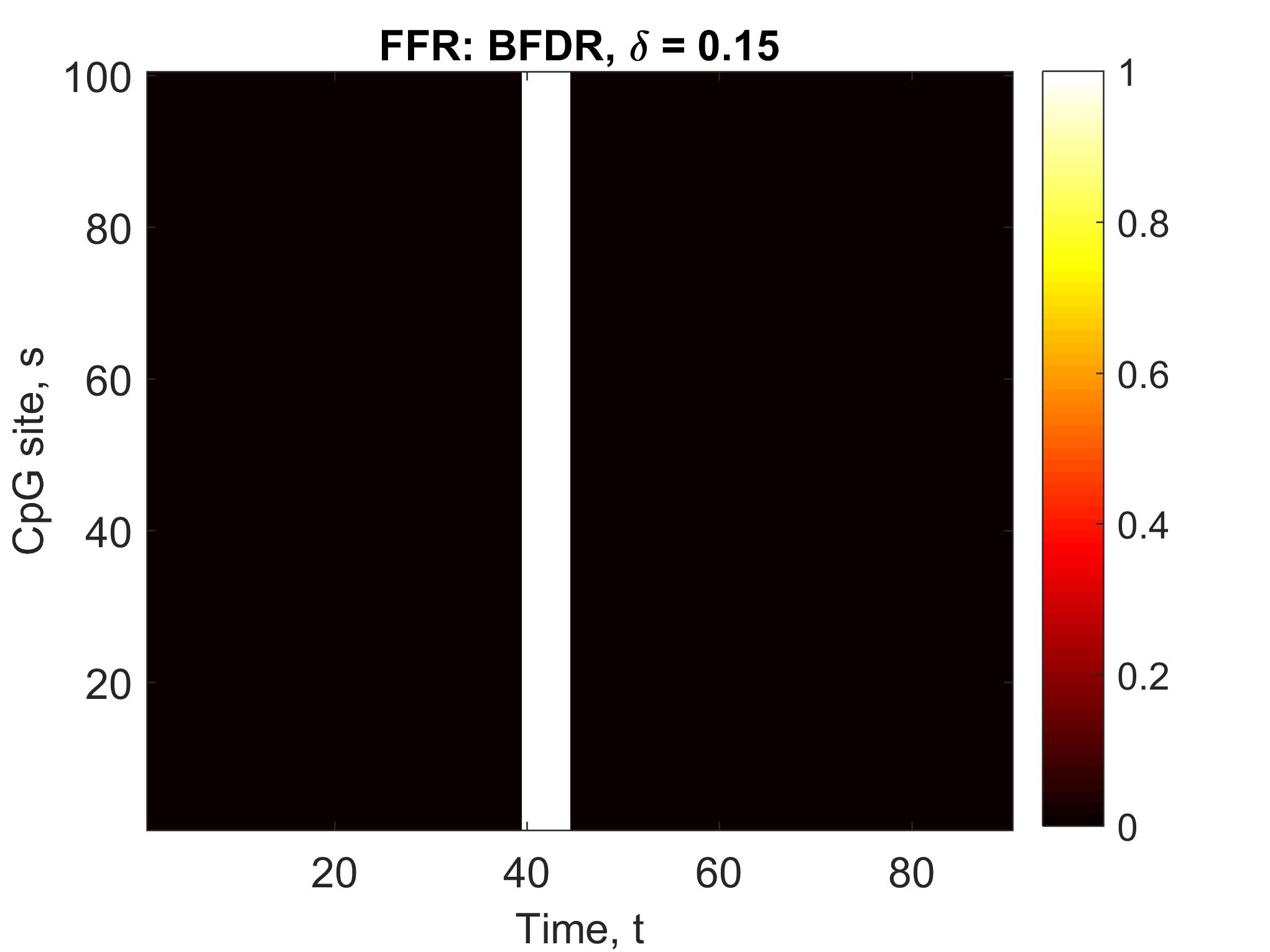} &
	\includegraphics[width=0.33\linewidth]{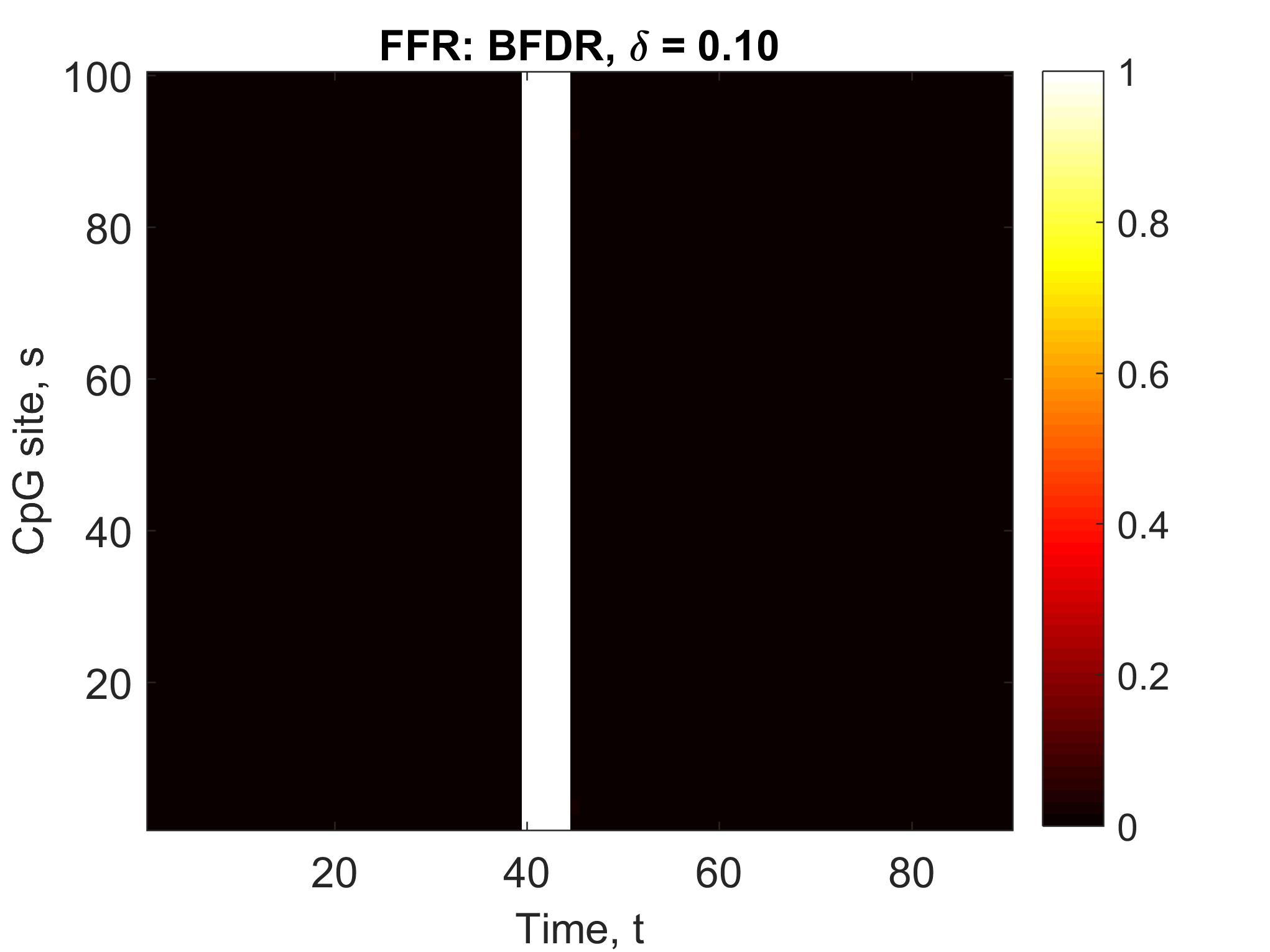} &
	\includegraphics[width=0.33\linewidth]{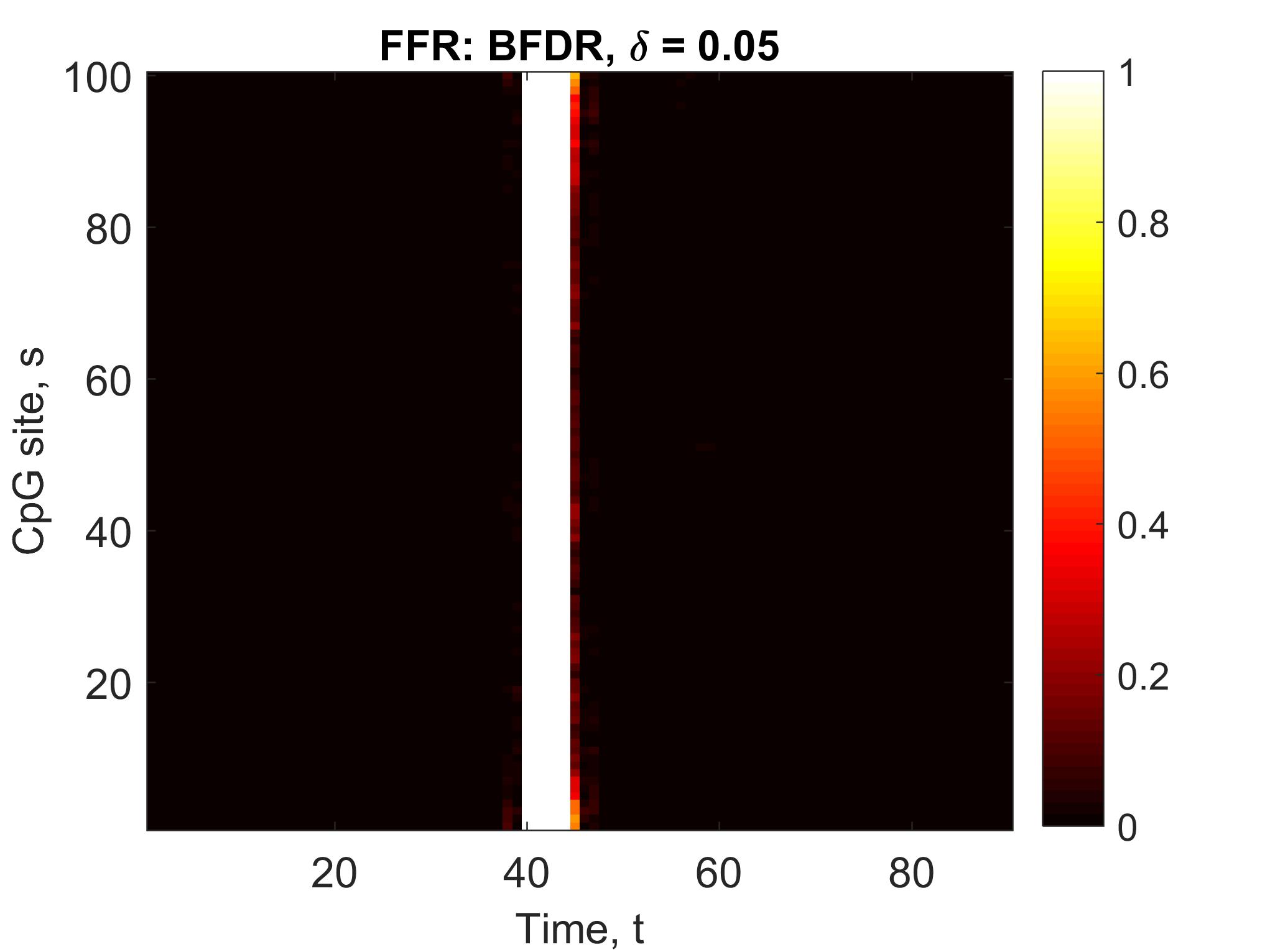} \\
	\includegraphics[width=0.33\linewidth]{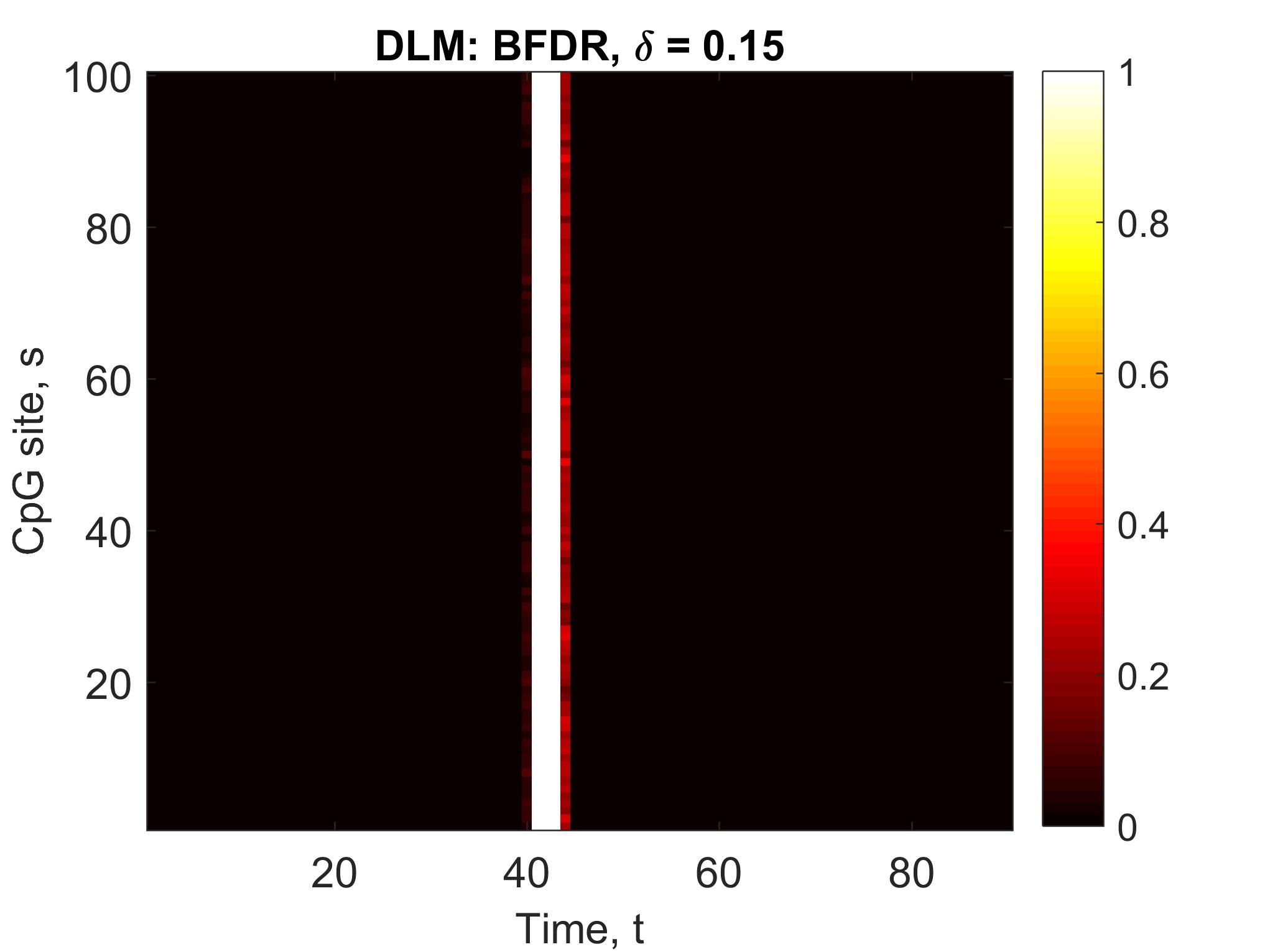} &
	\includegraphics[width=0.33\linewidth]{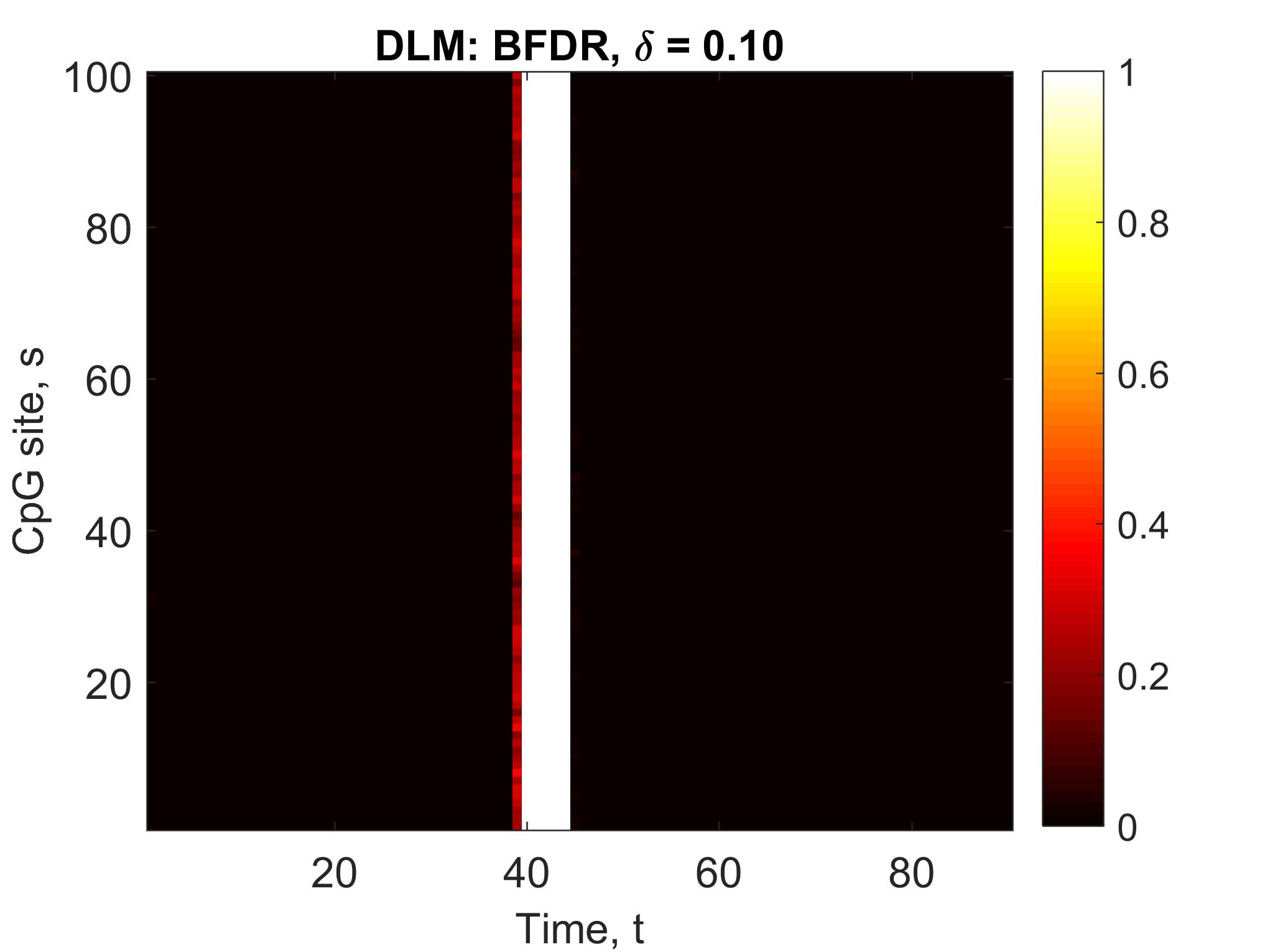} &
	\includegraphics[width=0.33\linewidth]{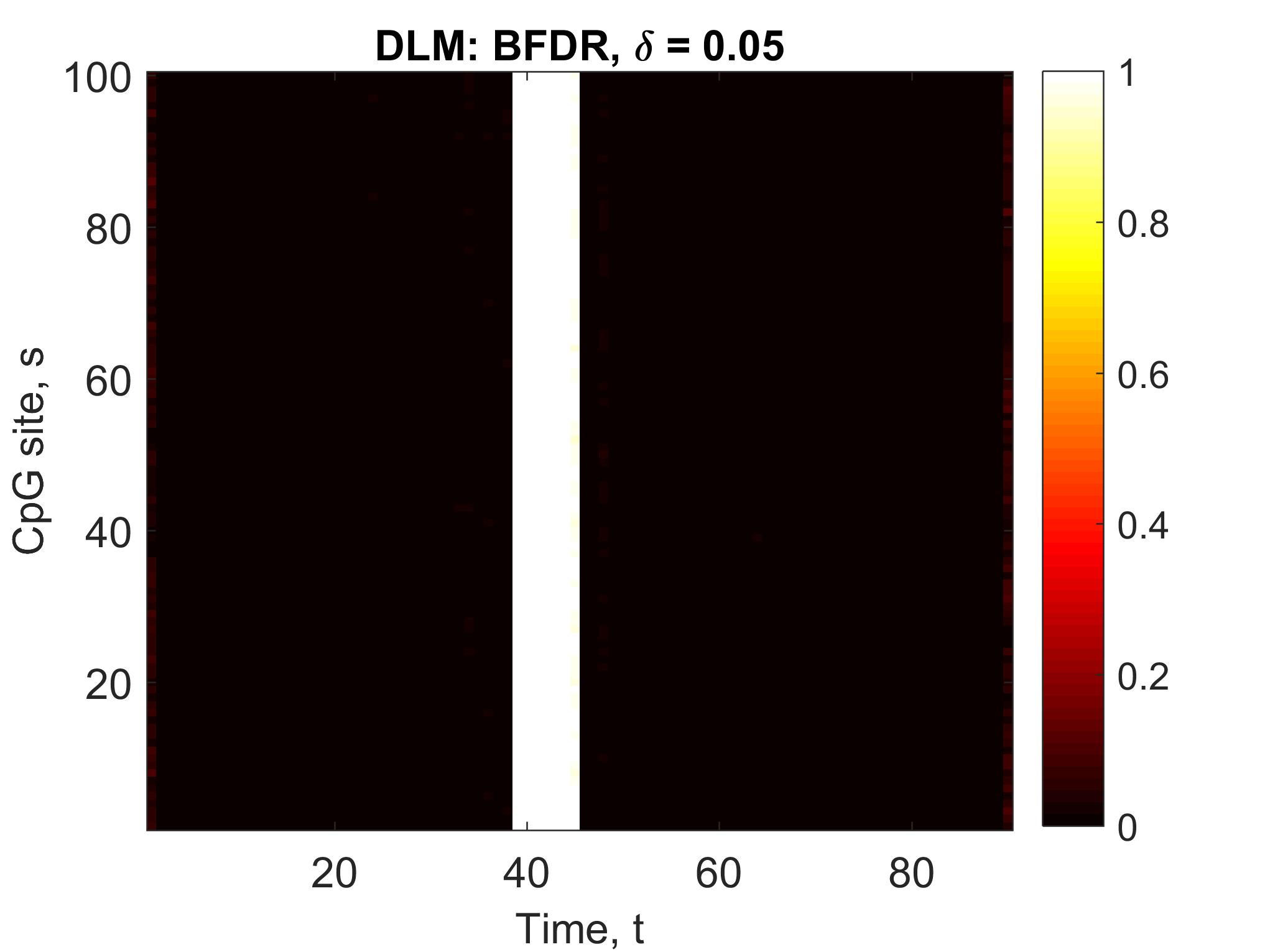}
\end{tabular}		
\caption{Heat maps of BFDR results at the STNR = 0.10 level for the FFR (top panel) and DLM (bottom panel) methods averaged over 100 simulations. Left: $\delta$ = 0.15 (75\% of true signal). Center: $\delta$ = 0.10 (50\% of true signal). Right: $\delta$ = 0.05 (25\% of true signal).}
\label{fig:fdr}
\end{figure}

We also compared the performance of the BFDR and SimBaS inferential procedures for the two methods, using $\alpha = 0.05$ to select significant locations for both procedures. For the BFDR, we used $\delta$-intensity changes of 0.15, 0.10, and 0.05 corresponding to 75\%, 50\% and 25\% of the true signal in the vertical band. We performed BFDR and SimBaS procedures on each estimated surface and then averaged over simulations. The heat maps in Figure \ref{fig:fdr} are shaded according to the proportion of simulations in which each location was flagged as significant by the BFDR procedure. Figure \ref{fig:fdr} shows heat maps for the STNR = 0.10 scenario across the three $\delta$ levels. Locations that were flagged as significant by all simulations are white, those that were never flagged are black, and locations that were occasionally flagged vary from red to yellow shading. The accompanying Table 1 displays the sensitivity and false discovery rate, FDR, for both methods at varying $\delta$ levels. Across $\delta$ levels, FFR performed well, flagging regions with a true signal as significant in all estimated surfaces, while maintaining a FDR below 5\%. This does not hold true for the DLM method. At $\delta = 0.15$, BFDR only flagged 66\% of the vertical band as significant, while at $\delta = 0.05$, the method flagged too many locations neighboring the true band as significant, pushing the FDR to 30\%.

\begin{table}[]
    \caption{Sensitivity and false discovery rate (FDR) for the BFDR procedure in the STNR = 0.10 setting over decreasing $\delta$ intensities.}
    \label{tab:fdrV2}
    \centering
    \begin{tabular}{l l c c c}
    \hline \hline
    Measure & Method & $\delta$ = 0.15 & $\delta$ = 0.10 & $\delta$ = 0.05 \\
    \hline 
    Sensitivity & FFR & 100.0\% & 100.0\% & 100.0\%  \\
                & DLM & 65.9\% & 100.0\% & 100.0\% \\
    FDR         & FFR & 0.0\% & 0.0\% & 4.7\% \\
                & DLM & 0.0\% & 4.8\% & 29.9\% \\
    \hline
    \end{tabular}
\end{table}

\begin{figure}
\begin{tabular}{ccc}
	\includegraphics[width=0.33\linewidth]{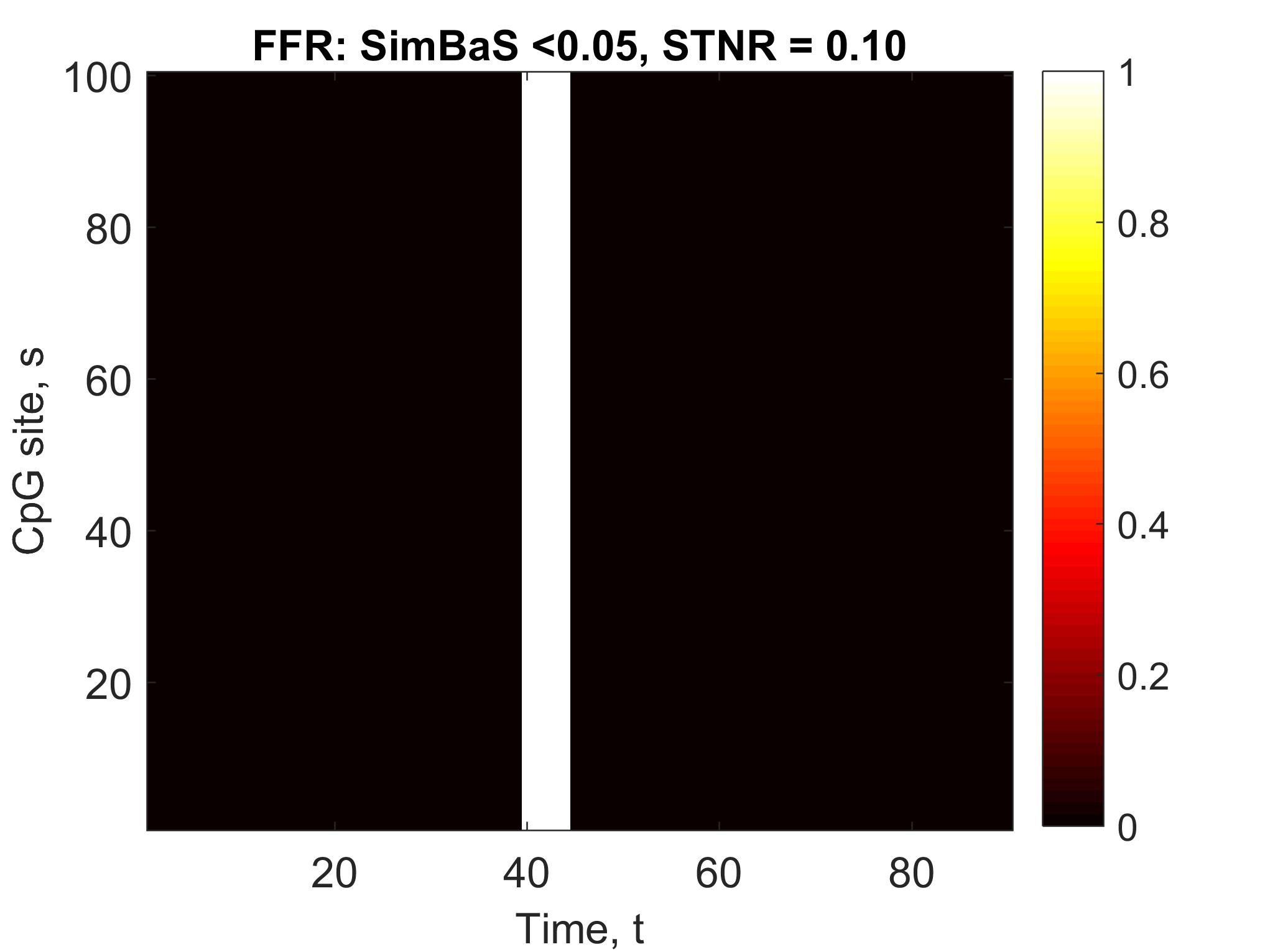} &
	\includegraphics[width=0.33\linewidth]{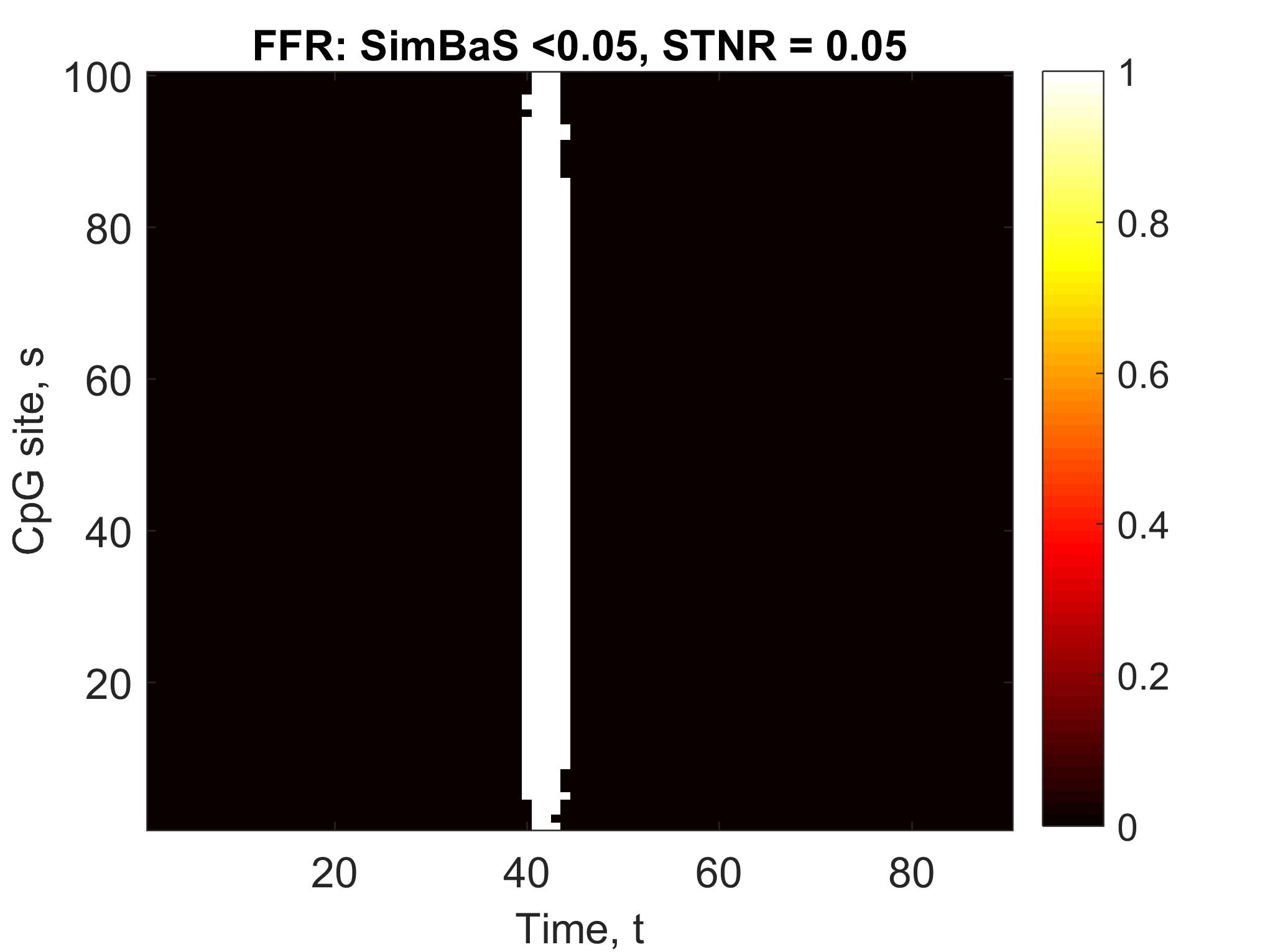} &
	\includegraphics[width=0.33\linewidth]{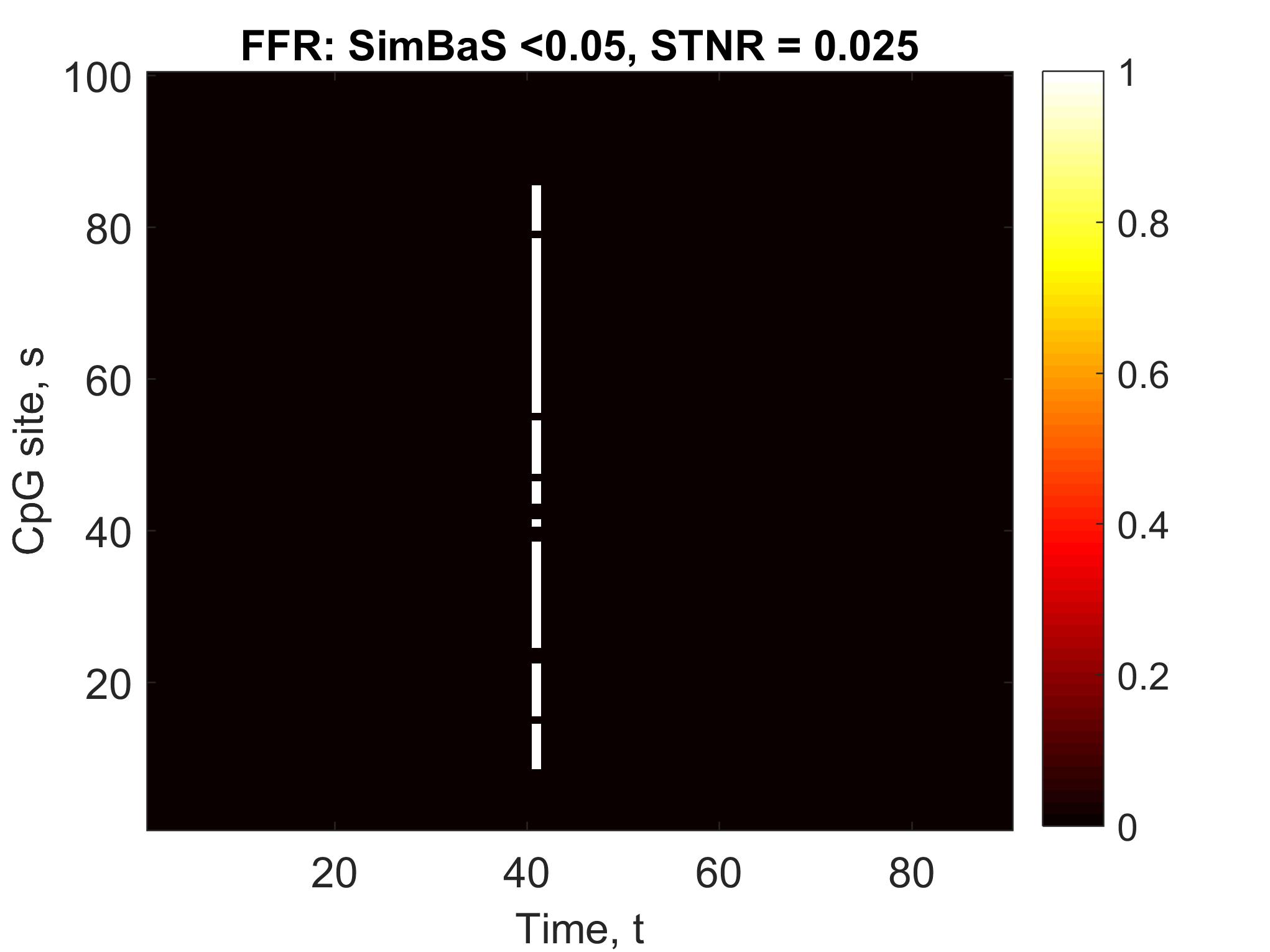} \\
	\includegraphics[width=0.33\linewidth]{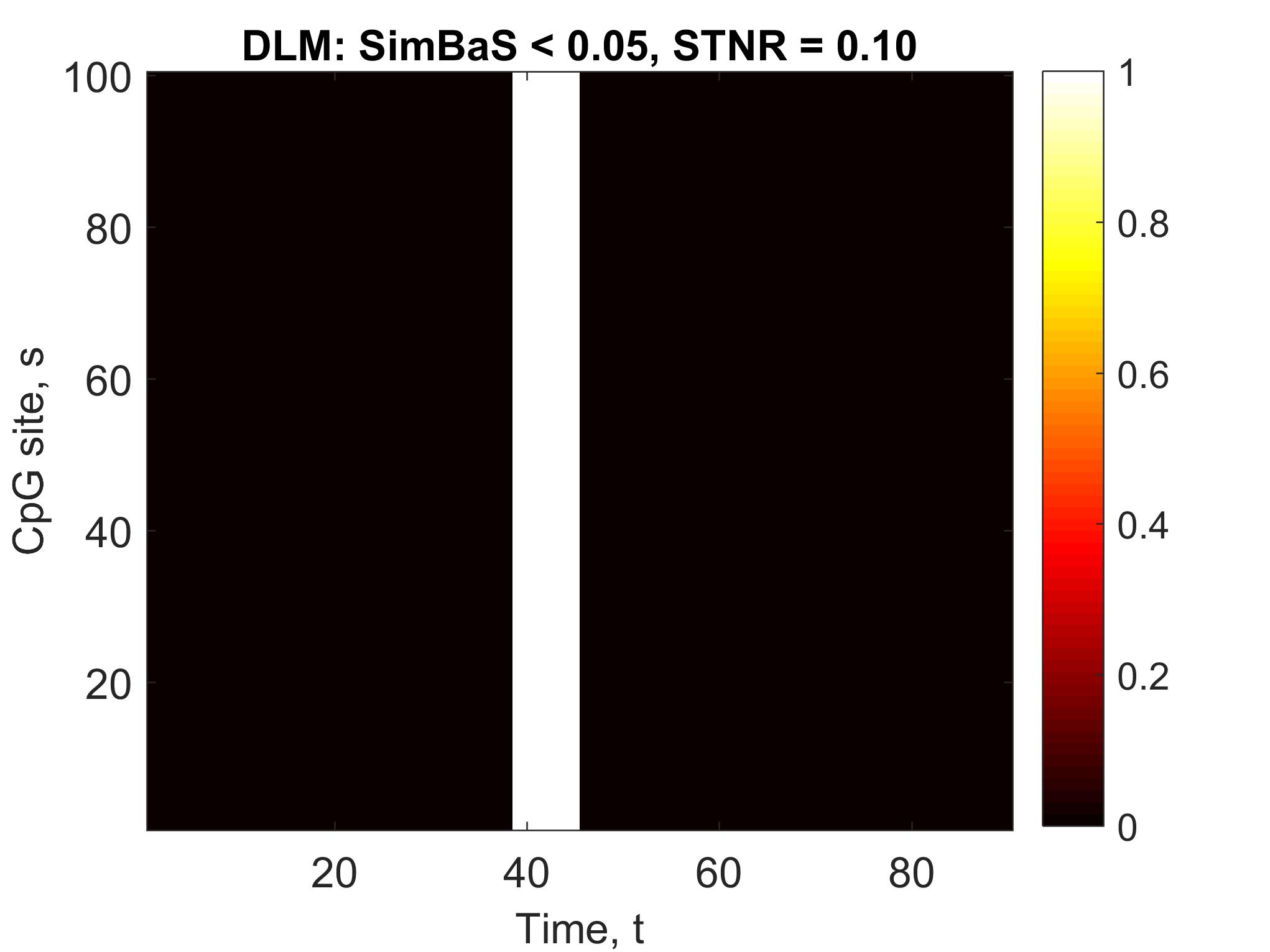} &
	\includegraphics[width=0.33\linewidth]{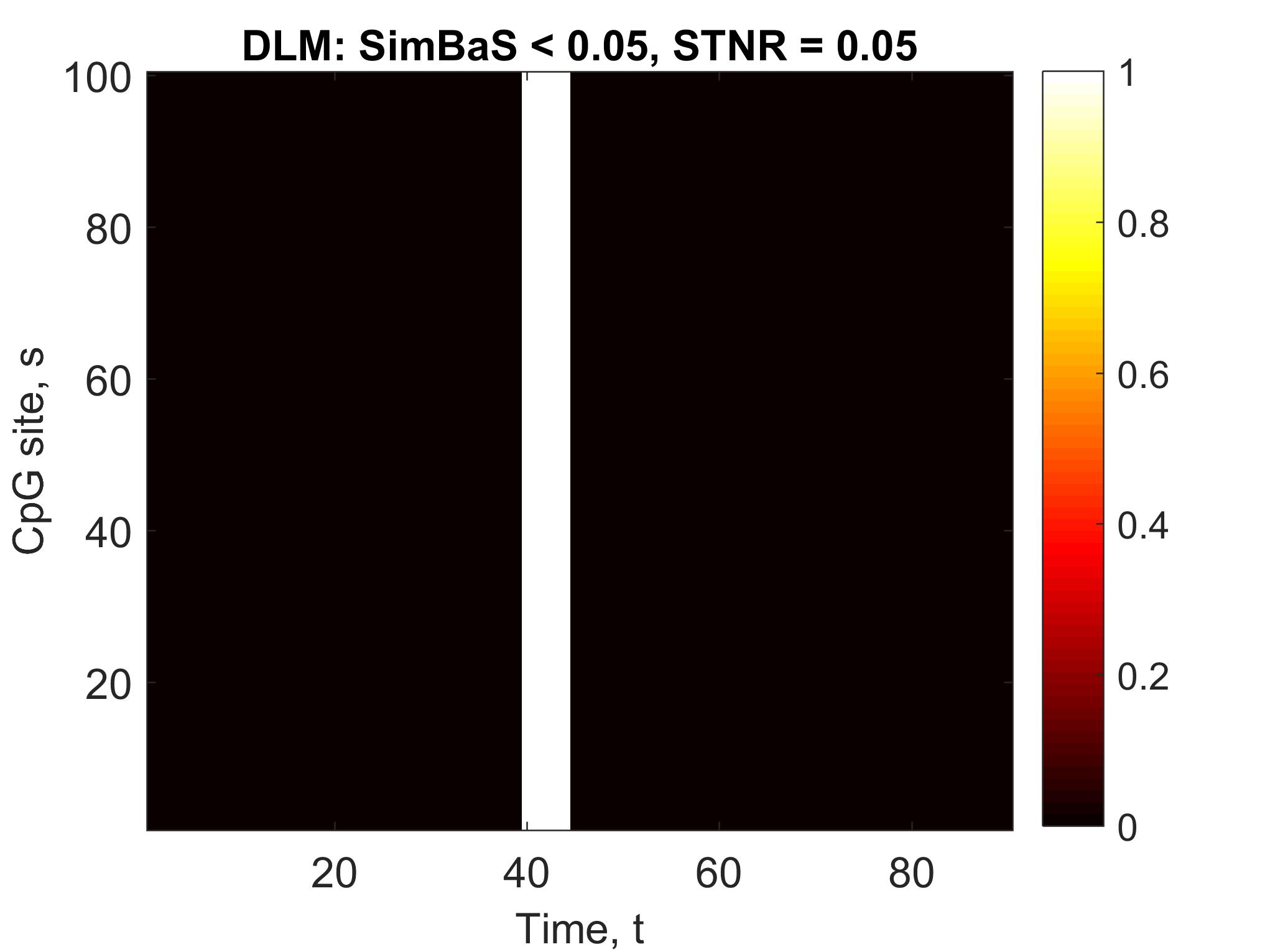} &
	\includegraphics[width=0.33\linewidth]{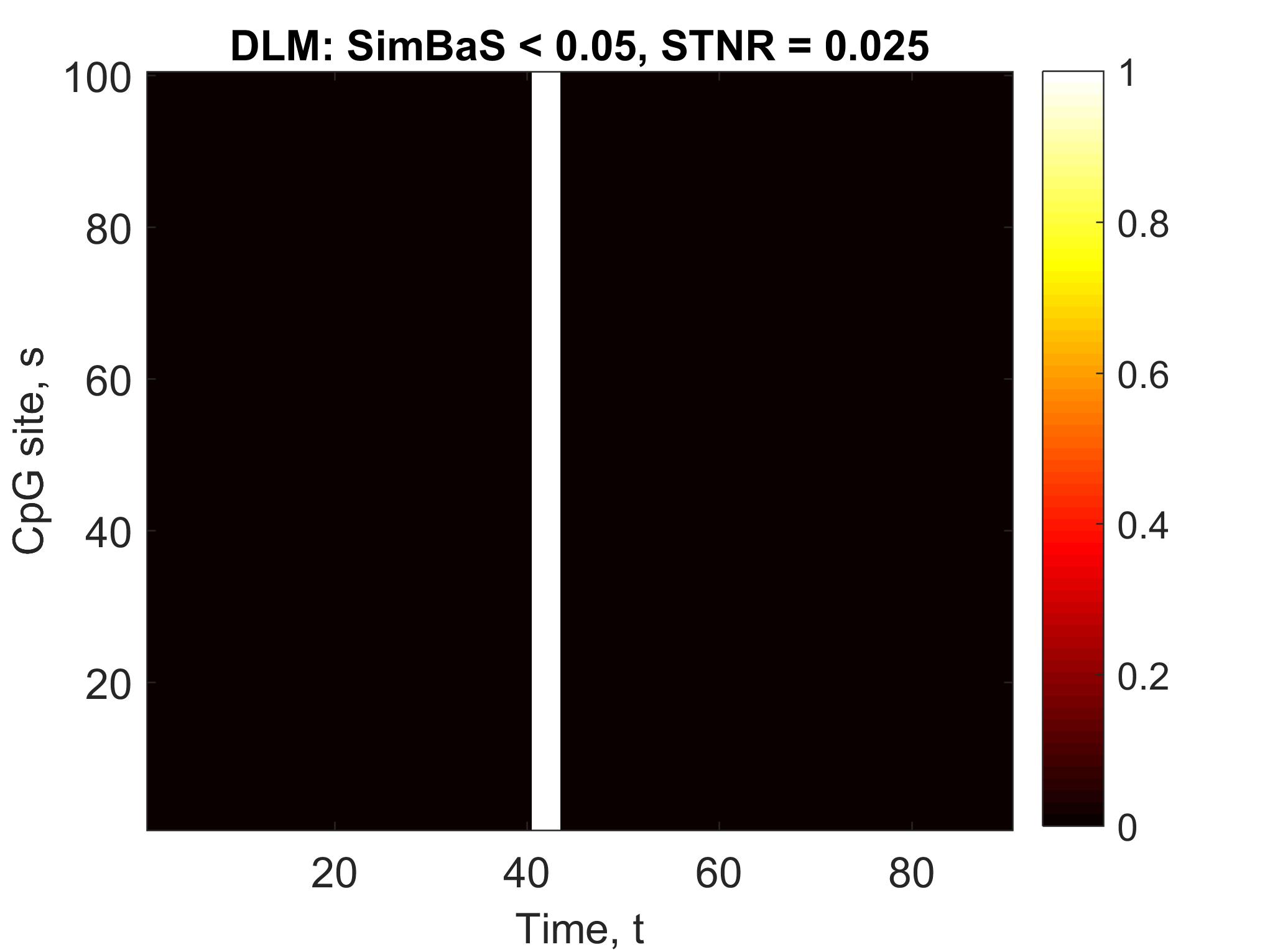}
\end{tabular}		
\caption{Heat maps of SimBaS results for the FFR (top panel) and DLM (bottom panel) methods averaged over 100 simulations and then thresholded at 0.05. }
\label{fig:simbas}
\end{figure}

Figure \ref{fig:simbas} shows heat maps obtained by averaging the SimBaS for each location over all simulations and then flagging locations with scores $\leq$ 0.05 in white. Similar to the BFDR results, at STNR = 0.10, FFR outperformed DLM by maintaining high sensitivity and low FDR, while the DLM had a high FDR of 29\%. However, as the STNR decreases, SimBaS for FFR was more conservative than SimBaS for the DLM; when STNR = 0.025, the FFR sensitivity dropped to just 13\% while the DLM maintained 60\% sensitivity.

\begin{table}[]
    \caption{Sensitivity and false discovery rate, FDR, for the SimBaS procedure over varying STNR settings.}
    \label{tab:simbasV2}
    \centering
    \begin{tabular}{l l c c c}
    \hline \hline
    Measure & Method & STNR = 0.10 & STNR = 0.05 & STNR = 0.025 \\
    \hline 
    Sensitivity & FFR & 100.0\% & 94.4\% & 13.4\%  \\
                & DLM & 100.0\% & 100.0\% & 60.0\% \\
    FDR         & FFR & 0.0\% & 0.0\% & 0.0\% \\
                & DLM & 28.6\% & 0.0\% & 0.0\% \\
    \hline
    \end{tabular}
\end{table}

\begin{figure}
\begin{tabular}{ccc}
	\includegraphics[width=0.33\linewidth]{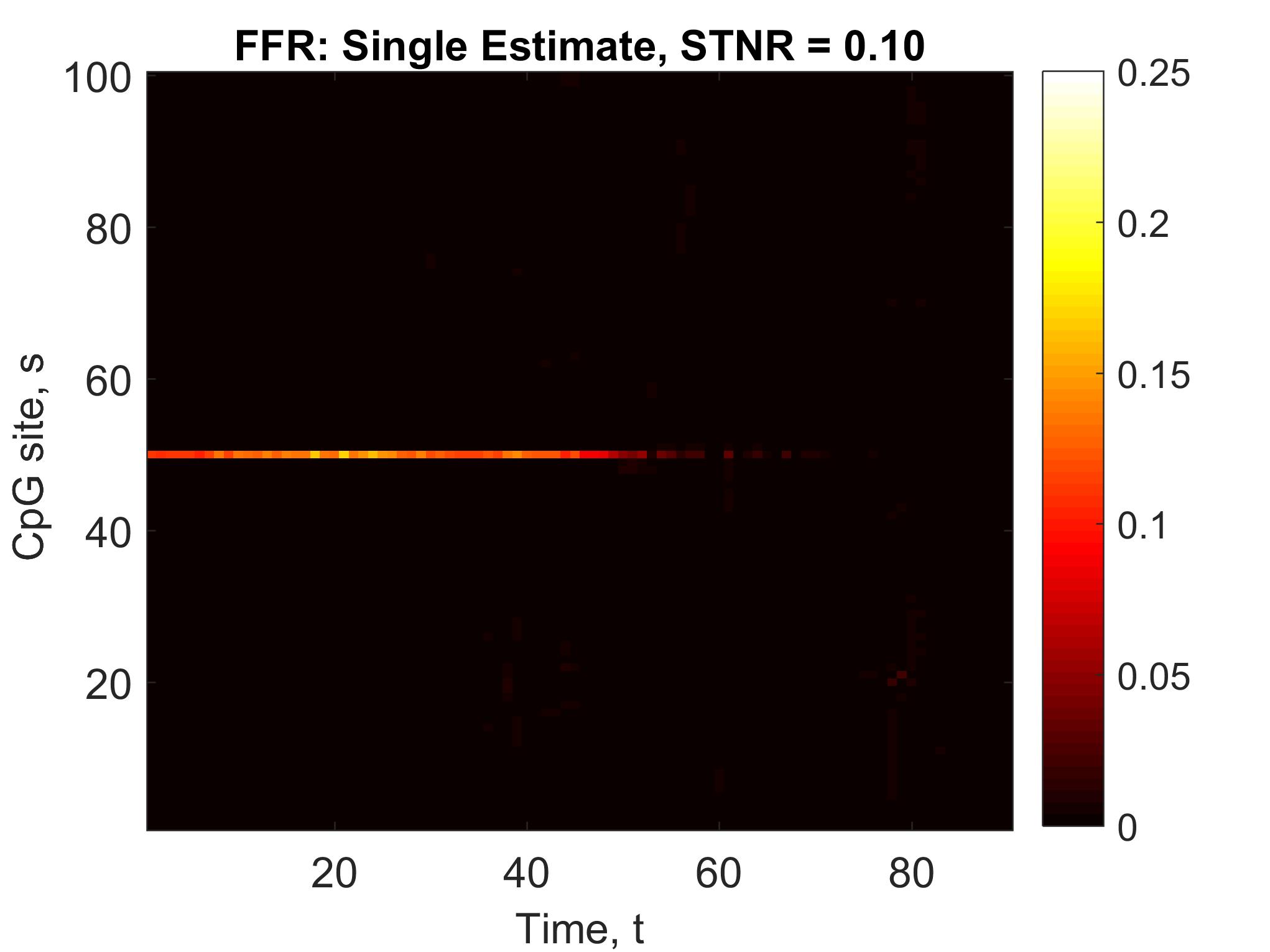} &
	\includegraphics[width=0.33\linewidth]{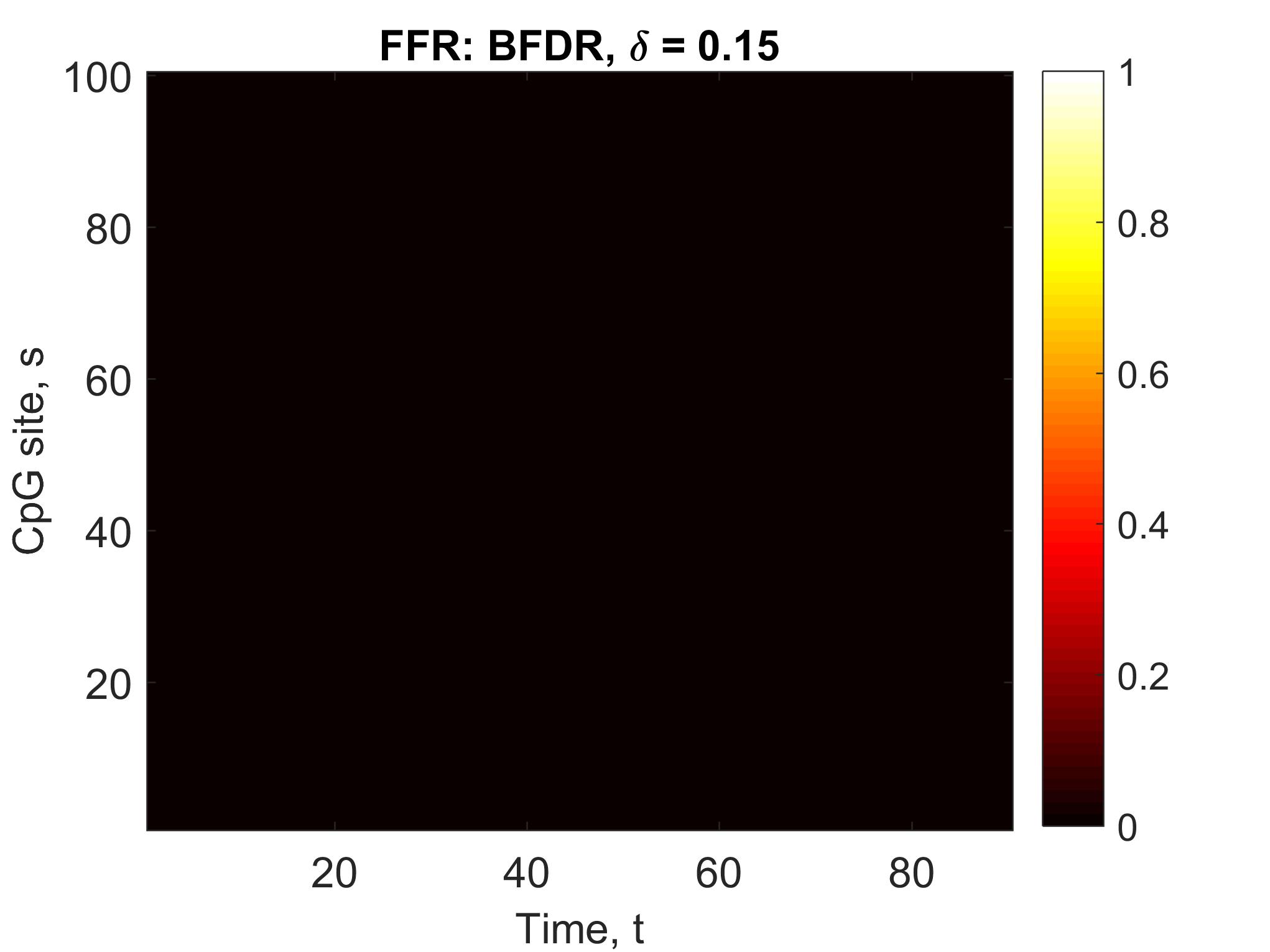} &
	\includegraphics[width=0.33\linewidth]{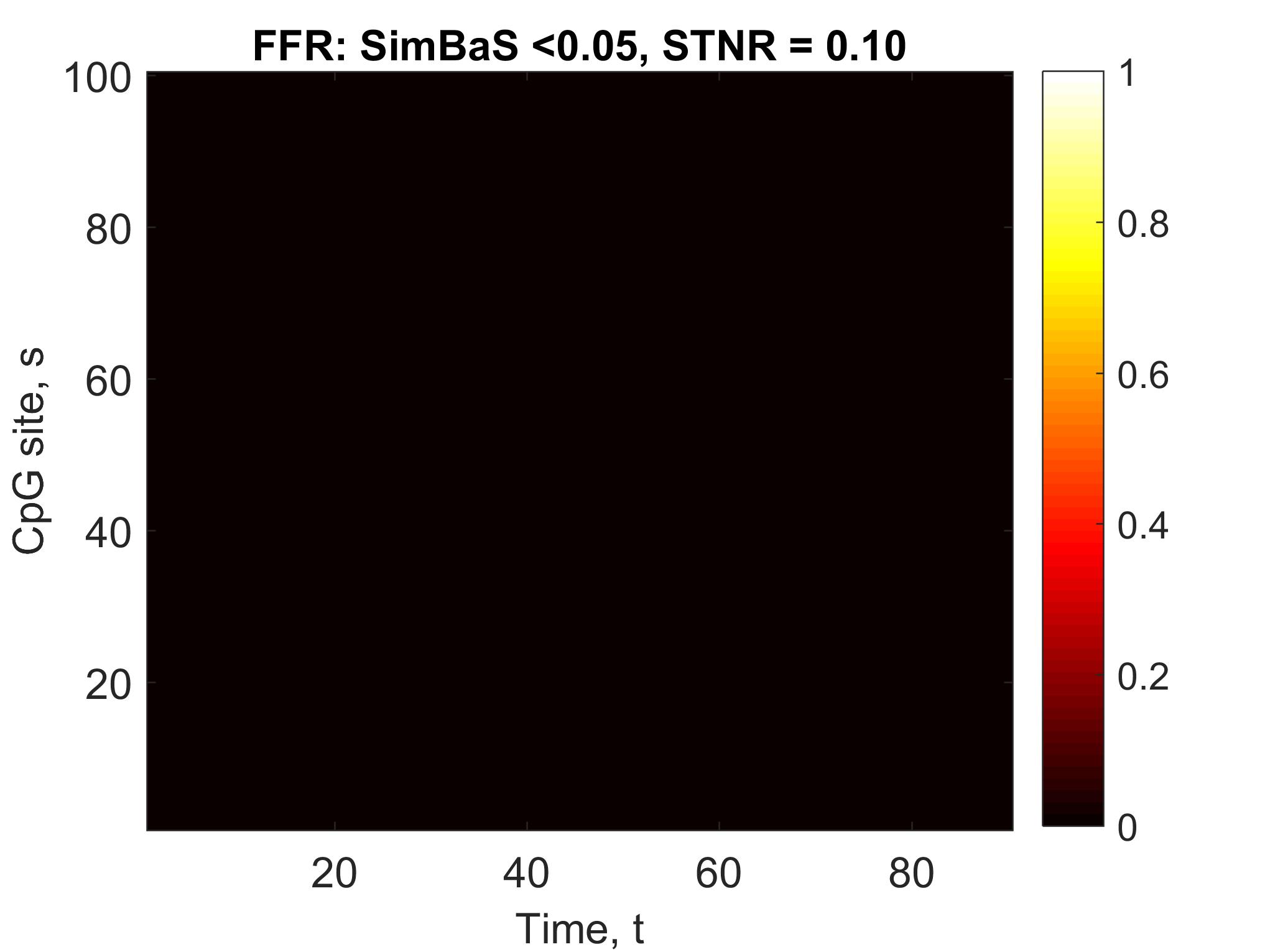} \\
	\includegraphics[width=0.33\linewidth]{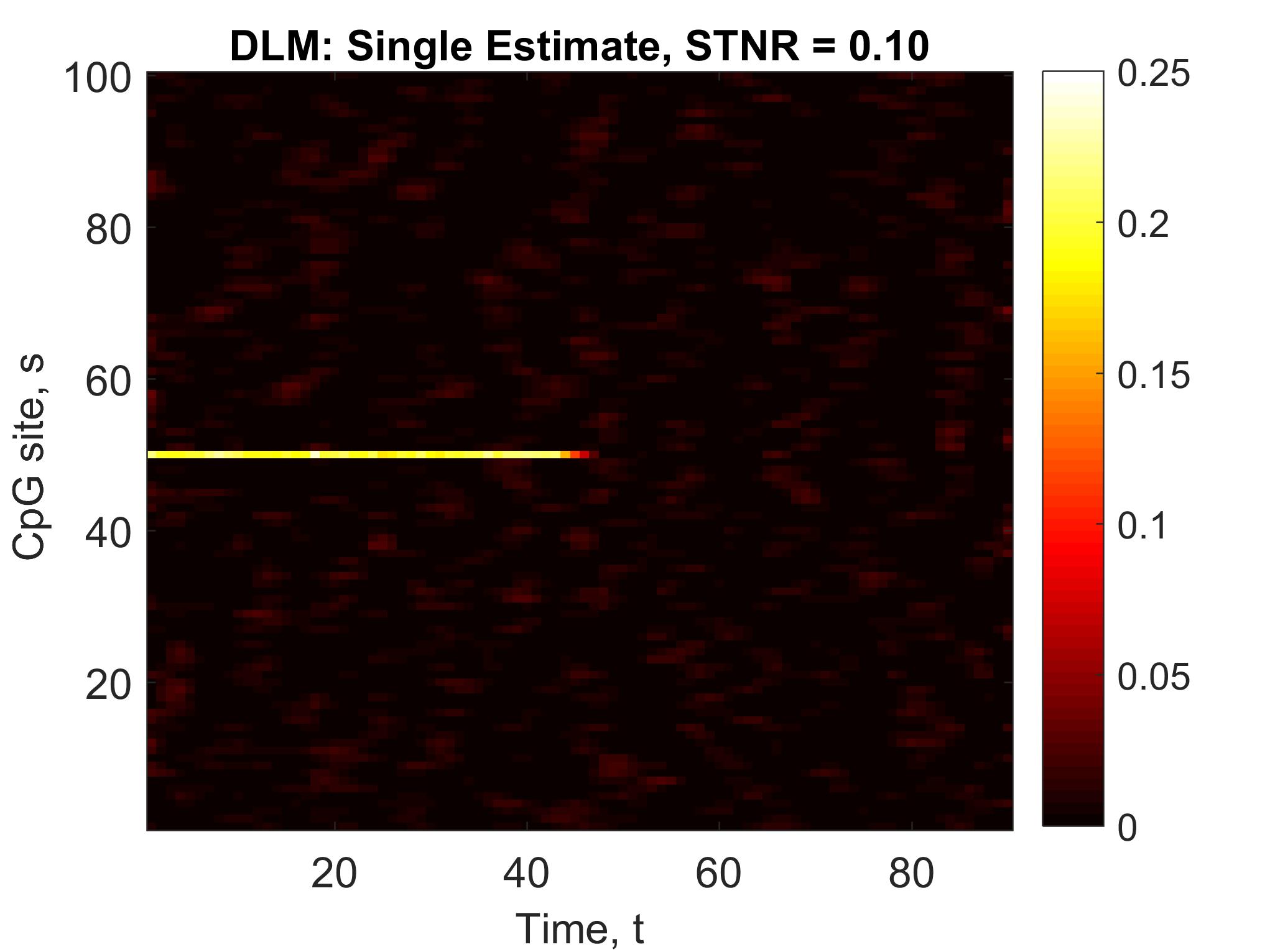} &
	\includegraphics[width=0.33\linewidth]{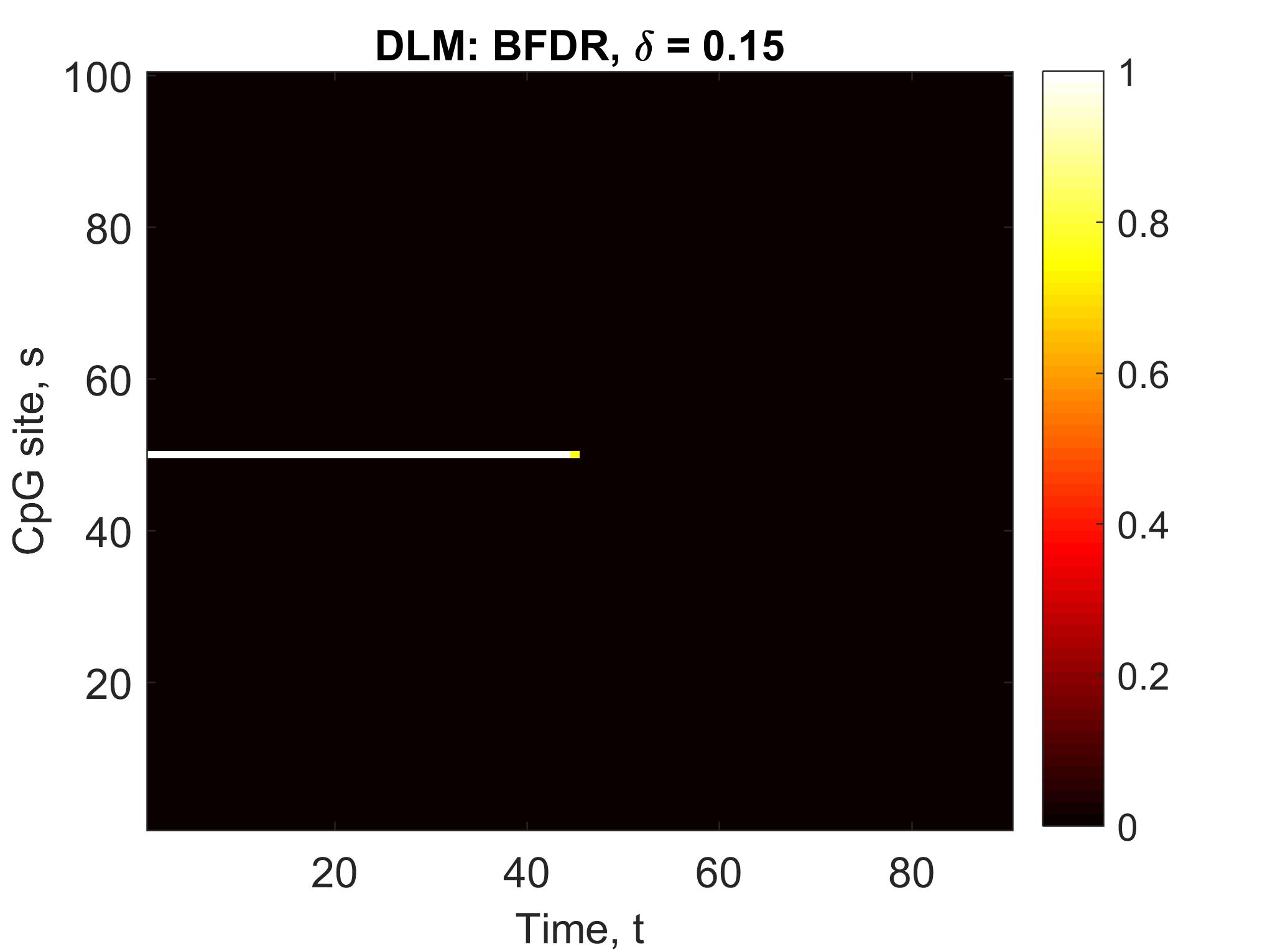} &
	\includegraphics[width=0.33\linewidth]{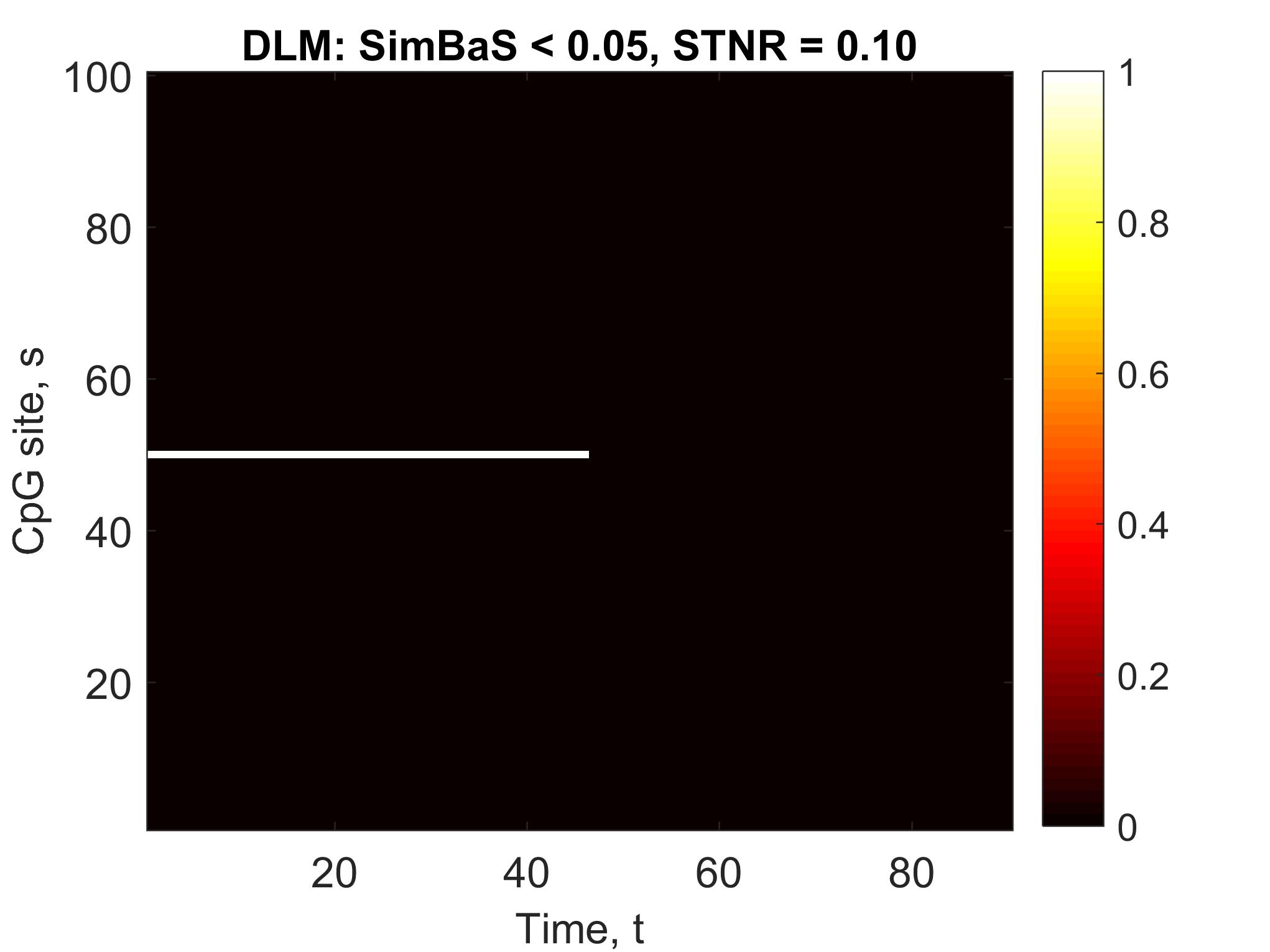}
\end{tabular}		
\caption{FFR (top panel) and DLM (bottom panel) estimated association surfaces, BFDR results for the horizontal band setting with STNR = 0.10, $\delta = 0.15$, and $\alpha = 0.05$, and SimBaS results thresholded at 0.05. $\beta = 0.20$ in the horizontal band signal region.}
\label{fig:H1}
\end{figure}

We also considered a second set of simulations for which the true association surface was a narrow, horizontal band with $\boldsymbol{\beta} = 0.2$ at probe $s = 50$ for $T \in \{1,\dots, 45\}$ and $\boldsymbol{\beta} = 0$ everywhere else (Figure \ref{fig:trueSurface}; right panel). In contrast to the vertical band setting, this association surface represented a sustained window of susceptibility at a single CpG site rather than across a genomic region.  Figure \ref{fig:H1} shows the estimated surfaces, BFDR, and SimBaS heat maps for the FFR and DLM approaches at the STNR = 0.10 level. In this setting with the signal confined to a single site, the shrinkage employed by the FFR hinders its ability to identify the probe affected by exposure. The magnitude of $\widehat{\boldsymbol{\beta}}_{FFR}$ in the signal region is about half the true value and the window fails to appear on either the BFDR or SimBaS plots. The DLM, however, successfully detects the true window of susceptibility for the site whose methylation is affected by exposure. 

\indent These simulations highlight the relative strengths and weaknesses of the FFR and DLM methods. For the global exposure effect setting in which an effect of exposure is shared across probes within a region of interest, FFR consistently outperforms DLM in terms of RMSE across the surface, as well as sensitivity and FDR for both the BFDR and SimBaS inferential procedures. On the other hand, for very small effect sizes (STNR = 0.025 scenario), both methods had low power, but the DLM site-by-site approach is more powerful than the FFR joint approach. 

For the situation in which the window of susceptibility is localized to a single probe, the DLM site-by-site approach is more powerful than the FFR joint approach. Thus, while the FFR's ability to borrow strength across probes is beneficial when there are shared windows of susceptibilty across a genomic region, the corresponding smoothing across the surface can be detrimental when the signal is sparsely distributed across probes or when the strength of the signal is low. These findings are unsurprising since bias-variance trade-offs are typical of shrinkage methods, particularly nonlinear shrinkage, like that used by the Bayesian FFR approach.  

\section{Results}
Using data from Project Viva, our goal was to characterize the time- and position-varying association between DNA methylation levels and air pollution exposure within the last trimester of gestation. To this end, we fit Model (\ref{fofscalar}) using DNA methylation level $y(s)$, a function of CpG site position $s$ (relative to other CpG sites on the same chromosome) as the outcome function. The exposure function was daily maternal PM$_{2.5}$ exposure during the 90 days prior to delivery. We included the following scalar covariates in the model:  maternal BMI, race, smoking status, education level, household income, child's race, child's sex, gestational age, season of birth, sample plate, as well as estimated cell proportions of leukocytes (CD8+, CD4+, natural killer cells, B-lymphocytes, monocytes, granulocytes, and nucleated red blood cells).

We used Debauchies wavelets with four vanishing moments, six levels of decomposition, and zero-padding for both the outcome and predictor functions. In order to reduce the dimension of the parameter space, we performed PCA on the continuous scalar covariates and retained components that preserved 95\% of their total variability. In simulation we observed that compression of the time-varying exposure led to signal distortion, so we only performed PC compression on the scalar covariates. 
We sampled 2000 posterior samples and discarded the first 1000. 

\begin{figure}
\begin{tabular}{cc}
\includegraphics[width=0.5\linewidth]{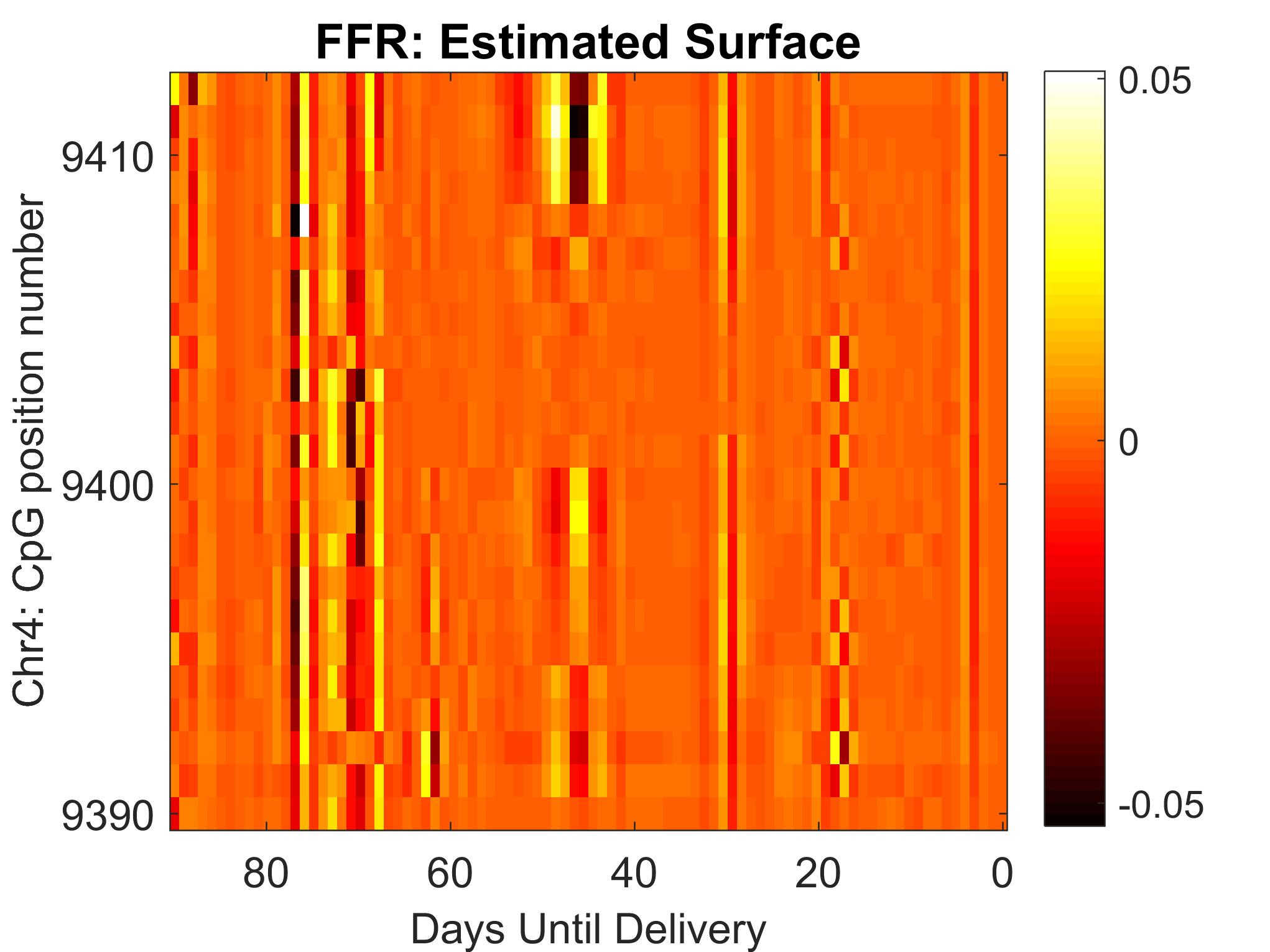} &
\includegraphics[width=0.5\linewidth]{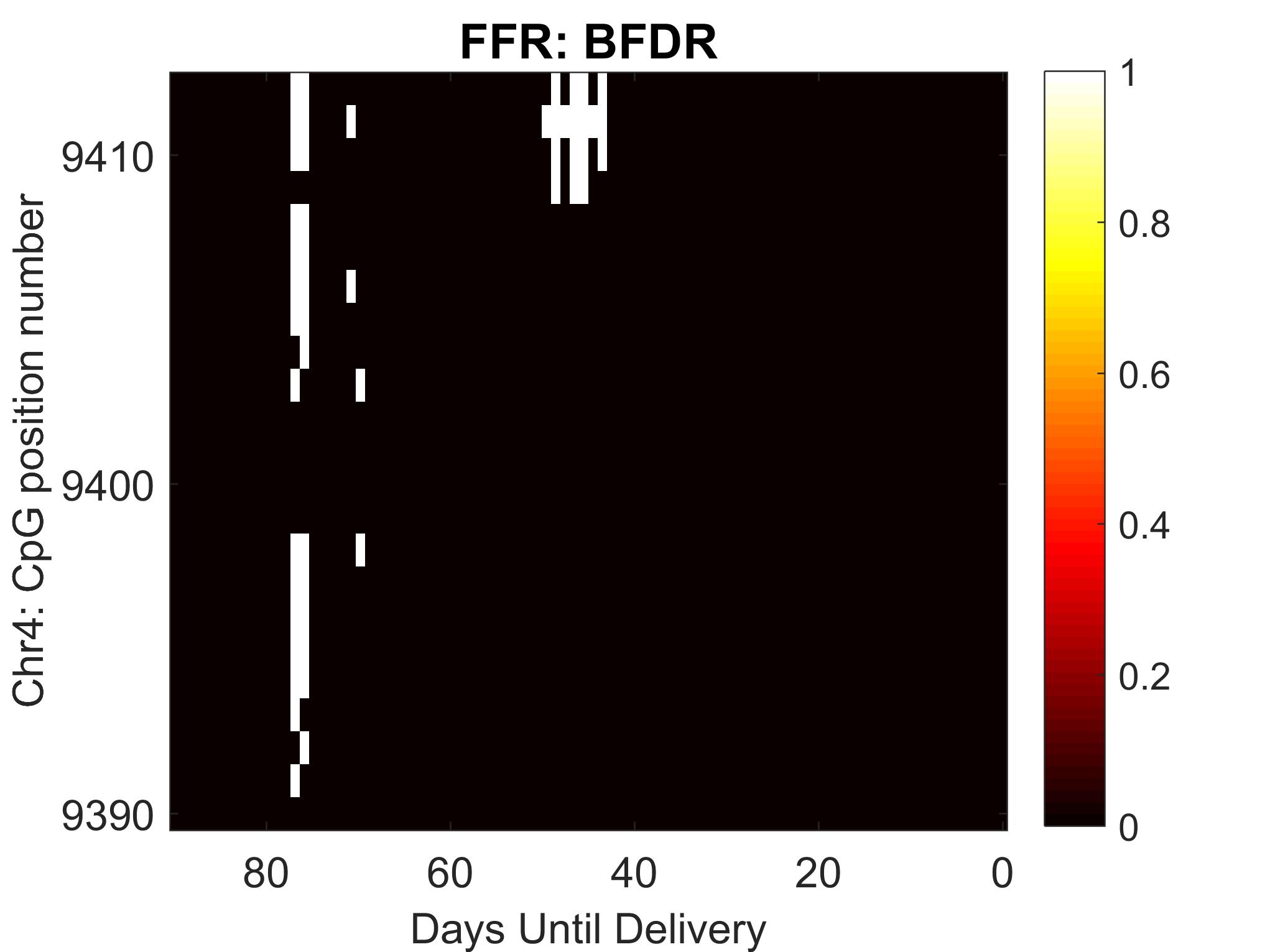} \\
\includegraphics[width=0.5\linewidth]{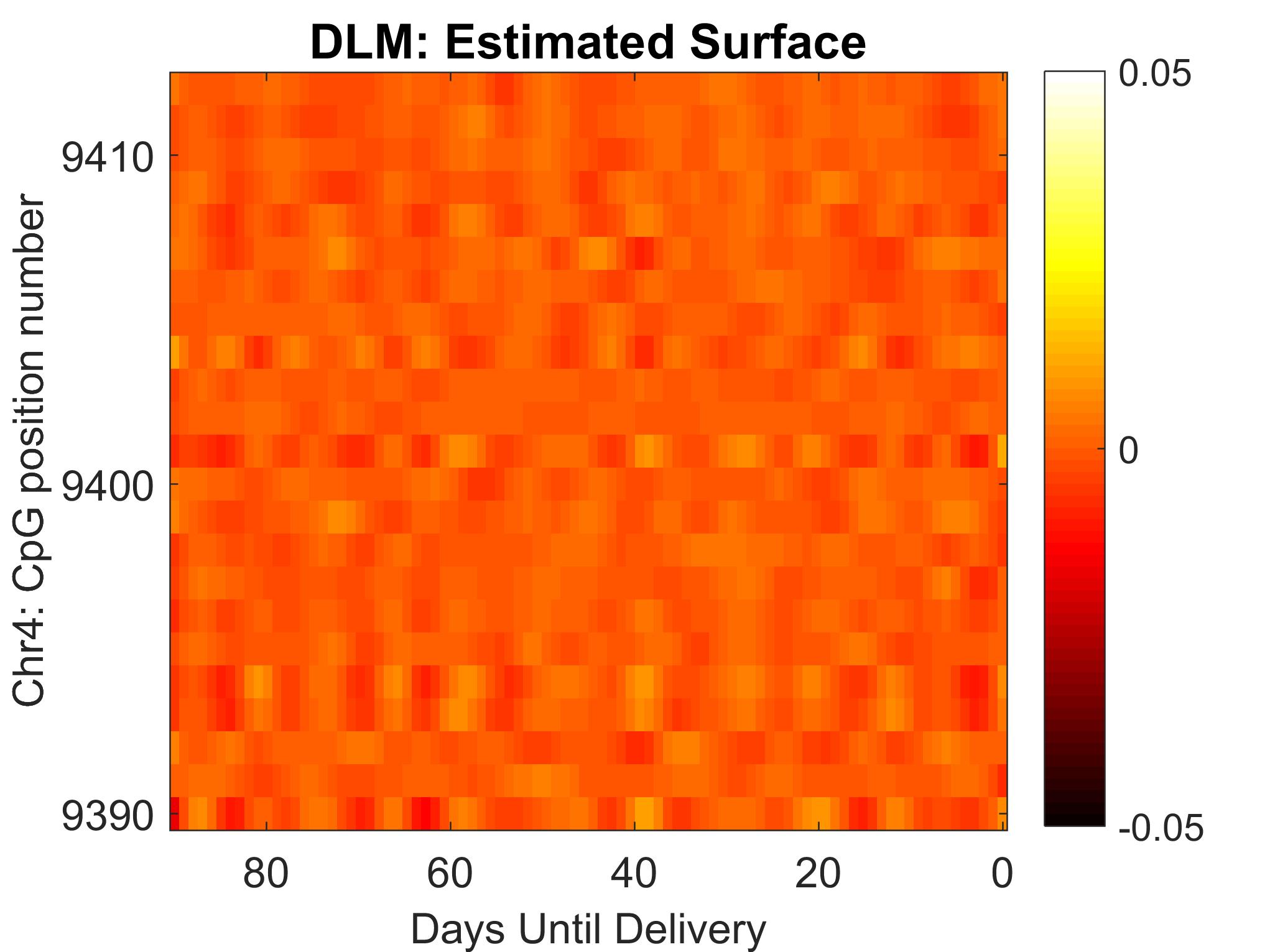} &
\includegraphics[width=0.5\linewidth]{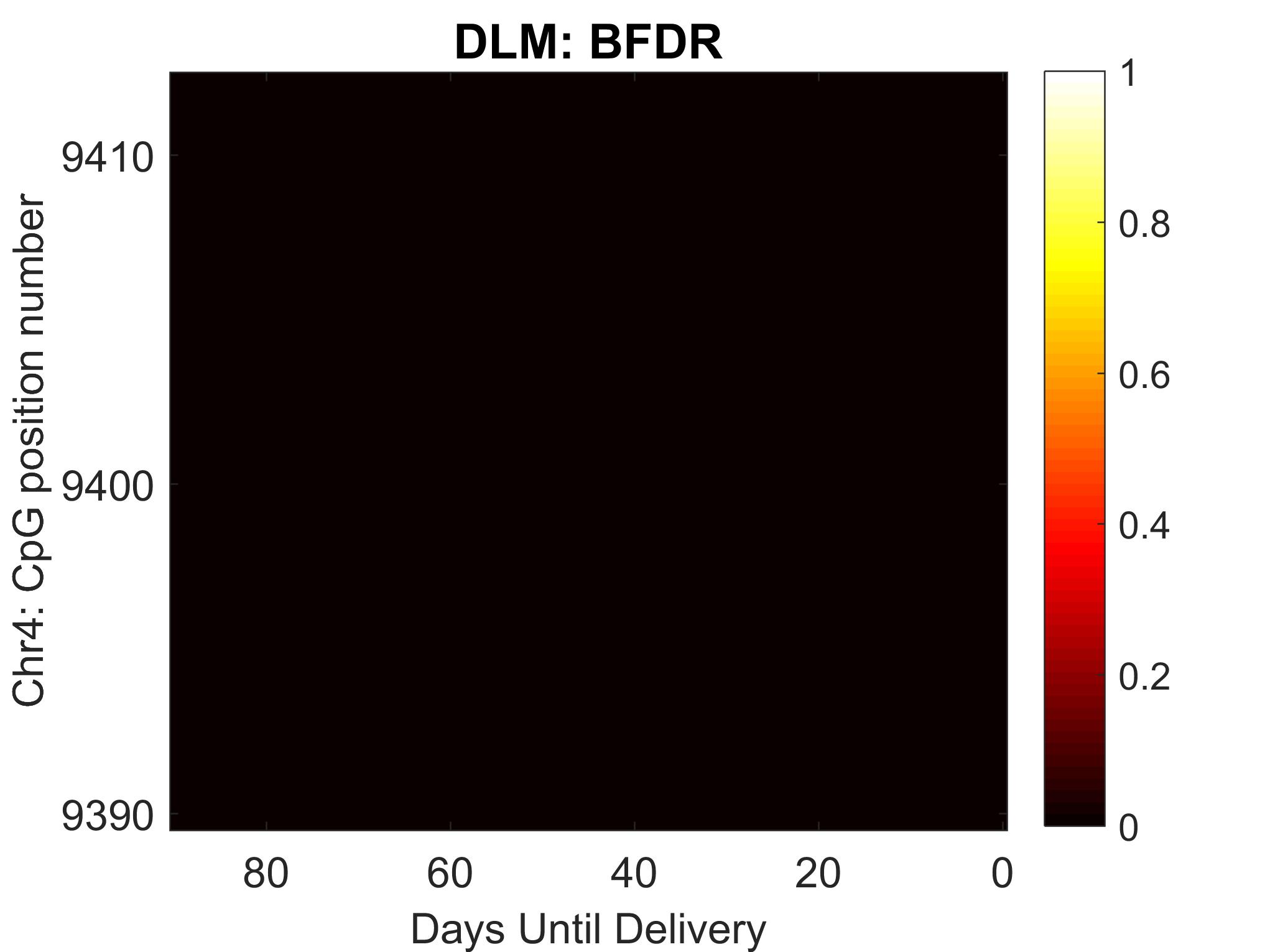}
\end{tabular}		
\caption{FFR analysis (top panel) and DLM analysis (bottom panel) for a region on chromosome 4 encompassing 23 CpG probes annotated to the \emph{FAM13A} gene. Position numbers for each probe correspond to their position relative to other CpG probes on chromosome 4. Right panel: BFDR results with $\delta = 0.01$ and $\alpha = 0.05$.}
\label{fig:chr4}
\end{figure}

\begin{figure}
\begin{tabular}{cc}
\includegraphics[width=0.5\linewidth]{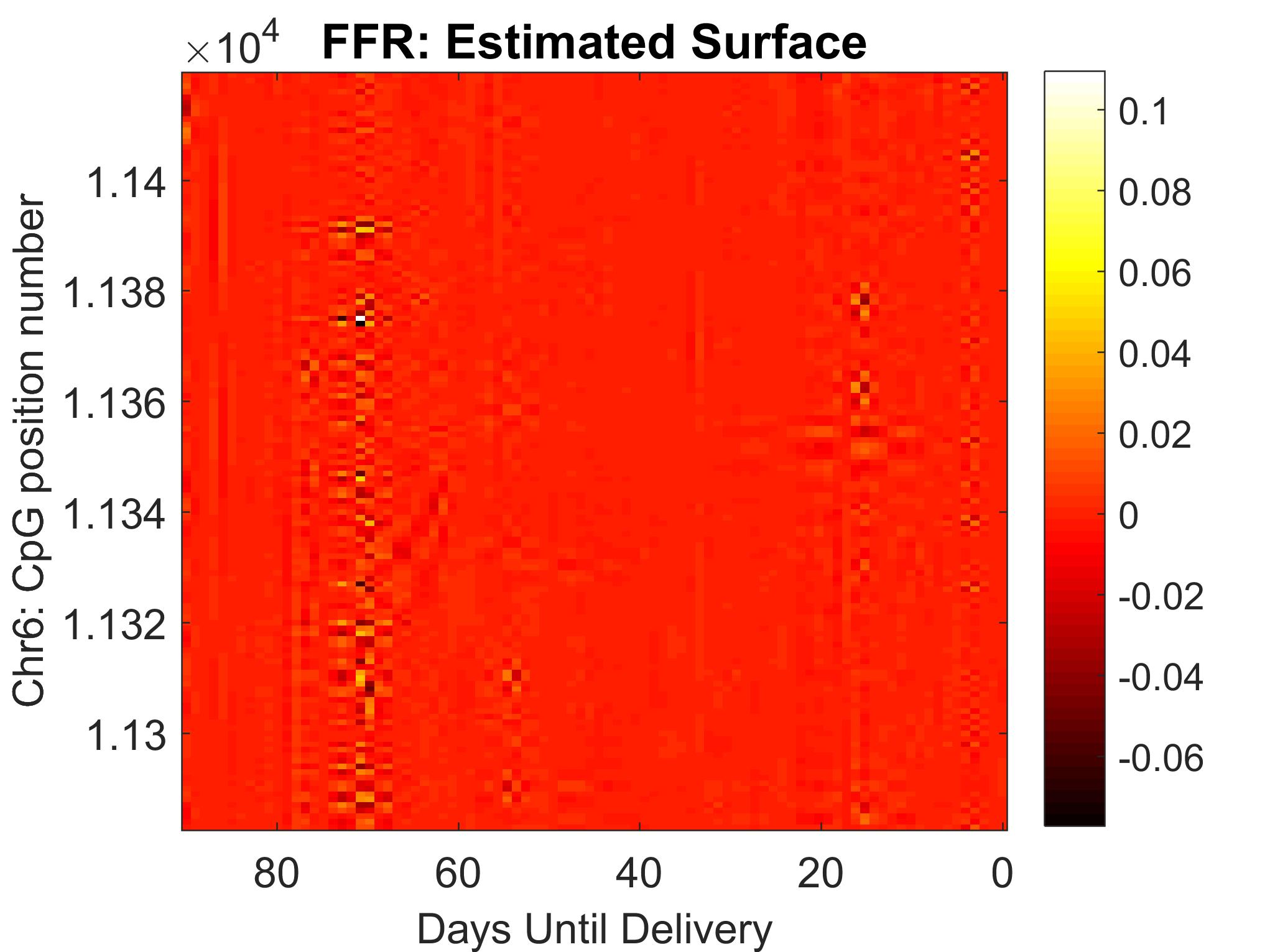} &
\includegraphics[width=0.5\linewidth]{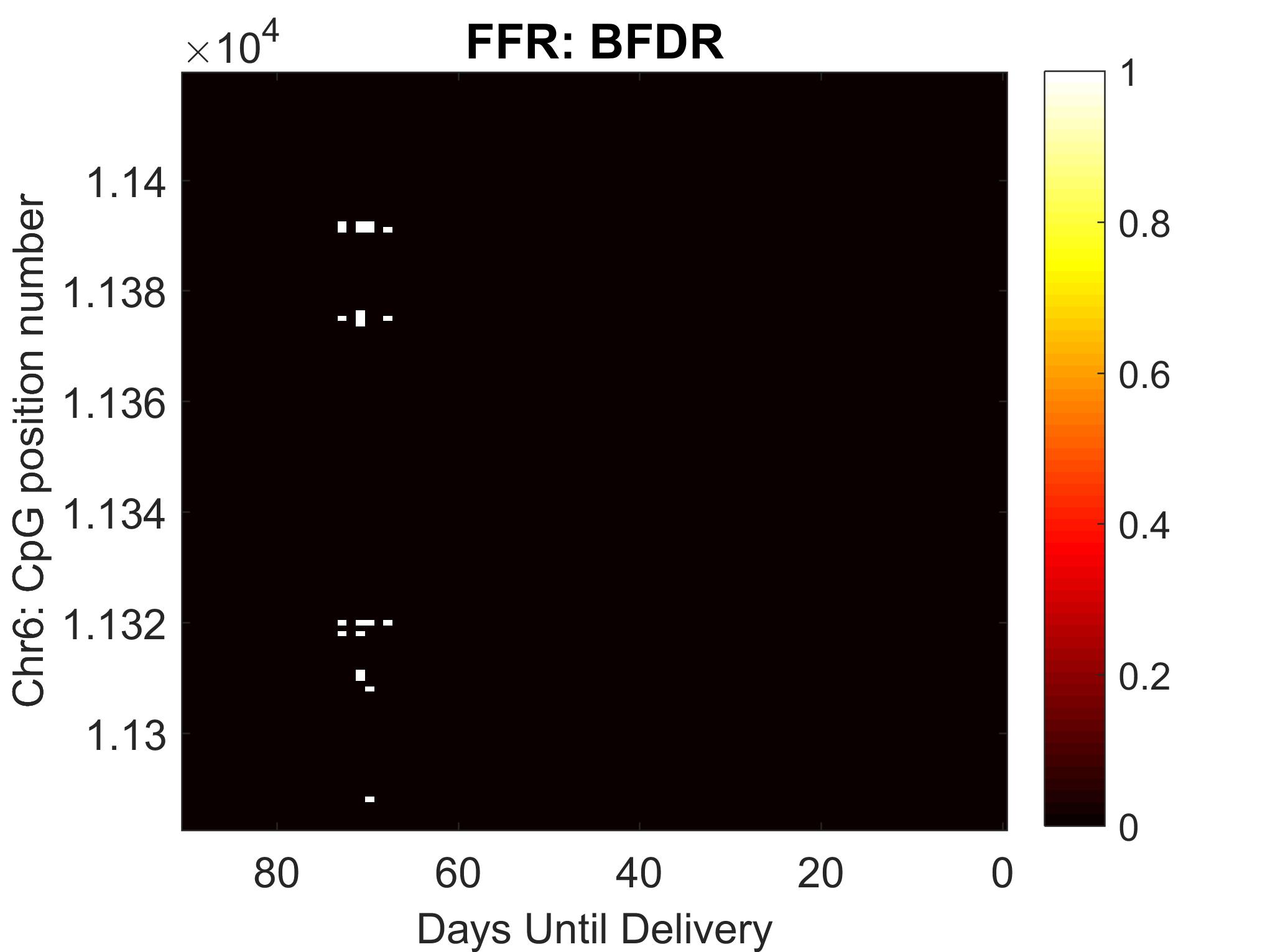} \\
\includegraphics[width=0.5\linewidth]{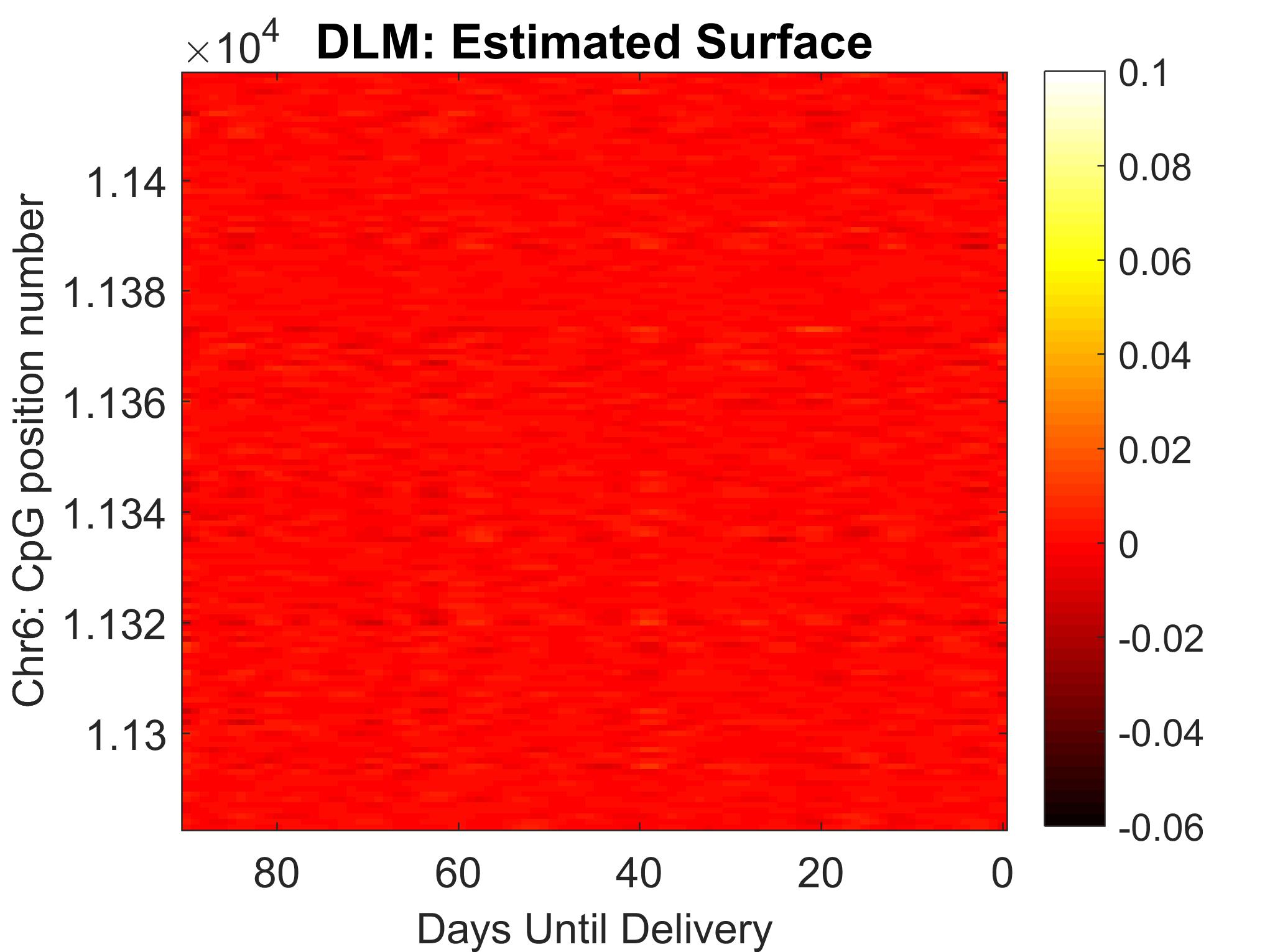} &
\includegraphics[width=0.5\linewidth]{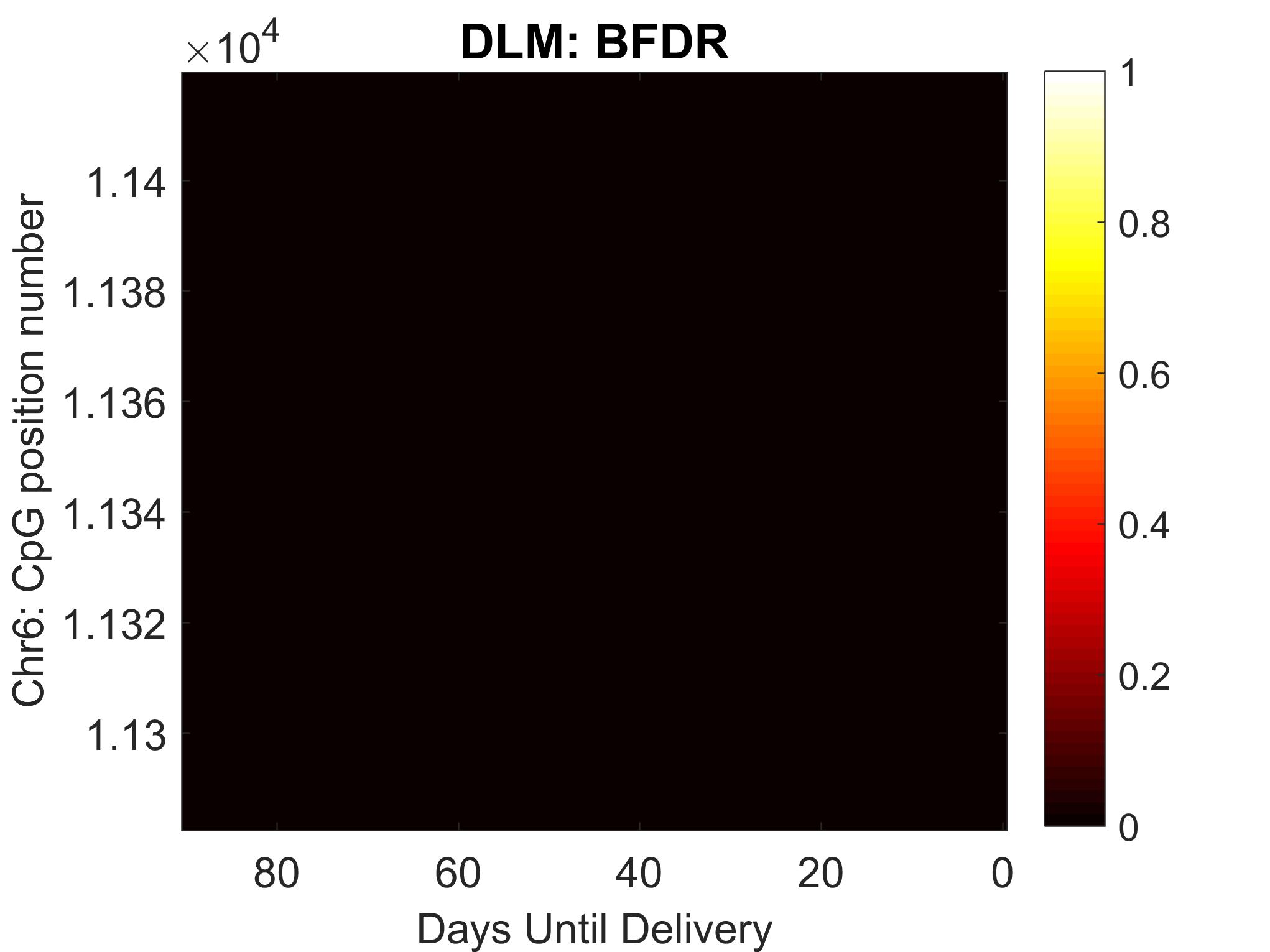}
\end{tabular}		
	\caption{FFR analysis (top) and DLM analysis (bottom) for a region on chromosome 6 encompassing 137 CpG probes annotated to the \emph{NOTCH4} gene. Position numbers for each probe correspond to their position relative to other CpG probes on chromosome 6. Right panel: BFDR results with $\delta = 0.01$ and $\alpha = 0.05$.}
 \label{fig:chr6}
 \end{figure}

As discussed in Section \ref{DNAmethylation}, we applied the FFR method to two regions encompassing CpG sites previously identified by \citealp{Gruzieva2019} where DNA methylation levels in cord blood were both significantly associated with PM exposure as well as implicated in respiratory-related outcomes: \emph{FAM13A} and \emph{NOTCH4}. Figure \ref{fig:chr6} shows the FFR- and DLM-estimated association surface for the 23 CpG probes annotated to \emph{FAM13A} on chromosome 4. These probes span 348,819 base pairs and are labeled on the heat map according to their position relative to other CpG probes on chromosome 4 (e.g. the first CpG probe on chromosome 4 corresponds to position 1, the next CpG probe corresponds to position 2, etc.). We set $\alpha = 0.05$ as the global BFDR bound and used $\delta = 0.01$ as the minimum practical effect size in the BFDR calculation. The areas flagged as significant in the BFDR analysis correspond to probes in \emph{FAM13A} coding regions and fall near the beginning of the third trimester, 78-79 days before delivery, as well as halfway through the third trimester for four probes (CpG positions 9409 - 9412 corresponding to cg17769793, cg06884401, cg25779483, cg04536922). The CpG identified by \citealp{Gruzieva2019}, cg00905156, corresponds to position 9402 in Figure  \ref{fig:chr4}. No windows of susceptibility were detected for this probe. When we performed the analogous analysis using a DLM approach, the BFDR procedure did not flag any areas of the surface. 

Little is currently known about what constitutes a biologically meaningful change in methylation level, but small changes in DNA methylation in some genomic regions have been shown to have a strong effect on transcriptional activity (\citealp{Breton2017});  we note that in sensitivity analyses with levels of $\delta > 0.03$, all of the flagged regions flagged in Figure \ref{fig:chr4} disappear. The region did not appear significant when we used the SimBaS procedure to control the experiment-wise error rate. 

Figure \ref{fig:chr6} shows the FFR- and DLM-estimated association surfaces for the 137 CpG probes annotated to \emph{NOTCH4} on chromosome 6. These probes span 28,306 base pairs and are labeled according to their position relative to other CpG probes on chromosome 6. For this region, FFR finds a band of association across much of the NOTCH4 gene 67-72 days before delivery. The CpG identified by \citealp{Gruzieva2019}, cg06849931, corresponds to position 11295 in Figure  \ref{fig:chr6} and does not exhibit a window of susceptibility on the BFDR heat map. The corresponding BFDR image for the DLM approach does not flag any regions as significant, suggesting that the power gains observed in the simulation study manifested in the analysis of this genomic region as well.

\subsection{Discussion}
Functional regression is a powerful method for analyzing and visualizing associations between different sources of functional data. While a number of methods already exist for identifying differentially methylated regions and a separate body of literature addresses identifying windows of susceptibility, here we accomplish both objectives within a unified modeling framework. By enabling us to analyze associations between a spatially-varying outcome function and a time-varying exposure, functional regression provides a means of identifying differentially methylated regions exhibiting windows of susceptibility to exposures measured at a fine temporal resolution. More generally, the flexible framework presented here could be useful in a variety of high-throughput genomic applications pertaining to the transcriptome, epigenome, or metabolome. In our setting, exposure was indexed by time, but this need not be the case. 

In simulation, we demonstrated that by jointly modeling epigenetic sites, FFR had greater power to identify regions associated with windows of susceptibility than the DLM approach that models sites independently. The FFR approach also maintained high sensitivity and low FDR under all but the lowest STNR scenario. In very low signal settings, the FFR lost power due to shrinkage across sites within the region of interest. This shrinkage also rendered the FFR approach less effective than the DLM at identifying windows of susceptibility when signal was confined to a single site.  Overall, the vertical and horizontal band simulation studies suggest that FFR is more effective at identifying sustained temporal effects across a genomic region, whereas a DLM is more effective at pinpointing sustained temporal effects at a spatially-localized site.  In the Project Viva data analysis, FFR showed a greater ability to highlight differentially methylated regions associated with PM$_{2.5}$ exposure during the last trimester of pregnancy than did the DLM. 

Because both sustained temporal effects across genomic regions and sustained temporal effects at individual sites could be biologically significant, we recommend running both DLM and FFR analyses in a staged analytic plan. Similar to the way in which site-by-site epigenome-wide association studies are often followed by region-finding methods like DMRcate or Bumphunter, when interest focuses on finding windows of susceptibility to an exposure, we suggest performing a site-by-site DLM analysis followed by the multivariate FFR approach.

There are several limitations of our work. First, the PM$_{2.5}$ daily measurements we used were estimated based on where the mothers enrolled in Project Viva lived, rather than by direct personal monitoring.  We do not account for the prediction error from the air pollution exposure model in our inferential procedures. Second, we used wavelet basis functions in our FFR implementation since simulations showed that these worked well for the simulated surfaces that we used, but it is possible that wavelets are not the optimal basis expansion. Just as we adapted the FFR implementation of Meyer et al. (2015) to our application by removing PC compression of the exposure data in the wavelet space, additional changes to the basis expansion could be made. Future work could explore whether a different basis expansion is preferable for modeling methylation profiles and air pollution exposures.
An additional limitation is that we restricted our investigation to CpG probes annotated to two genes. Our aim here was to perform the first analysis of prenatal windows of susceptibility driving the two most noteworthy associations reported by \citealp{Gruzieva2019}. Future work will involve a more comprehensive exploration of the genome and additional air pollutants. Identifying additional windows of susceptibility can direct attention to specific biologic mechanisms underlying associations and ultimately inform interventions to improve children's health. This work suggests function-on-function regression is a valuable tool to achieve these objectives.

\section*{Acknowledgements}
This work was supported by NIH grants ES007142-36, ES028811, UH3 OD023286, ES000002, and US EPA grant RD-83587201. Its contents are solely the responsibility of the grantee and do not necessarily represent the official views of the USEPA. Further, USEPA does not endorse the purchase of any commercial products or services mentioned in the publication. 
\bibliography{documentrefs}
\appendix 
\section*{Appendices}
\section{Additional Simulation Results }

\begin{figure}[H]
\begin{tabular}{ccc}
	\includegraphics[width=0.33\linewidth]{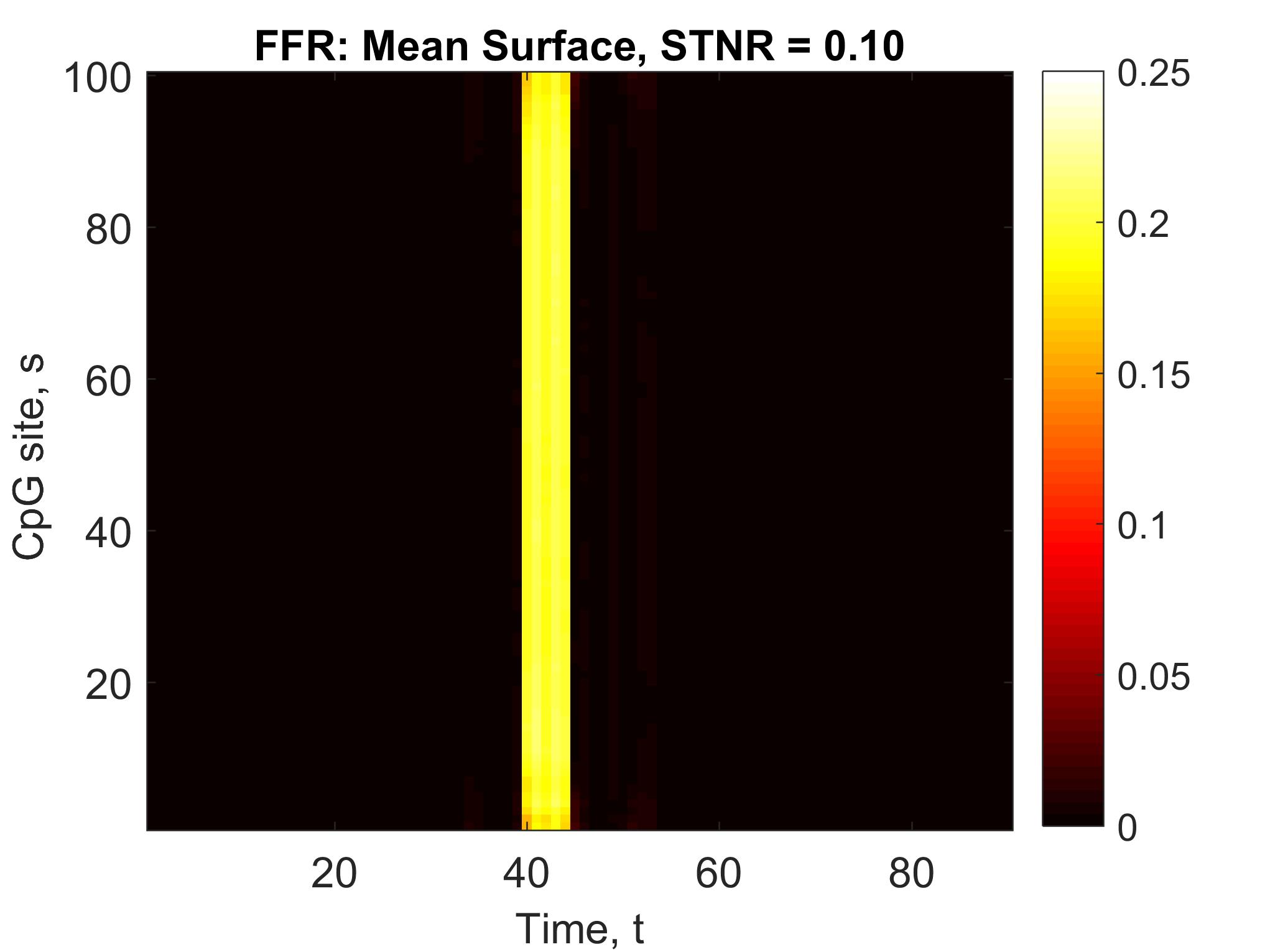} &
	\includegraphics[width=0.33\linewidth]{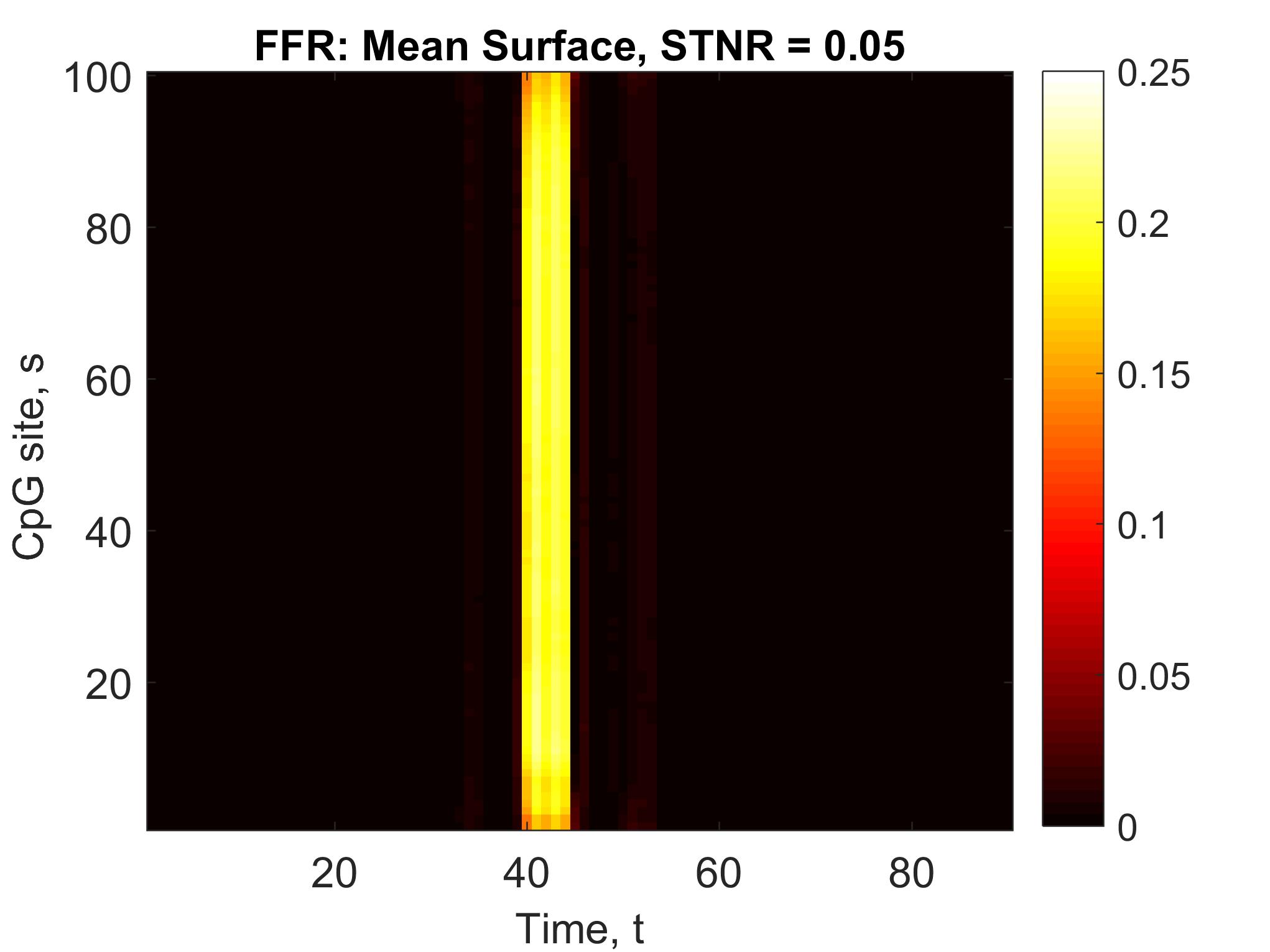} &
	\includegraphics[width=0.33\linewidth]{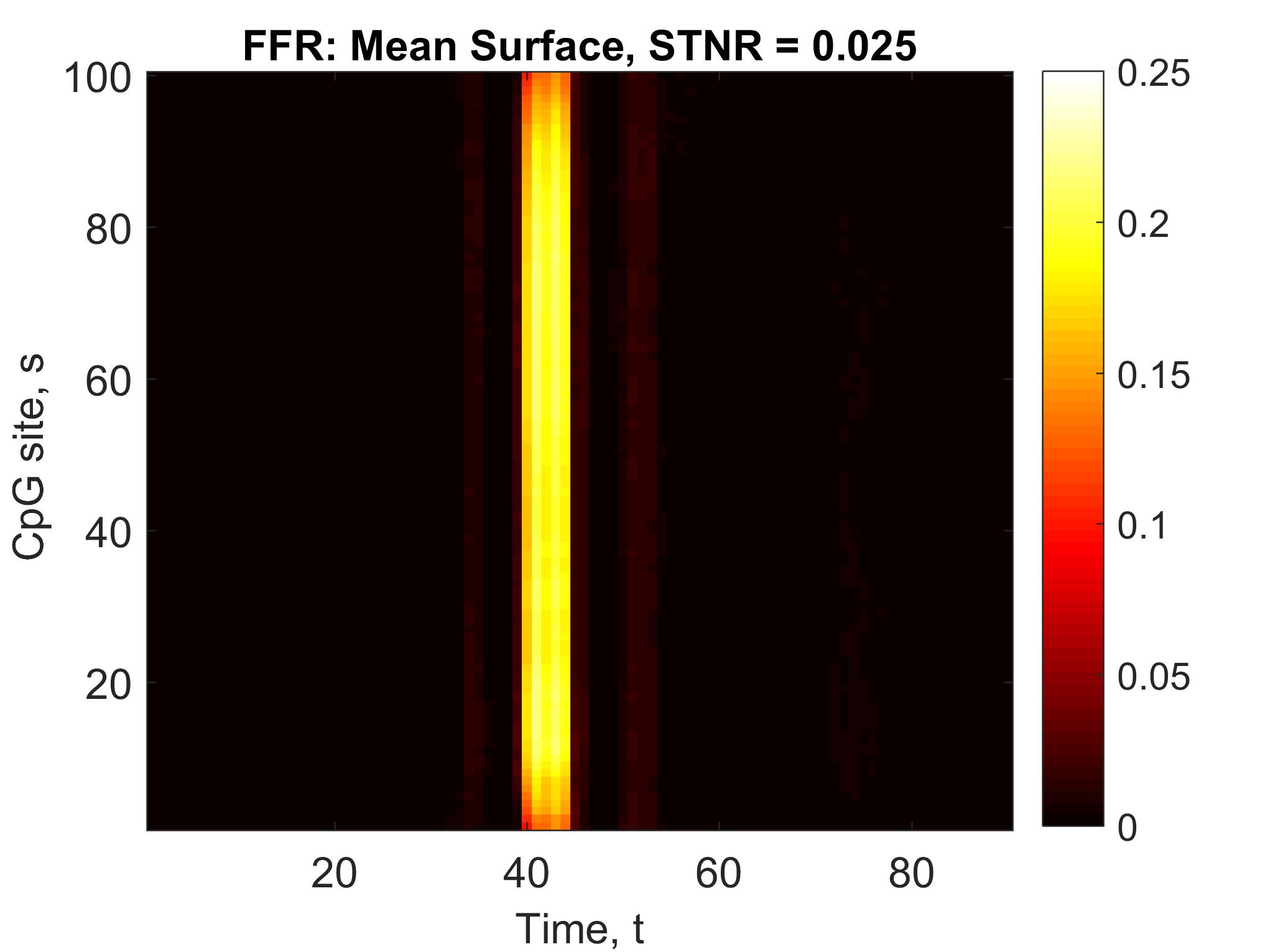} \\
	\includegraphics[width=0.33\linewidth]{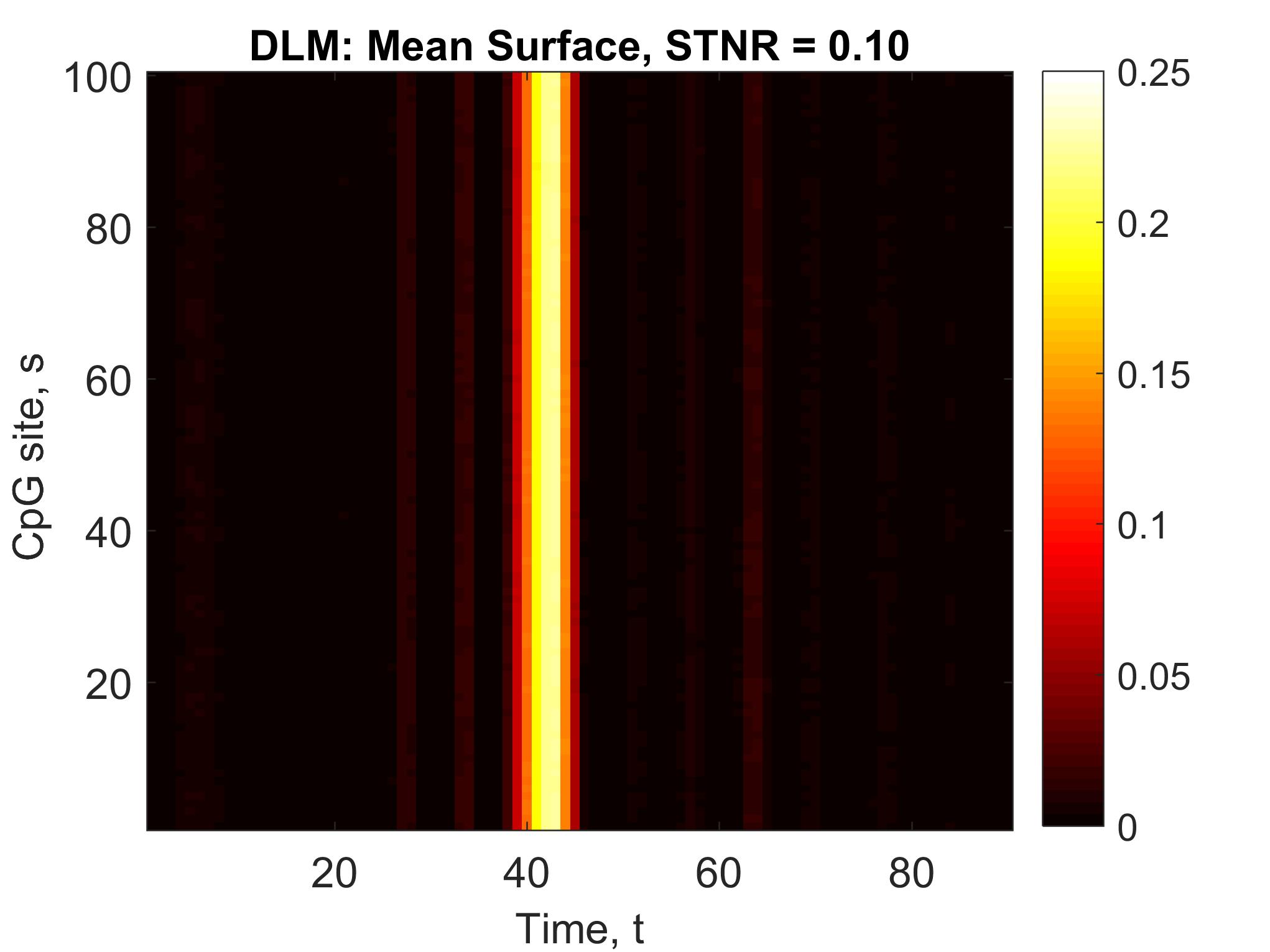} &
	\includegraphics[width=0.33\linewidth]{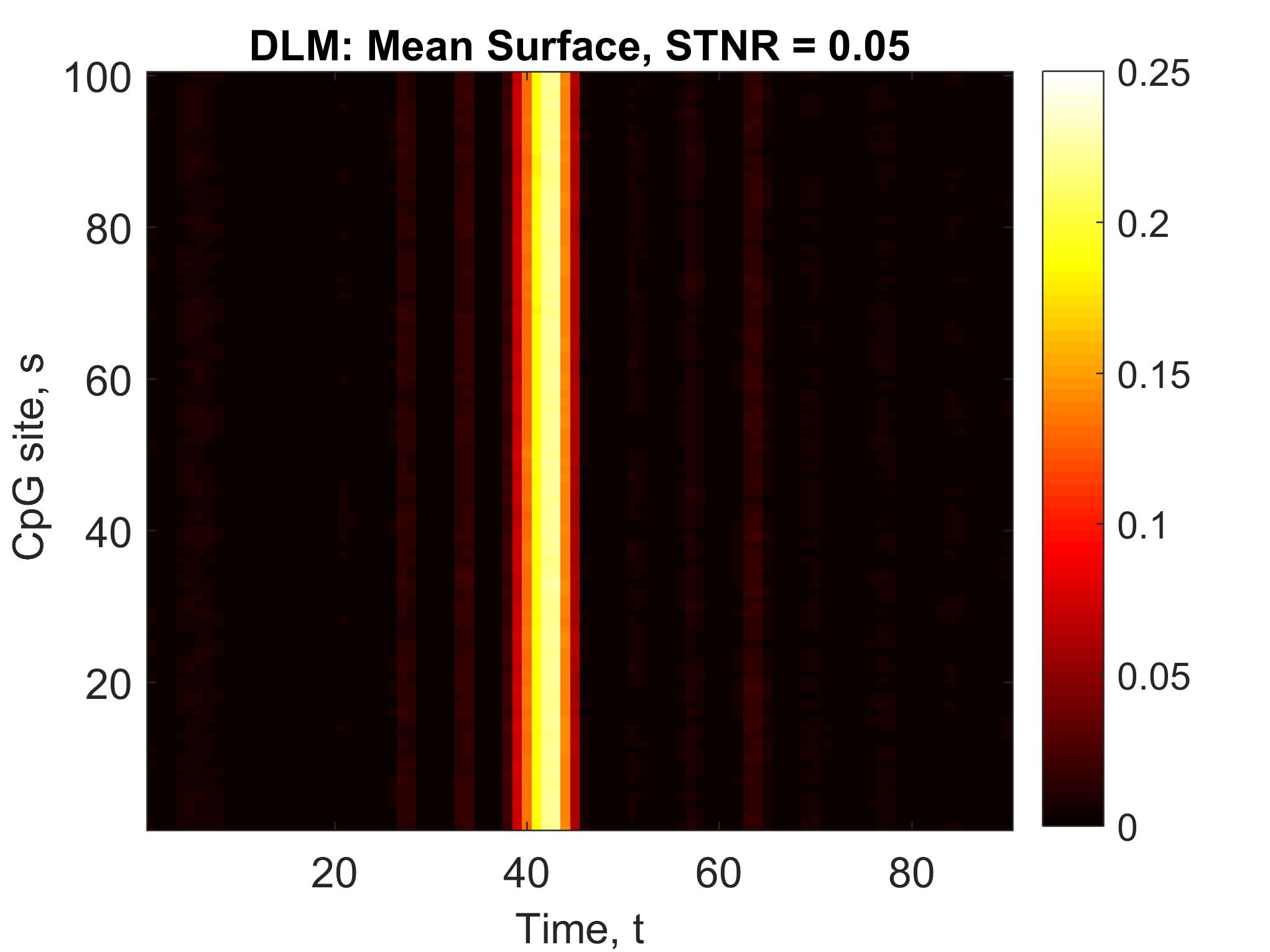} &
	\includegraphics[width=0.33\linewidth]{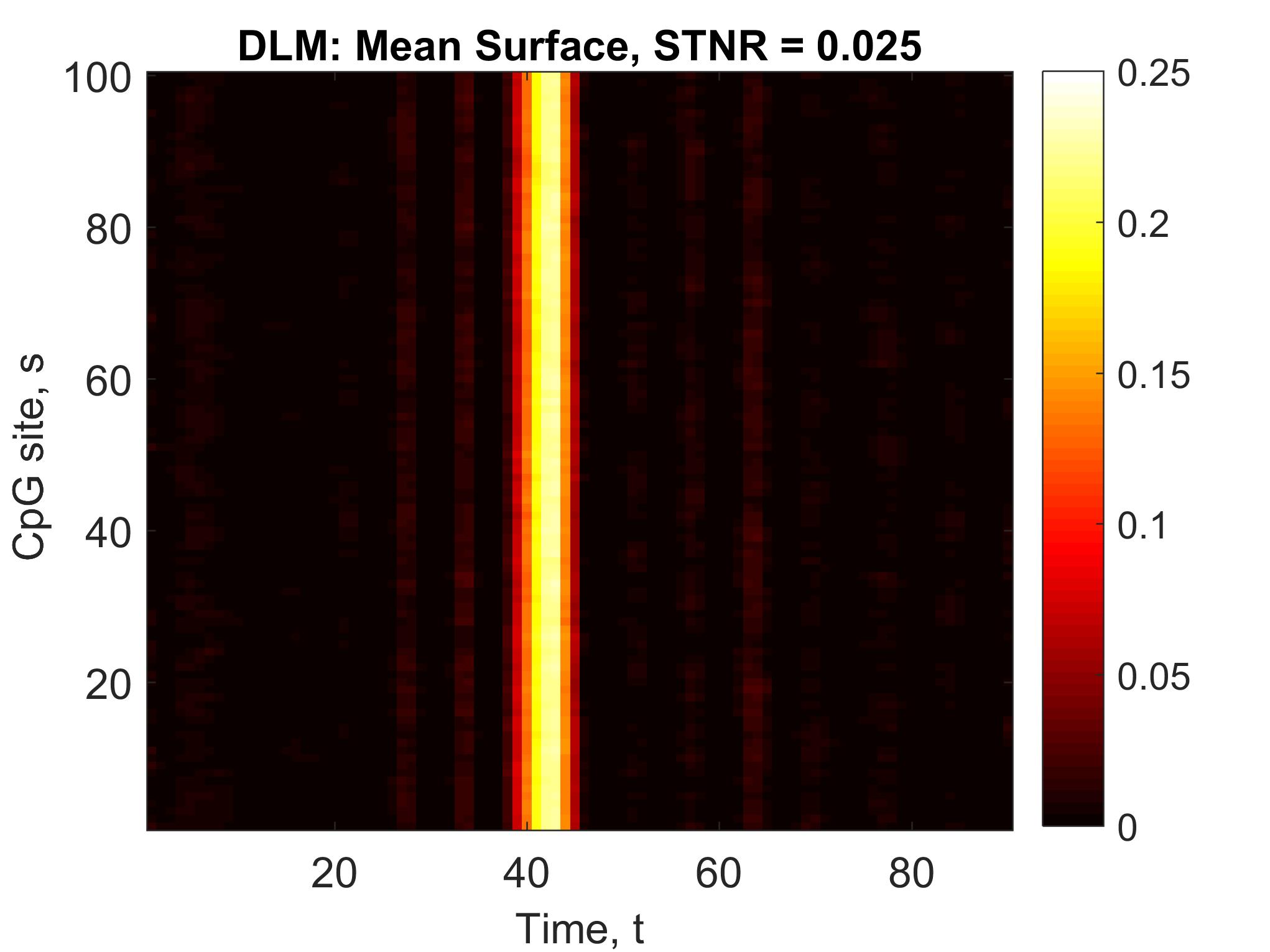}
\end{tabular}		
\caption{Heat maps of the estimated surface averaged over 100 datasets. Top panel: FFR estimates. Bottom panel: DLM estimates concatenated across the surface. }
\label{fig:avgEstimate}
\end{figure}

\section{PM$_{2.5}$ exposure}\label{app}
\begin{figure}[H]
    \centering
    \includegraphics[width=1.0\linewidth]{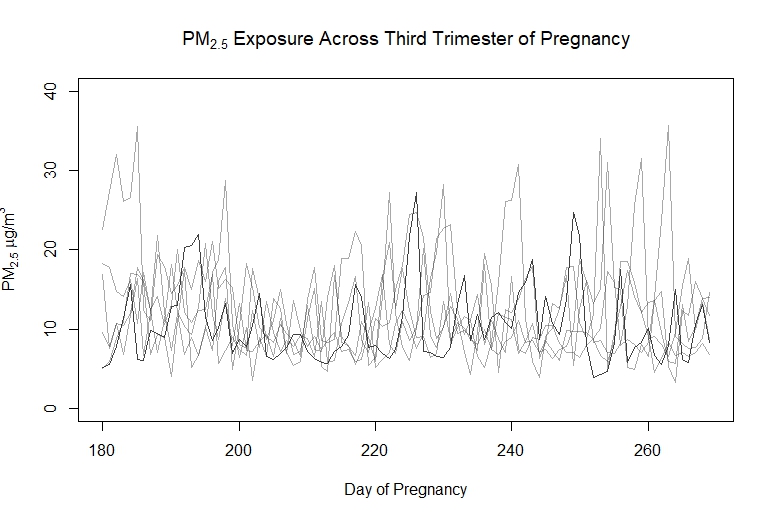}
    \caption{PM$_{2.5}$ exposure over the third trimester of pregnancy for a randomly selected subject in Project Viva (black) and five additional subjects (grey).} 
    \label{fig:PM}
\end{figure}

\end{document}